\documentclass[11pt,twoside]{book} 
\usepackage[hyperindex,breaklinks]{hyperref}
\usepackage{asp2008s}
\usepackage{times}
\usepackage{lscape}
\usepackage{epsf}

\usepackage{graphicx}
\usepackage{amssymb}
\usepackage{longtable}
\usepackage[figuresright]{rotating}
\usepackage{multirow}

\newcommand{\simless}{\mathbin{\lower 3pt\hbox {$\rlap{\raise 5pt\hbox{$\char'074$}}\mathchar"7218$}}}

\mainmatter

\newlength{\deftabcolsep}
\begin{document}

\title{Young Nearby Loose Associations }   %%% Fill in title
\author{Carlos A. O. Torres, Germano R. Quast}   %%% Fill in author names
\affil{Laborat\'orio Nacional de Astrof\'{i}sica/ MCT \\
Rua Estados Unidos 154, 37504-364 Itajub\'a (MG), Brazil}
 %%% Fill in author affiliations

\author{Claudio H. F. Melo, Michael F. Sterzik}   %%% Fill in author names
\affil{European Southern Observatory, Casilla 19001, Santiago 19, Chile}    %%% Fill in author affiliations

\begin{abstract}
A significant population of stars
with ages younger than the Pleiades exists in the solar neighborhood.
They are grouped in loose young associations, sharing similar
kinematical and physical properties, but, due to their vicinity to the
Sun, they are dispersed in the sky, and hard to identify.  Their
strong stellar coronal activity, causing enhanced X-ray emission,
allows them to be identified as counterparts of X-ray sources.  The
analysis presented here is based mainly on the SACY project, aimed to
survey in a systematic way counterparts of ROSAT all-sky X-ray sources
in the Southern Hemisphere for which proper motions are known.  We
give the definition, main properties, and lists of high-probability
members of nine confirmed loose young associations that do not belong
directly to the well-known Oph-Sco-Cen complex.  The youth and
vicinity of many members of these new associations make them ideal
targets for follow-up studies, specifically geared towards the
understanding of planetary system formation.  Searches for very
low-mass and brown dwarf companions are ongoing, and it will be
promising to search for planetary companions with next generation
instruments.

\end{abstract}

%%% MAIN BODY OF TEXT GOES HERE. CONSULT "INSTRUCTIONS FOR AUTHORS USING
%%% LATEX2E MARKUP", SECTIONS 2.3-2.6 FOR HELP WITH EQUATIONS, FIGURES,
%%% AND TABLES.

%\section{}   %%% Top level section head (remove "%" symbol)
%\subsection{}   %%% Second level section head (remove "%" symbol)
%\subsubsection{}   %%% Lowest level section head (remove "%" symbol)
%\section*{}	%%% Unnumbered top level section head (remove "%" symbol)
%\subsection*{}   %%% Unnumbered second level section head (remove "%" symbol)
\section{Introduction}
\subsection{Overview}
In a seminal paper, \cite{Herbig1978} proposed the existence of post-T~Tauri stars (pTTS)
and established a list of candidates. pTTS follow classical and weak line
T~Tauri stars (cTTS, wTTS) in an evolutionary sequence.
Although they were expected to outnumber the TTS, their discovery or
identification remained difficult for a long time.
Their relation with another category of young stars -- the isolated TTS (young low mass stars
found spatially far away from any apparent dark or parental molecular cloud) --
remained an open problem.
Interestingly, Herbig's list of pTTS and the one of isolated TTS
\citep{Quast1987} were actually quite similar.
The most intriguing example was TW~Hya, a high Galactic latitude
TTS \citep{RucinskiKrautter1983} at a distance of at least 13$^\circ$ from the nearest  dark clouds
and located in a region characterized by the absence of any cloudlets from which it could have originated.

A systematic search for more isolated TTS was pursued with the  optical spectroscopic
Pico dos Dias Survey (PDS) among  optical counterparts of the  IRAS Point Source Catalog
\citep{gregorio92, Torres1995, Torres1999}.
One of the first results of the PDS was the discovery of four additional TTS
around TW~Hya  \citep{delareza89, gregorio92}.
They concluded that this group was likely a very young association relatively close to the sun.

Only very few good candidates for isolated TTS were found within the PDS.
But while the IR-excess selection criterion  effectively finds young stellar objects
embedded in their placental material or with circumstellar disks,
it fails to signal older objects whose disks have already been dissipated.
Therefore, most stars  with ages between $\ga10-70$~Myr (i.e. wTTS and pTTS)
escaped discovery by this method.

Due to the enhanced X-ray activity of young stars \citep{Walter86}, more efficient selection
criteria for post and isolated TTS candidates were developed.
The high sensitivity and full sky coverage of the ROSAT all-sky survey
\citep{truemper} revealed thousands of new X-ray sources projected in the
direction of nearby star forming regions \citep{Guillout98}.
Ground-based spectroscopic follow-up studies showed that a large fraction
of these X-ray sources were indeed wTTS  together with older pTTS and ZAMS
stars \citep[e.g.,][]{Alcala00}.
Surprisingly, many of the newly found weak-line TTS were {\it not} obviously
connected to any molecular cloud region,
raising again many questions about their origin \citep{Sterzik95}.

Based on the similarity of the ROSAT X-ray fluxes, radial velocities, astrometry (Hipparcos)
and spectroscopic characteristics of the four stars around TW~Hya, \cite{Kastner97}
confirmed that these stars formed a physical association with TW~Hya,
about 20~Myr old and at a distance of 40 to 60 pc from Earth, which they called
the TW~Hya Association.

Immediately after the existence of the TW~Hya Association was confirmed,
several research groups became interested in this association and other members
were found (see Section~4).
Two main approaches were explored, a first one
aiming to study the properties of its members, whereas other
groups started to look for similar nearby young associations
hidden among the ROSAT X-ray sources.

In 2000, as a result of this effort to find new associations, two new adjacent
and similar associations were proposed,
in Tucana \citep{zuckerman00} and in Horologium \citep{torres00}.
To examine the physical relation between them and to search for other associations,
we started the SACY (Search for Associations Containing Young stars) survey \citep{torres06}.
The SACY sample contains stars:\\
{\it (i)} later than G0, in order to be able to use the Li $\lambda$6707 line as a youth indicator \citep{martin97};\\
{\it (ii)} belonging to the TYCHO-2 or Hipparcos catalogs in order to have access to proper motions;\\
{\it (iii)} that are candidate  optical counterparts of sources from the ROSAT All-Sky
Bright Source Catalogue.

\citet{torres06} presented a catalog of 1626 spectroscopically
observed stars in the Southern Hemisphere.  In
Figure{~\ref{fig:thpiz}} we show the celestial distribution of the
observed SACY sample, with additional stars observed from 2006 to
2008.  The SACY survey is now complete in the Southern Hemisphere (but
for four stars) and the updated catalog has 2093 stars.  The survey
enables us to define properties and membership probabilities for stars
in these putative associations.  Preliminary results appear in
\citet{torres03a, torres03}.  In \citet{torres06}, the prototypical
methodology and analysis of the $\beta$~Pic Association is presented
and in forthcoming papers (in preparation) a more detailed analysis of
other associations found will be given.  The Lithium abundances of the
nine associations presented in this chapter are studied in
\citet{silva08} where they present the lists of the association
members.

A similar survey is being pursued in the Northern Hemisphere and the
first results are appearing now \citep{guillout08}.
Only a very low frequency of
stars younger than the Pleiades is identified, more than one order of magnitude less than in the SACY survey.
This strong hemispheric anisotropy could be explained, at least partially, by
taking into account the different survey biases and completeness limits.
Anyway, with only five young stars at this moment, the Northern survey does not help to find young associations.

\subsection{Method of Analysis}

An association is a group of stars appearing {\it concentrated}
together in a small volume in space sharing some common properties
such as age, chemical composition, distance and kinematics\footnote{
In this sense we prefer to use the term association and not moving
group.}.  However, if such a group is close enough to the Sun, its
members will appear to cover a large extent in the sky (as an example,
Orion at 50~pc would cover almost the whole sky).  Thus, to find a
group, projected spatial concentrations (i.e., in terms of right
ascension and declination only) and proper motions may not be enough.
A better criterion is to look for objects sharing similar heliocentric
space motions (UVW) all around the sky (U positive towards the
Galactic center, V positive in the direction of Galactic rotation).

\citet{torres06} describe the convergence method developed
to search for members of an association and a corresponding membership probability model in detail.
Both convergence method and probability model examine the stars in the hexa-dimensional space,
UVWXYZ, as defined by the space motions relative to the Sun  and the physical space coordinates centered
on the Sun (XYZ, in the same directions as UVW).
We represent with m$_v$, M$_v$ and M$_{v,iso}$ the apparent visual magnitude,
the resultant absolute magnitude with the distance obtained from the convergence
method,  and the absolute magnitude
given by  the adopted isochrone for the $(V-I)_C$ stellar color;
and   $\mu_\alpha$, $\mu_\delta$ and V$_r$ are the proper motions and the radial velocity.
Briefly explained, if there is no reliable
trigonometric distance available\footnote{We consider the trigonometric
parallaxes as unreliable if they have errors larger than 2~mas and we do not use them.},
the convergence method finds the distance (d) for each star in the sample
that minimizes the F value of Equation~\ref{eq:f}.
The first term is a photometric distance modulus and the second is a kinematical one.
The method needs, as input, an assumed age and initial velocity values $(U_0,
V_0, W_0)$ for the proposed association,
and a cutoff value for F above which stars should be considered spurious.
This cutoff value varies for each association but usually we begin with 3.5
(this approximately means 0.7 magnitudes
for the distance modulus and 3~km~s$^{-1}$ for the velocity modulus).
The method is iterative, and for each iteration a list of stars with new  $(U_0, V_0, W_0)$ is obtained.
The process ends when the list of stars and the velocities $(U_0, V_0, W_0)$ do
not change significantly.
\begin{eqnarray}
\label{eq:f}
\nonumber \lefteqn{F(m_v,\mu_\alpha,\mu_\delta,V_r;d) =}\\
&[p\times(M_v-M_{v,iso})^2+(U-U_0)^2+(V-V_0)^2+(W-W_0)^2]^{1/2}
\end{eqnarray}
where $p$ is a constant weighting the importance of the evolutionary
distance with respect to the kinematic distance.
Actually we use $p>0$ only for the Oct Association, since it has no stars
with trigonometric parallax. This means that in general our distances are only
kinematical.

The  list of candidates serves as a training set for the probability model
\cite[k--NN model,][]{Sterzik95}.
In this model we define around each star of the entire sample 6-dimensional
spheres that contain a certain number $k$ of stars.
A membership probability is then defined
by the proportion of stars in these spheres that belong to the training set.
The probabilities depend on the compactness of the association and the field density.
Thus, for each association we can define a cutoff probability
where we consider the stars as  probable members.
Using this list of (high) probability members, we return to the convergence
method until both lists match.
Finally, a possible kinematical member becomes an actual good candidate if its
Li content is compatible with the Li depletion for its age \citep{neu97}.

As all the nearby loose associations proposed at this moment
can be defined through their properties using SACY stars,  membership
probabilities for  stars suggested elsewhere can be calculated whenever
their basic kinematic data are available.

\begin{figure}[!ht]
\centering
\includegraphics[draft=False,width=0.49\textwidth]{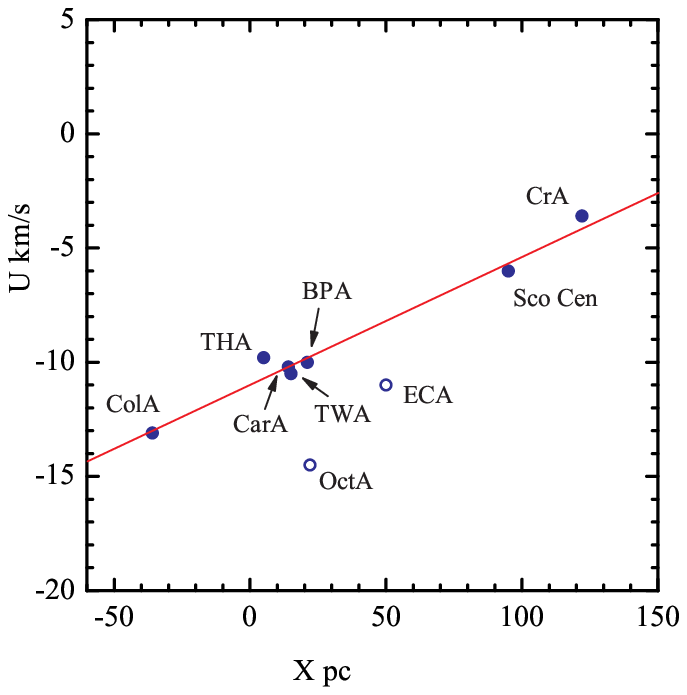}
\includegraphics[draft=False,width=0.49\textwidth]{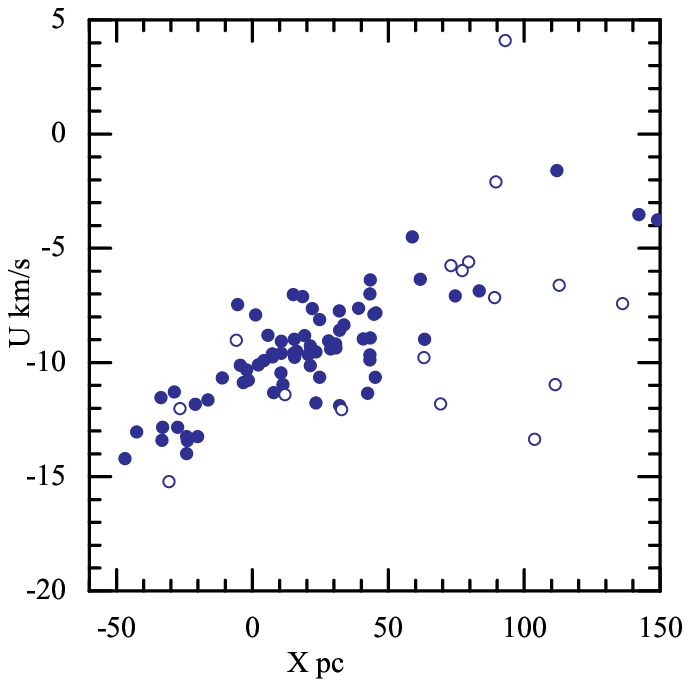}
\caption{{\it Left:} The expansion in X direction for the young nearby associations, including the Sco-Cen
and the R CrA (CrA) associations using the SACY data.
We could not find expansion for the $\epsilon$~Cha and the Carina associations, as
they are very compact, and any expansion of the Oct Association should be confirmed -- see Section~6.
The other abbreviations in the figure mean the associations:
Columba, Tucana-Horologium, TW~Hya, $\beta$~Pic, $\epsilon$~Cha and Octans.
{\it Right:} The expansion of all young stars in the SACY sample with reliable parallaxes.
Open circles are stars which are not members of any association.}
\label{fig:exp}
\end{figure}

When analyzing the entire SACY sample,
we discovered an unexpected kinematical phenomenon that seems to be a general
property of the young stars in the
solar neighborhood and affects the application of the convergence method
-- a positive correlation (r\,=\,0.99) between  U and X for stars younger than $\sim$30~Myr.
The correlation is seen in Figure~\ref{fig:exp} where, in the left panel,
we plot the mean values of U and X  derived from the SACY sample for these associations.
This correlation can be interpreted as a physical expansion in the X direction.

This is not an artifact from the convergence method.
The right panel of Figure~\ref{fig:exp} shows the observed U and X values
for all young SACY stars that have good quality Hipparcos parallaxes.
While a strong correlation (r\,=\,0.98) of U and X persists for those stars that belong to
young associations (filled circles), the correlation is only weak (r\,=\,0.46)
for stars that do not belong to any association.
Maybe one reason for the difference is the presence of not identified single lined spectroscopic
binaries.
The expansion is not only seen in the sample of nearby young associations as a
whole, but also within each of these associations, with a similar
spread of U and X values \citep{torres06}.
Some young associations have only a small extension in the X axis, and do not
allow to determine their expansion ($\epsilon$ Cha and Car associations).
The Oct Association does not seem to follow the behavior shown in
Figure~\ref{fig:exp}, perhaps due to its higher Galactic latitude (see Table~\ref{table:finalb}).
For associations older than $\sim$30~Myr
this expansion is not present any more (AB Dor and Argus associations; see
Table~\ref{table:final}).

A similar phenomenon has been reported for a few individual
associations, see for example \citet{mamajek05} and \citet{bobylev07},
but the global expansion must be considered first.  Unfortunately we
have no explanation for the presence of this expansion.  It might
reflect a more global motion like the Galactic arm epicycle movement
which is still conserved from a recent local star formation event, and
which has not yet been lost by higher dispersion acquired through
Galactic dynamics on a longer timescale.

Regardless of the cause of this expansion, it must be taken into account
when searching for kinematical young associations.
It can be represented by the relation of  Equation~\ref{eq:exp} that is
incorporated in the convergence method in the appropriate cases:

\begin{equation}
\label{eq:exp}
U = 0.05 (X) - U_0
\end{equation}

Reliable isochrones are essential as input to obtain the photometric distance modulus.
However, none of the observed sequences can be represented by the published
theoretical isochrones for these associations.
Thus, to test the star membership, ad-hoc
observational evolutionary sequences were used in the convergence method,
represented by third degree polynomials:\\
\\
For 5~Myr:
\begin{equation}
\label{eq:iso5}
M_v = 0.60 + 4.98(V-I)_C - 1.16(V-I)_C^2 +0.193(V-I)_C^3
\end{equation}
For 8~Myr:
\begin{equation}
\label{eq:iso8}
M_v = 1.20 + 4.98(V-I)_C - 1.16(V-I)_C^2 +0.193(V-I)_C^3
\end{equation}
For 10~Myr:
\begin{equation}
\label{eq:iso10}
M_v = 1.50 + 4.98(V-I)_C - 1.16(V-I)_C^2 + 0.193(V-I)_C^3
\end{equation}
For 30~Myr:
\begin{equation}
\label{eq:iso30}
M_v = 1.18 + 6.28(V-I)_C - 1.68(V-I)_C^2 + 0.248(V-I)_C^3
\end{equation}
For 70~Myr:
\begin{equation}
\label{eq:iso70}
M_v = 0.64 + 7.14(V-I)_C - 2.05(V-I)_C^2 + 0.314(V-I)_C^3
\end{equation}
valid in the color interval $-0.1 < (V-I)_C < 3.1$
\\
\\
These heuristic {\sl absolute} ages obtained partially in comparison  with
pre-main sequence models \citep{siess00} must obviously be taken with caution.
Nevertheless, the {\sl relative} ages  are real: for example,
the TW~Hya Association is older than the  $\epsilon$~Cha Association and younger
than the $\beta$~Pic Association.

\subsection{General Results}

Almost half of the young stars in the SACY sample belong
to the large Oph-Sco-Cen Association (see the chapters by Wilking et al. and Preibisch \& Mamajek).
In this chapter we discuss  nine nearby young loose associations,
which are kinematically well defined, but do not belong directly to the Oph-Sco-Cen Complex.
Their main characteristics are summarized in Tables~\ref{table:final} and \ref{table:finalb}.
Their distribution in the sky  can be seen in  polar projection
in  Figures{~\ref{fig:thpiz} to ~\ref{fig:abpiz}}.

\begin{table}[!ht]
\caption{Heliocentric space motions and expansion of the nearby associations}
\smallskip
\begin{center}
{
\label{table:final}
\begin{tabular}{lrrrcr
}
\tableline
\noalign{\smallskip}
Association &
U~~~~~~~&
V~~~~~~~&
W~~~~~~~&
Expansion&
N\\
 &[km s$^{-1}]$&[km s$^{-1}]$&[km s$^{-1}]$&&\\
\noalign{\smallskip}
\tableline
\noalign{\smallskip}
$\beta$~Pic   &-10.1$\pm$2.1&-15.9$\pm$0.8& -9.2$\pm$1.0&yes&48\\
Tuc-Hor&       -9.9$\pm$1.5&-20.9$\pm$0.8& -1.4$\pm$0.9&yes&44\\
Col    &       -13.2$\pm$1.3&-21.8$\pm$0.8& -5.9$\pm$1.2&yes&41\\
Car    &       -10.2$\pm$0.4&-23.0$\pm$0.8& -4.4$\pm$1.5&no?&23\\
TW~Hya           & -10.5$\pm$0.9&-18.0$\pm$1.5& -4.9$\pm$0.9&yes&22\\
$\epsilon$~Cha           & -11.0$\pm$1.2&-19.9$\pm$1.2&-10.4$\pm$1.6&no?&24\\
Oct          &  -14.5$\pm$0.9& -3.6$\pm$1.6& -11.2$\pm$1.4&no?&15\\
Argus          & -22.0$\pm$0.3&-14.4$\pm$1.3& -5.0$\pm$1.3&no&64\\
AB~Dor&        -6.8$\pm$1.3&-27.2$\pm$1.2&-13.3$\pm$1.6&no&89\\
\noalign{\smallskip}
\tableline
\end{tabular}
}
\end{center}
\end{table}

\begin{table}[!h]
\caption{Space distribution, mean distances and ages of the nearby associations}
\smallskip
\begin{center}
{
\label{table:finalb}
\begin{tabular}{l@{\hskip8pt}r@{\hskip8pt}r@{\hskip8pt}r@{\hskip8pt}r@{\hskip8pt}r@{\hskip8pt}r@{\hskip8pt}r@{\hskip8pt}r}
\tableline
\noalign{\smallskip}

Assoc. &
X~~~& %$\sigma$&
X Range& %$\sigma$&
Y~~~&
Y Range& %$\sigma$&
Z~~~&
Z Range&
D~~~~~& %$\sigma$&
Age
\\
&[pc]&[pc]~~~~&[pc]&[pc]~~~~&[pc]&[pc]~~~~&[pc]~~~~&[Myr]
\\
\noalign{\smallskip}
\tableline
\noalign{\smallskip}
$\beta$~Pic&20&  -32/76  &-5   &-33/21   &-15 &-29/-1&   31$\pm$21  &10\\
Tuc-Hor&  3 &    -61/43  &-24  &-47/-4   &-35 &-44/-30&  48$\pm$7~~~&30\\
Col    &-42 &    -106/9  &-56  &-168/1   &-47 &-99/6& 82$\pm$30  &30\\
Car    & 14 &     -2/33  &-94  &-154/-39 &-17 &-33/5&    85$\pm$35  &30\\
TW~Hya & 15 &      2/34  &-44  &-61/-26  &21  &10/27&    48$\pm$13  & 8\\
$\epsilon$~Cha&50& 34/60 &-92  &-105/-78 &-28 &-44/-12& 108$\pm$9~~~& 6\\
Oct&     22 &     -79/142&-106 &-138/-60 &-68 &-85/-38& 141$\pm$34  &20?\\
Argus&    5 &     -55/64 &-115 &-154/-6  &-18 &-67/8&  106$\pm$51  &40\\
AB~Dor&  -6 &     -94/73 &-14  &-131/58  &-20 &-66/23&   34$\pm$26  &70\\
\noalign{\smallskip}
\tableline
\end{tabular}
}
\end{center}
\end{table}

\begin{sidewaysfigure}
\label{fig:sacyxyz}
\centering
%\begin{center}
% \begin{tabular}{r}
% \resizebox{0.8\hsize}{!}{
\includegraphics[width=0.28\textwidth]{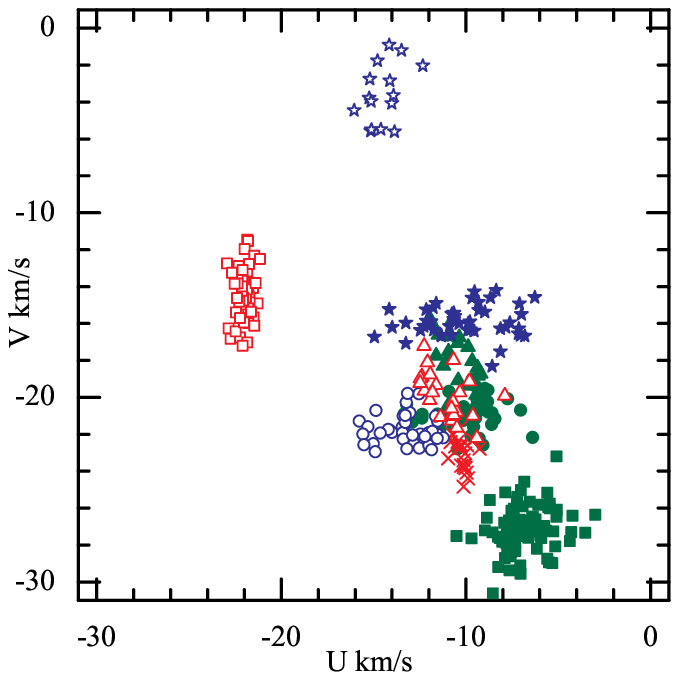}
\includegraphics[width=0.28\textwidth]{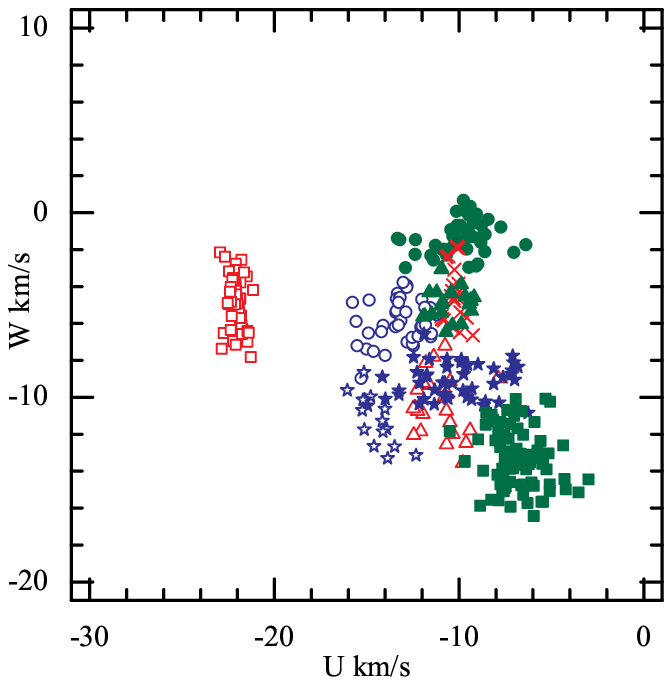}
\includegraphics[width=0.28\textwidth]{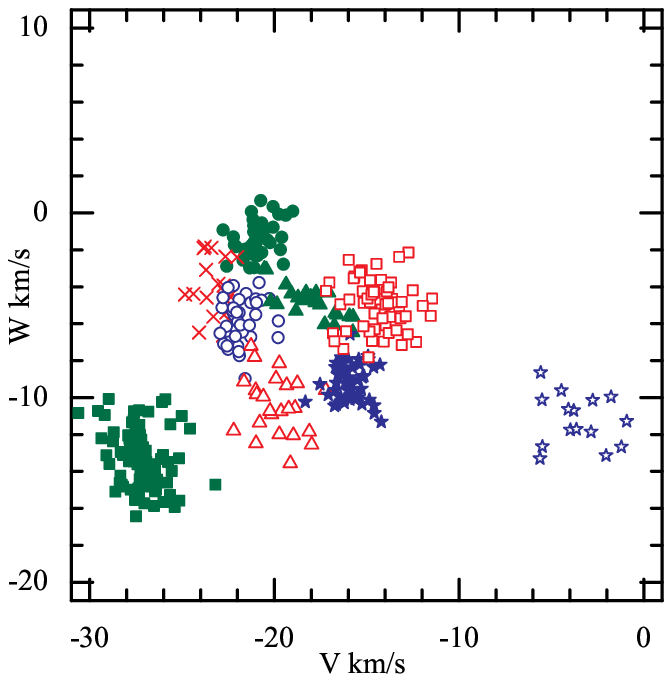}\\
%}\\
% \resizebox{0.8\hsize}{!}{
\includegraphics[width=0.28\textwidth]{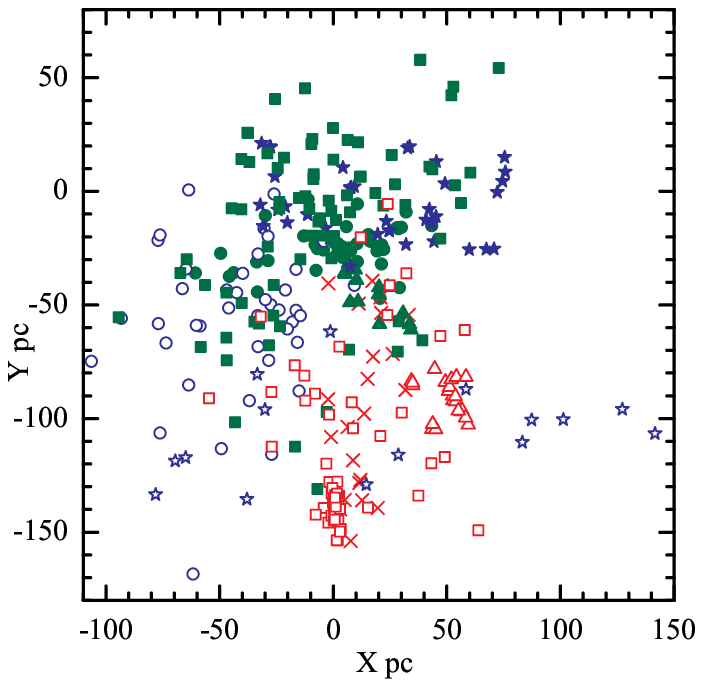}
\includegraphics[width=0.28\textwidth]{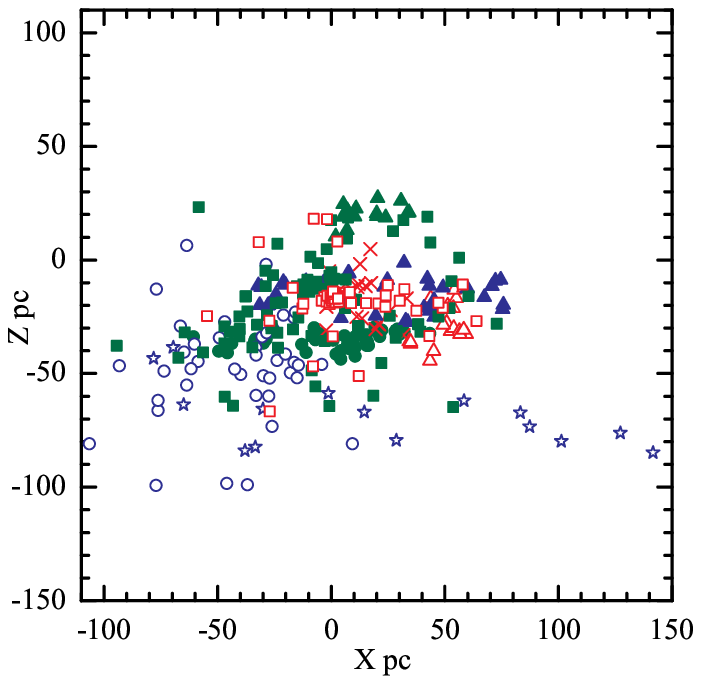}
\includegraphics[width=0.28\textwidth]{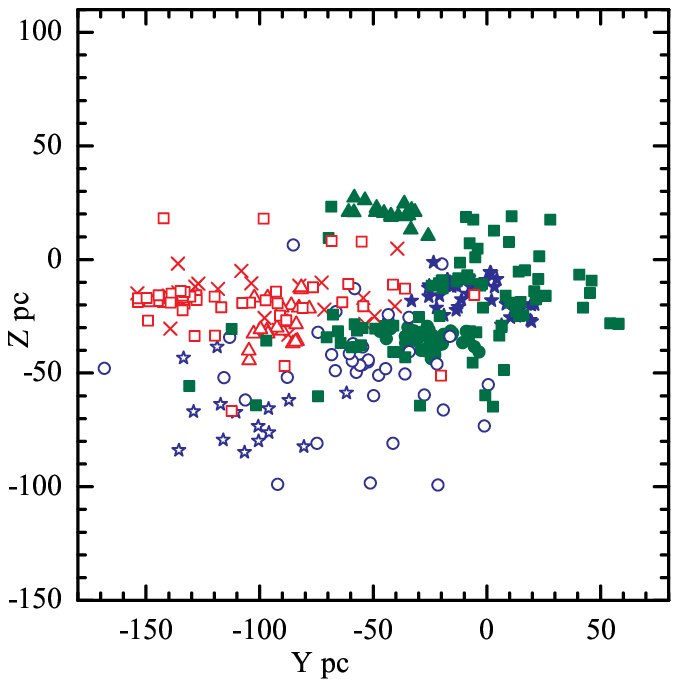}\\
%}\\
% \end{tabular}
%  \end{center}
\caption{Combinations of the sub-spaces of the UVWXYZ--space for the SACY Associations
showing the clusters in both kinematical and spatial coordinates.
Associations symbols: filled stars - $\beta$~Pic; filled circles - Tuc-Hor; open circles - Columba; crosses - Carina;
filled triangles - TW~Hya; open triangles - $\epsilon$~Cha; open stars - Octans; open squares - Argus;
filled squares - AB~Dor.}
\end{sidewaysfigure}

There are interpretations connecting the local young associations with
some more global stellar populations, like the Oph-Sco-Cen Complex.  A
particularly interesting paper is the one by \citet{fernandez08},
in which they propose that both the Sco-Cen Complex and the young local
associations originated by the impact of the spiral shock wave against
a giant molecular cloud.  To clarify the questions opened by these
kinds of models, it is fundamental to have a precise definition of
each association and a very good way to define their star memberships
and, also, their age.

A  distinct and independent way to obtain the age and the origin of  nearby
young stars is proposed by  \citet{makarov07}.
He traces back in time the Galactic orbits and evaluates near approaches in order to infer close conjunctions
and clustering in the past of the stars.
He concludes that the majority of nearby
young stars were formed during close passages or encounters of their natal
clouds with other cloud complexes today located at somewhat larger distances.
The method requires excellent kinematical data and depends
on the true membership of the proposed star for each association.
The ages of the Tuc-Hor and TW~Hya associations agree reasonably
well for both approaches.

These young associations have remarkably small velocity dispersions
($\approx$1~km per sec, see Table~\ref{table:final}). Their sizes are
larger than implied from their dispersion velocities and their ages,
but they are fully consistent with low-mass star forming regions and
OB associations.  Most of these associations have a non-spherical
distribution, and show distortions that seem age-dependent: the
younger ones are almost spherical ($\epsilon$~Cha and TW~Hya
associations), the ones with intermediate age are extended in X
direction ($\beta$~Pic, Oct, Tuc-Hor associations) while the older
ones in Y direction (Car, Col, Argus and AB~Dor associations).  If we
approximate them with spheres, they will be within a radius of 25~pc
for the $\epsilon$~Cha and the TW~Hya associations, 40~pc for the
Tuc-Hor Association, 60~pc for the $\beta$~Pic, the Car and the Col
associations, and 100~pc for the Oct, the Argus, and the AB~Dor
associations.

The important review on nearby young stars and their properties by
\citet{zuckerman04} presents five associations known at that time
($\beta$~Pic, Tuc-Hor, TW~Hya,\linebreak $\eta$~Cha and AB~Dor associations) and
introduces a new one, "Cha-Near" \footnote{It is very similar to the
$\epsilon$~Cha Association (that includes the $\eta$~Cha cluster)
defined here, see Section 5.}.  The focus of the present review is the
re-definition of their nearby associations and the definition of new
ones with the method described above and the significantly increased sample.
We emphasize that the strength of our method is the availability of
high quality radial velocities and proper motions together with
spectral information, which is essential to find the associations
presented here.  For that the SACY sample is of great importance to
better define these associations.  Nevertheless it is of almost no
help for the $\epsilon$~Cha association, which is at the limit of the
SACY sample possibilities, with most of the candidates coming from the
literature.  Therefore, in the next sections, we will present the high
probability members of these nine associations and their main
characteristics, and, when pertinent, we compare our new definitions
with those of \citet{zuckerman04}.

\begin{figure}[]
  \begin{center}
  \begin{tabular}{c}
 \resizebox{1.0\hsize}{!}{
\includegraphics[draft=False]{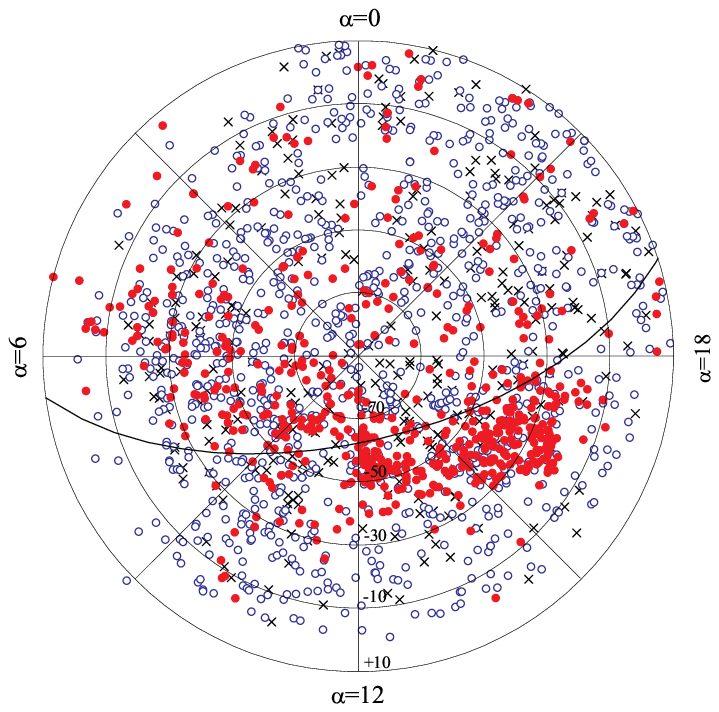}
\includegraphics[draft=False]{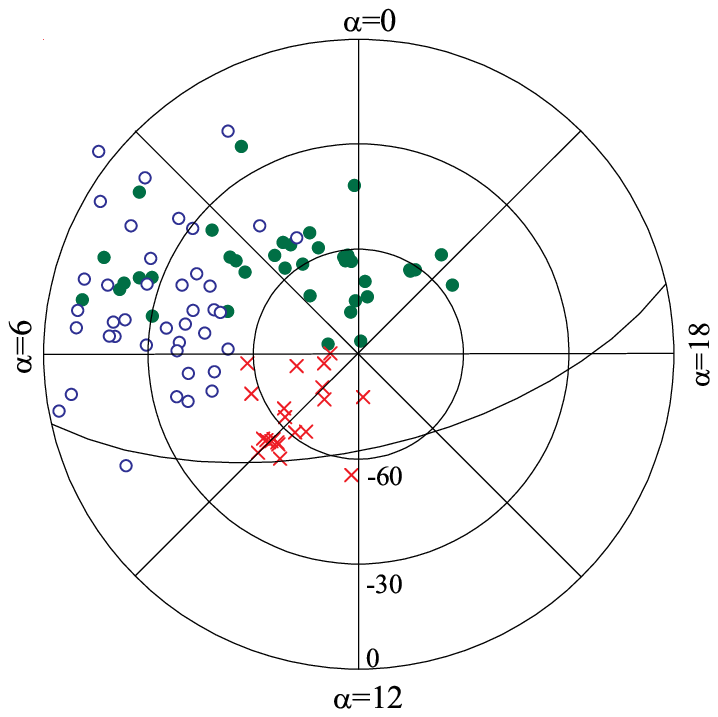}
}
   \end{tabular}
    \end{center}
\caption{{\it Left:} Celestial polar projection of the actual SACY observed sample.
This projection reaches out to +10$\deg$. Young stars are in filled circles.
Note the concentration at the Sco-Cen complex (from $\alpha$ 12 to 18).
Crosses are giant stars observed in the survey.
{\it Right:} Celestial polar projection of the associations in the GAYA complex (see
Section~3): the Tuc-Hor (filled circles),
the Col (open circles)  and the Car  (crosses) associations.
The transverse curve represents the Galactic plane.
}
\label{fig:thpiz}
\end{figure}

\begin{figure}[]
  \begin{center}
  \begin{tabular}{r}
\resizebox{1.0\hsize}{!}{
\includegraphics[draft=False]{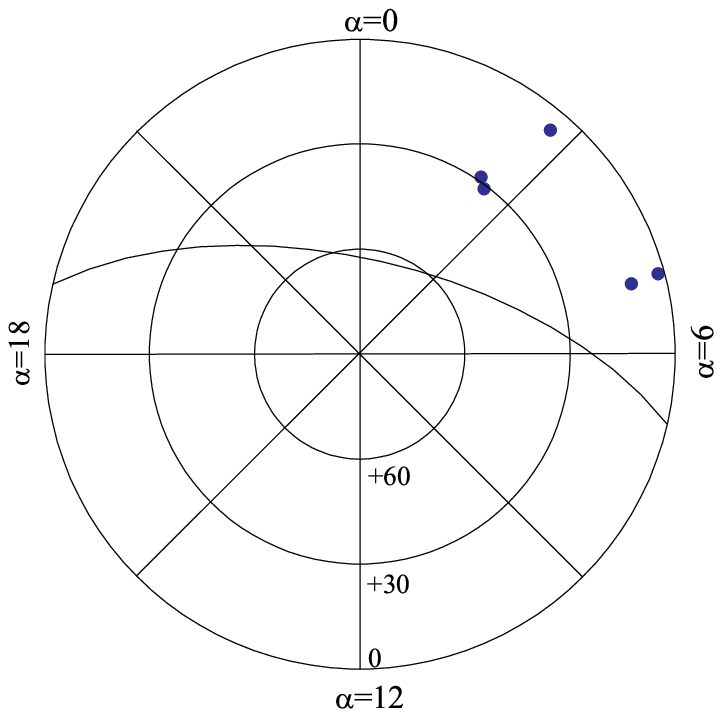}
\includegraphics[draft=False]{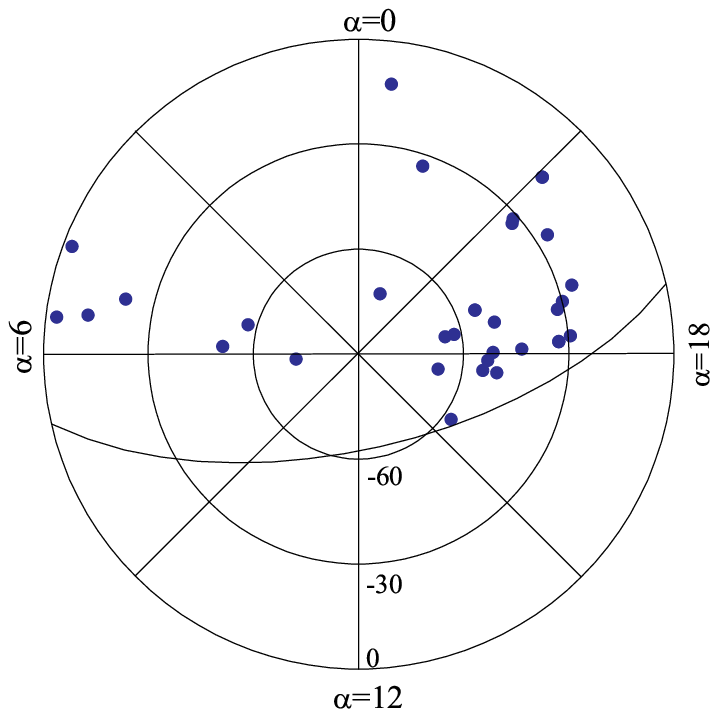}
}
   \end{tabular}
    \end{center}
\caption{Celestial polar projections of the $\beta$~Pic Association. {\it Left:}
Northern Hemisphere.
{\it Right:} Southern Hemisphere.
The transverse curve represents the Galactic plane.}
\label{fig:bppiz}
\end{figure}

\begin{figure}[!ht]
   \begin{center}
  \begin{tabular}{r}
\resizebox{1.0\hsize}{!}{
\includegraphics[draft=False]{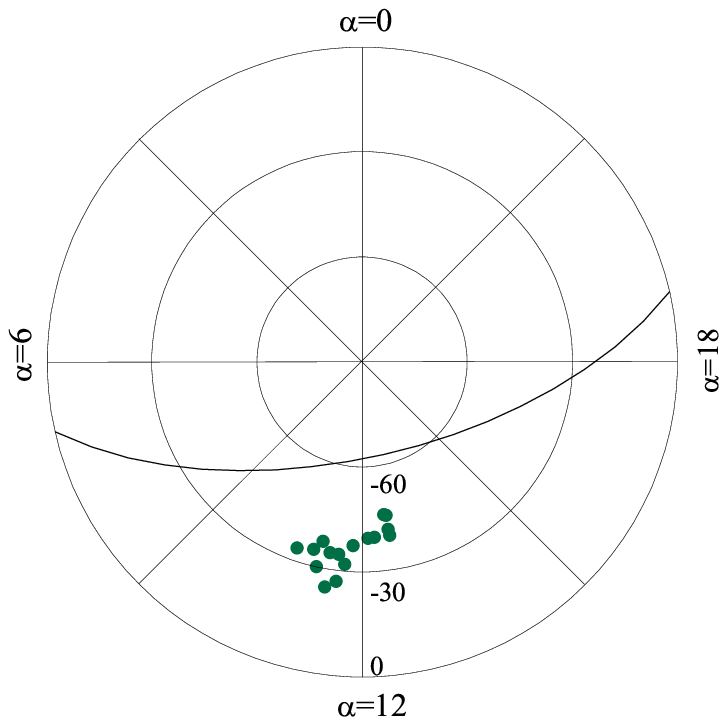}
\includegraphics[draft=False]{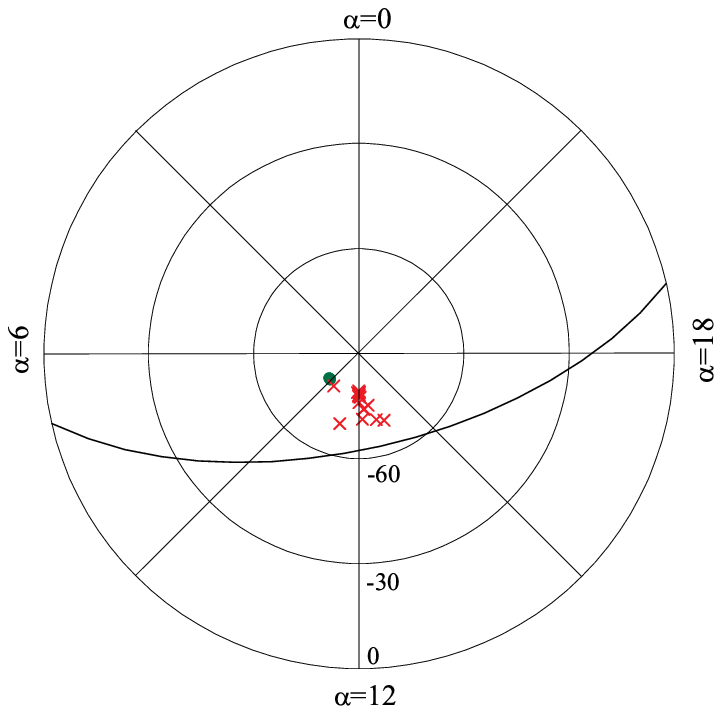}
}
    \end{tabular}
    \end{center}
\caption{{\it Left:} Celestial polar projection of the
TW~Hya association.
{\it Right:} Celestial polar projection of the  $\epsilon$~Cha Association -- the
crosses represent the field  $\epsilon$~Cha Association members
and the filled circles the $\eta$ Cha cluster members.
The transverse curve represents the Galactic plane.
}
\label{fig:echpiz}
\end{figure}

\begin{figure}[]
  \begin{center}
  \begin{tabular}{r}
\resizebox{1.0\hsize}{!}{
\includegraphics[draft=False]{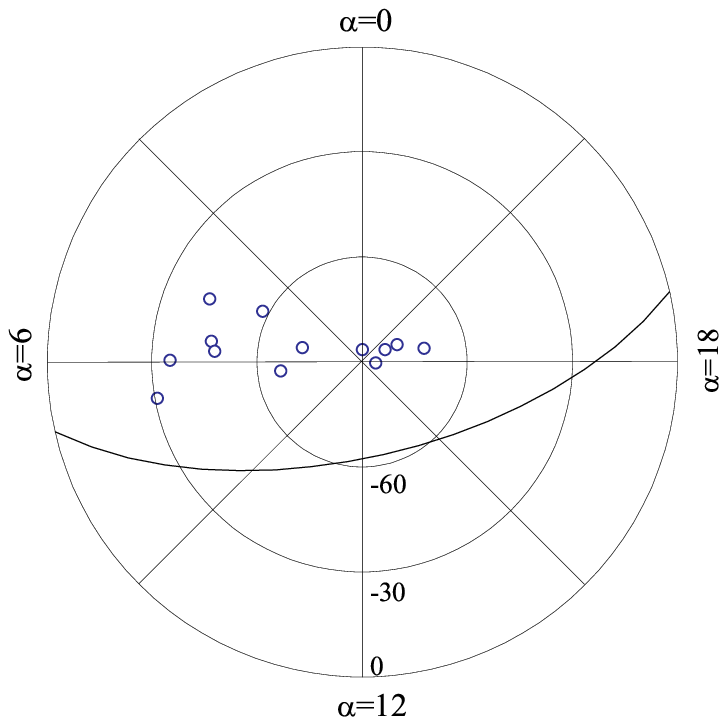}
\includegraphics[draft=False]{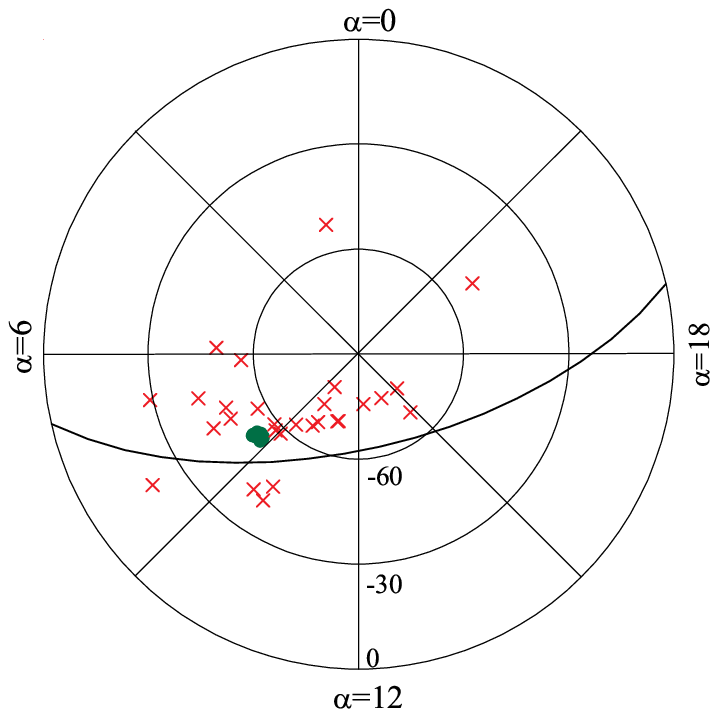}
}
   \end{tabular}
    \end{center}
\caption{{\it Left:} Celestial polar projection of the
Octans Association.
{\it Right:} Celestial polar projection of the Argus Association -- the crosses
represent the field Argus Association members and the filled
circles the IC~2391 members.
The transverse curve represents the Galactic plane.
}
\label{fig:argpiz}
\end{figure}

\begin{figure}[]
  \begin{center}
  \begin{tabular}{r}
 \resizebox{1.0\hsize}{!}{
\includegraphics[draft=False]{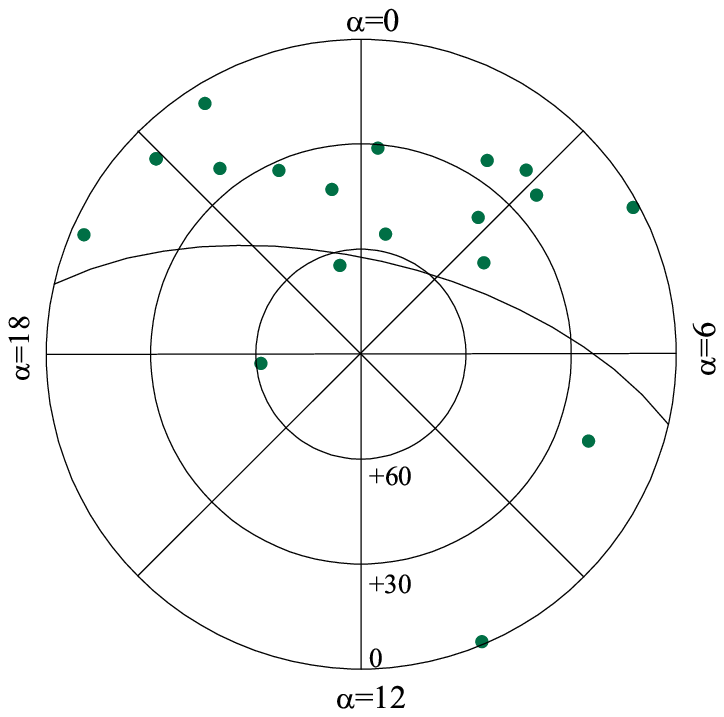}
\includegraphics[draft=False]{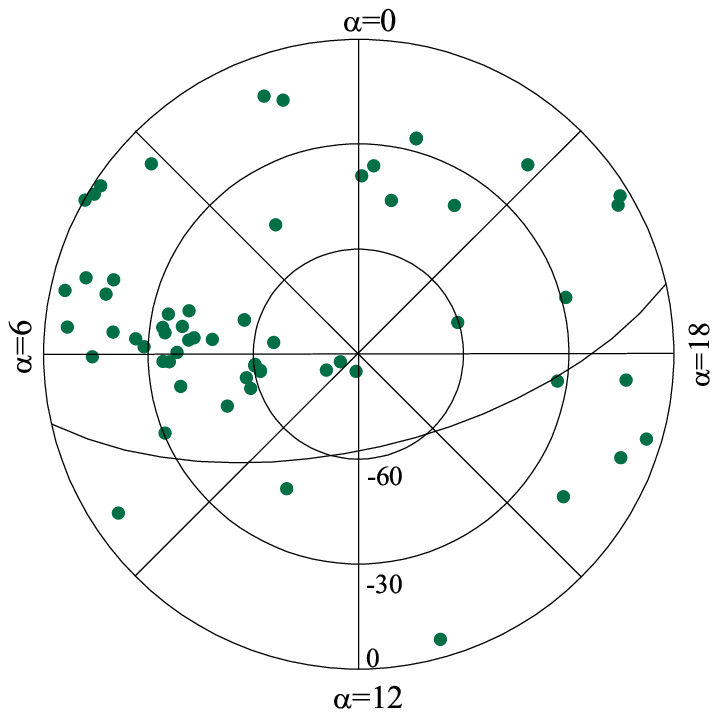}
}
     \end{tabular}
    \end{center}
\caption{Celestial polar projections of  the AB~Dor Association. {\it Left:} Northern
Hemisphere.
{\it Right:} Southern Hemisphere.
}
\label{fig:abpiz}
\end{figure}

As will be more clearly explained in the next sections, the
definition of the associations, especially their age, is strongly dependent
on their low mass star population.  The depth limit of the SACY sample
can give low mass candidates only for the associations nearer than
$\sim$50~pc.  For the more distant associations for which we found no other
way to obtain low-mass candidates, their definitions are less reliable.

From Figures~\ref{fig:thpiz} to ~\ref{fig:abpiz} we can see that the
Columba, $\beta$~Pic and AB~Dor associations have members in the
Northern Hemisphere and surveys in this part of the sky can reveal new
members for these associations (and perhaps for the Tuc-Hor
Association too).  Also other authors are trying to obtain associations or
new members, mainly by the convergence point method \cite[see, for
example,][]{debruijne99}.  Our experience shows that this method has
low reliability.  \citet{song02b} arrive at a similar conclusion in a
fruitless search for new TW~Hya Association members in a list of
candidates proposed by \citet{makarov01} using the convergence point
method.  Another example is the moving group in Carina-Vela proposed
by \citet{makarov00}, as it is discussed in Sections 3.3 and 7 that
seems to be, at least partially, a mixture of Car and Argus
associations.

In some cases, visual binary components (or cluster members)
have distinct proper motions or radial velocities.
This may result in distinct convergence values, membership probabilities and, sometimes, distances.
In the worst cases, only one component is a  high probability member,
and only this one is presented.
Evidently this is an indication that the kinematical data of the system needs to be improved.

Close visual binaries or spectroscopic binaries may deteriorate the quality of the kinematical data.
An instructive example is HD~202947 (BS~Ind),  proposed by \citet{zuckerman04}
as a member of the Tuc-Hor Association.
It has been observed 11 times in SACY and  found to be a double line spectroscopic binary.
The Hipparcos catalogue detected an eclipsing binary light curve with a period of 0.435~days.
\citet{guenther05} studied the radial velocities of BS~Ind and found a period of 3.3~years,
and the true systemic velocity could finally be derived.
(There are also broad peaks in the cross correlation function both from SACY and
\citet{guenther05} that can be assigned to the eclipsing binary.)
Due to its strong Li line BS~Ind is considered a young object,
but for its space motions (U=+0.1, V=--25.3, W=--8.6) it can not belong to the
Tuc-Hor Association or to any of the nearby associations presented here.

The young loose associations, being near to the Sun and having an age
distribution from 5\,Myr to about 100\,Myr,
enable studies of stellar physics that depend on the initial phases of the
stellar evolution, for example multiplicity, abundances, rotation, stellar activity, etc.
Examples of papers on rotation and activity using these associations are
\citet{delareza04} and \citet{scholz07}.
A preliminary study \citep{kastner03} of the X-ray emission properties of members of the
TW~Hya, $\beta$~Pic and Tuc-Hor associations shows trends in the hardness ratio
distributions when compared to TTS, the Hyades or older stars.

At a glance, looking at the figures of the Lithium distribution in the next sections,
we can see that the Lithium depletion appears to be more scattered in the older associations.
The present collection of several associations with ages from 6 up to 70 Myr (Table~\ref{table:finalb})
furnishes an excellent opportunity to investigate the Lithium depletion
during this time interval in the PMS evolution \citep{silva08}.
The Lithium depletion in associations can be compared with that in open clusters
(see \citet{jeffries06} for a review on this subject).
The effect of rotation on the Lithium depletion in the PMS phase is also important.
In fact, stars that are fast rotators (vsin(i)~$\ga$~20\,km~s$^{-1}$)
have higher Li line equivalent widths (upper points of the Lithium distribution
in the figures of the next sections).
This is in agreement with what is found for cluster stars,
at least for K type stars, in the sense that fast rotators
appear to have high Lithium abundances \citep{jeffries06}.

Most of the stars of the lists presented in the next sections stem from the
SACY sample, that is, X-ray selected low mass stars.
Any study using our sample must be aware of this bias.
The data presented in Tables~\ref{table:final} and \ref{table:finalb}
define the main characteristics and properties of these associations and can be
used as starting points to find additional members.
For example, GAIA will be an excellent tool to construct unbiased samples.

Stars in the young nearby associations  are ideal targets for studies of
planetary formation and very low mass sub-stellar objects.
As a matter of fact, many members of the associations presented here
have already been  studied in the context of the characterization of their protoplanetary disks
and the search for sub-stellar companions.
These works are reviewed in the last section.

\section{The $\beta$~Pic Association}

\begin{table}[]
\caption{The high probability members proposed for the $\beta$~Pic Association}
\smallskip
{
\label{table:bpa}
\begin{tabular}{lcclllrl}
\tableline
\noalign{\smallskip}
Name&$\alpha_{2000}$   & $\delta_{2000}$    &\hspace{3mm}V & Sp.T.    & D  & P.&Ref.\\
&&&&&[pc]&\%&\\
\noalign{\smallskip}
\tableline
\noalign{\smallskip}
\it{HD 203} &00 06 50.1     &-23 06 27      &\hspace{2mm}6.19  &F3V  &39H&75&Z\\
HIP 10679   &02 17 24.7     &+28 44 31      &\hspace{2mm}7.75  &G2V   &39&100&Z\\
HD 14062    &02 17 25.2     &+28 44 43      &\hspace{2mm}6.99  &F5V   &40&95&Z\\
\it{HD 15115}&02 26 16.2   &+06 17 33      &\hspace{2mm}6.79  &F4IV  &45H&60&M\\
BD+30 397B  &02 27 28.1     &+30 58 41      &12.44   &M2Ve  &43   &90&Z\\
AG Tri      &02 27 29.3     &+30 58 25      &10.12   &K6Ve  &43  &100&Z\\
BD+05 378   &02 41 25.9     &+05 59 18      &10.37   &K6Ve  &39  &95&Z,S\\
HD 29391    &04 37 36.1     &-02 28 24      &\hspace{2mm}5.22  &F0V   &30H&100&Z\\
GJ 3305     &04 37 37.5     &-02 28 28      &10.59   &M1Ve &      30H   &100&Z\\
V1005 Ori   &04 59 34.8     &+01 47 01      &10.05   &M0Ve  &24  &100&S\\
CD-57 1054  &05 00 47.1	    &-57 15 25	    &10.00   &M0Ve  &26H    &100&Z,S\\
HIP 23418   &05 01 58.8     &+09 59 00      &11.95*  &M3Ve  &34 &100&Z\\
BD-21 1074BC&05 06 49.5     &-21 35 04      &11.61*  &M3Ve  &18&100&S,B\\
BD-21 1074A &05 06 49.9     &-21 35 09      &10.29   &M1Ve  &18& 100&S,B\\
AF Lep      &05 27 04.7     &-11 54 03      &\hspace{2mm}6.56* &F7V   &27H &100&Z\\
V1311 Ori   &05 32 04.5     &-03 05 29      &11.52   &M2Ve&  36 &   95  &S,B\\
$\beta$~Pic &05 47 17.1	    &-51 04 00	    &\hspace{2mm}3.77  &A3V   &19H &100&Z\\
AO Men      &06 18 28.2	    &-72 02 41	    &\hspace{2mm}9.80  &K4Ve  &39H &95&Z,S\\
HD 139084B  &15 38 56.8	    &-57 42 19	    &14.80   &M5Ve  &40H     &95&Z,S\\
V343 Nor    &15 38 57.6	    &-57 42 26	    &\hspace{2mm}7.97  &K0V   &40H&90&Z,S\\
%hip 79881\footnote{non-members? see text}$^1$ &16 18 17.9&-28 36 50      &4.80  &A0     &23.2H&85\\
V824 Ara    &17 17 25.5	    &-66 57 04	    &\hspace{2mm}7.23* &G7IV  &31H &100&Z,S\\
HD 155555C  &17 17 31.3	    &-66 57 06	    &12.82   &M3Ve  &31H     &100&Z,S\\
GSC8350-1924&17 29 20.7     &-50 14 53      &13.47*  &M3Ve  &76       &95&B\\
CD-54 7336  &17 29 55.1	    &-54 15 49	    &\hspace{2mm}9.55  &K1V   &66 &90&S\\
HD 161460   &17 48 33.8	    &-53 06 43	    &\hspace{2mm}9.61* &K0IV   &74&90&S\\
HD 164249   &18 03 03.4     &-51 38 56      &\hspace{2mm}7.01  &F6V   &47H&100&Z\\
HD 164249B  &18 03 04.1     &-51 38 56      &12.5    &M2Ve  &    47H&100&T\\
HD 165189   &18 06 49.9     &-43 25 31      &\hspace{2mm}5.67* &A5V    &44H&100&Z\\
V4046 Sgr   &18 14 10.5	    &-32 47 33	    &10.94*  &K6Ve  &73             &95&T\\
GSC7396-0759&18 14 22.1	    &-32 46 10	    &12.78   &M1Ve  &73&          90&T\\
HD 168210   &18 19 52.2	    &-29 16 33	    &\hspace{2mm}8.89  &G5V   &75H &90&S\\
HD 172555   &18 45 26.9     &-64 52 15      &\hspace{2mm}4.78  &A6IV  &29H&100&Z\\
CD-64 1208  &18 45 36.9     &-64 51 48      &\hspace{2mm}9.54  &K5Ve  &29H &100&Z\\
TYC9073-0762&18 46 52.6	    &-62 10 36	    &12.08   &M1Ve  &54&        100&S\\
CD-31 16041 &18 50 44.5	    &-31 47 47	    &11.20   &K7Ve  &51&        95&S\\
PZ Tel      &18 53 05.9	    &-50 10 49	    &\hspace{2mm}8.29  &G9IV  & 50H&100&Z,S\\
TYC6872-1011&18 58 04.2	    &-29 53 05	    &11.78   &M0Ve  &79&        95&S\\
CD-26 13904 &19 11 44.6	    &-26 04 09	    &10.39*  &K4V(e)&80&        95&S\\
$\eta$ Tel  &19 22 51.2     &-54 25 25      &\hspace{2mm}5.02  &A0V   &   48H&100&Z\\
HD 181327   &19 22 58.9     &-54 32 17      &\hspace{2mm}7.03  &F6V   &  51H&100&Z\\
HD 191089   &20 09 05.2     &-26 13 27      &\hspace{2mm}7.18  &F5V   &  53H&100 &M\\
AT MicB     &20 41 51.1	    &-32 26 10	    &11.09*  &M4Ve  &           9.5&100&Z,S\\
AT MicA     &20 41 51.2	    &-32 26 07	    &10.99*  &M4Ve	&       9.5&100&Z,S\\
\\
\smallskip
(Continued)
\end{tabular}
}
\end{table}

\begin{table}[]
\hspace{20pt}  Table~\ref{table:bpa}. ~~ (Continued)
\smallskip

{
\begin{tabular}{lcclllrl}
\tableline
\noalign{\smallskip}
Name&$\alpha_{2000}$   & $\delta_{2000}$    &\hspace{3mm}V & Sp.T.    & D  &P.&Ref.\\
&&&&&[pc]&\%&\\
\noalign{\smallskip}
\tableline
\noalign{\smallskip}

AU Mic      &20 45 09.5	    &-31 20 27	    &\hspace{2mm}8.73  &M1Ve  &   10H&100&Z,S\\
HD 199143   &20 55 47.7     &-17 06 51&\hspace{2mm}7.35*&F7V&48H& 75&Z\\
AZ Cap      &20 56 02.7	    &-17 10 54	    &10.62*  &K6Ve  & 47& 95& Z,S\\
CP-72 2713  &22 42 49.0	    &-71 42 21	    &10.60   &K7Ve  &36& 100& S\\
WW PsA      &22 44 58.0	    &-33 15 02	    &12.07   &M4Ve  &20 &100&Z,S\\
TX PsA      &22 45 00.0	    &-33 15 26	    &13.36   &M5Ve  &20 &100&Z,S\\
BD-13 6424  &23 32 30.9	    &-12 15 52	    &10.54   &M0Ve  &28 &100& S\\

\noalign{\smallskip}
\tableline
\noalign{\smallskip}
\end{tabular}
}
\smallskip
{\hspace{5mm}
{(*) the photometric values are corrected for duplicity.}\\
The distances are from Hipparcos (H in the table) or kinematical ones,
calculated with the convergence method.\\
italics = possible members (see text).\\
Z=\citet{zuckerman04}; S=in the SACY survey; M=\citet{moor06};
T=\citet{torres06},
not in the SACY sample; B=proposed in this Handbook.
}
\end{table}

The $\beta$~Pic Association was first proposed by \citet{zuckerman01a}
and new members were suggested by \citet{song03} and by
\citet{moor06}.  \citet{zuckerman01a} and \citet{kaisler04} noted that
the so-called Capricornus Association formerly proposed by
\citet{ancker00} is part of the $\beta$~Pic Association.  A list of 33
proposed members is given in \citet{zuckerman04}.  Only one of the
proposed members, the brown dwarf HD~181296B \citep{lowrance00}, has
no kinematical data published and its membership can not be determined
by the methods presented here.

\begin{figure}[]
  \begin{center}
  \begin{tabular}{r}

\resizebox{1.0\hsize}{!}{
\includegraphics[draft=False]{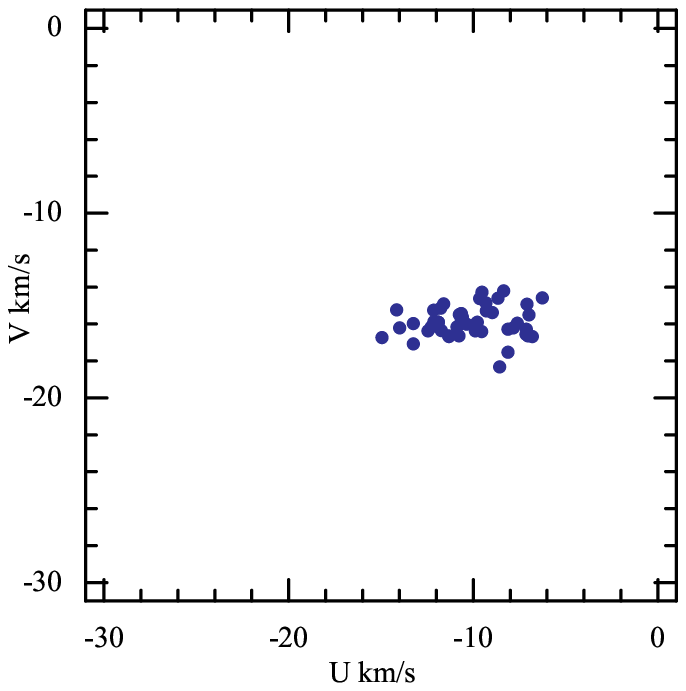}
\includegraphics[draft=False]{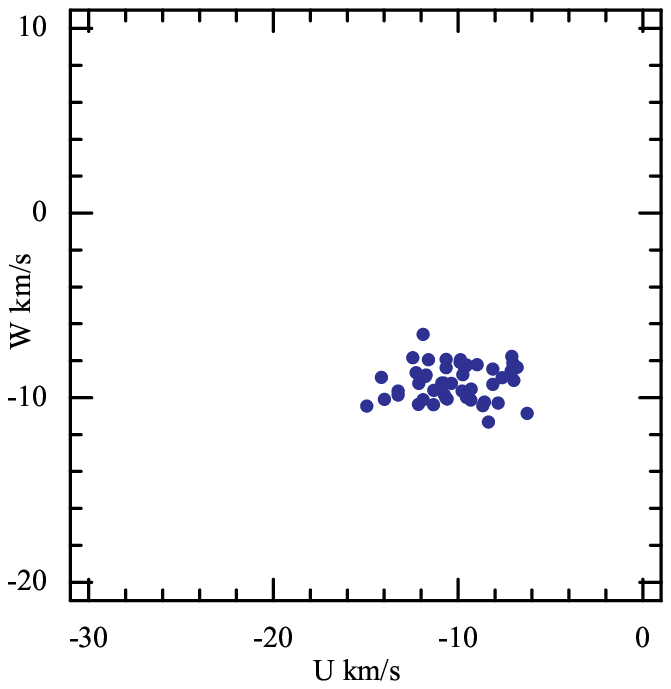}
\includegraphics[draft=False]{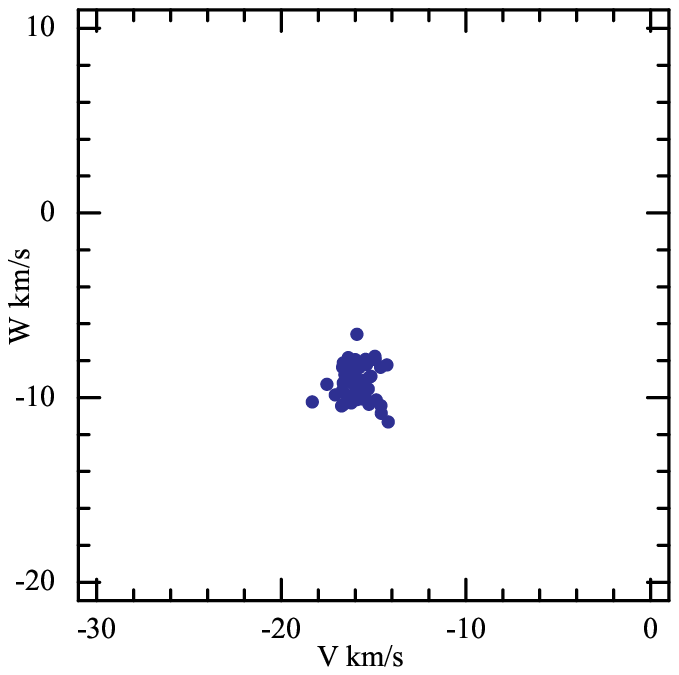}
}\\

\resizebox{1.0\hsize}{!}{
\includegraphics[draft=False]{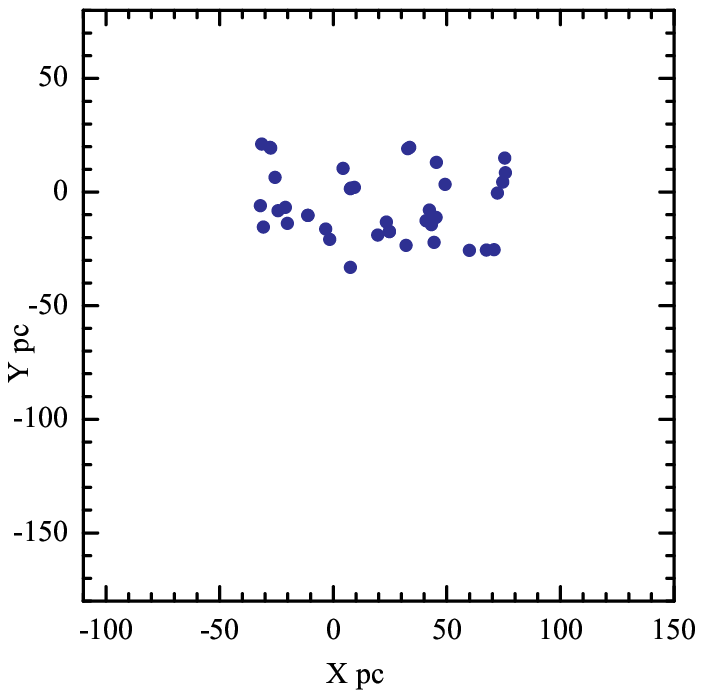}
\includegraphics[draft=False]{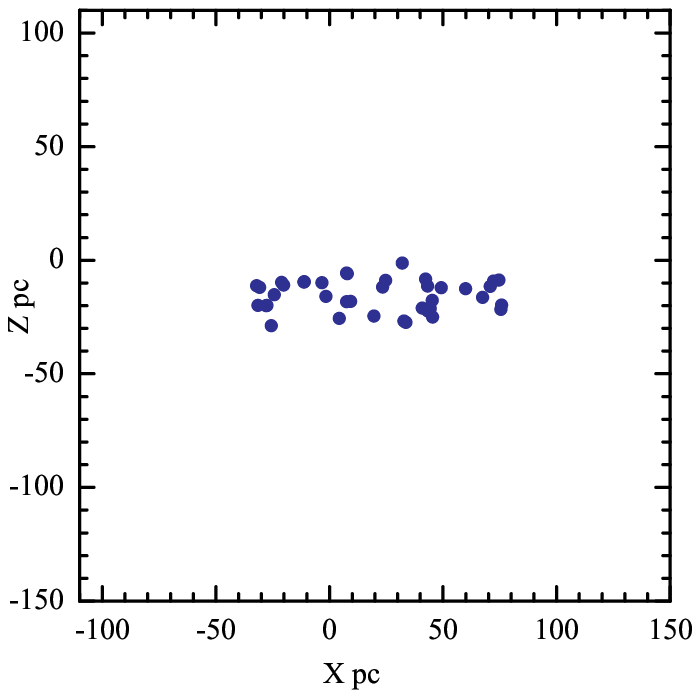}
\includegraphics[draft=False]{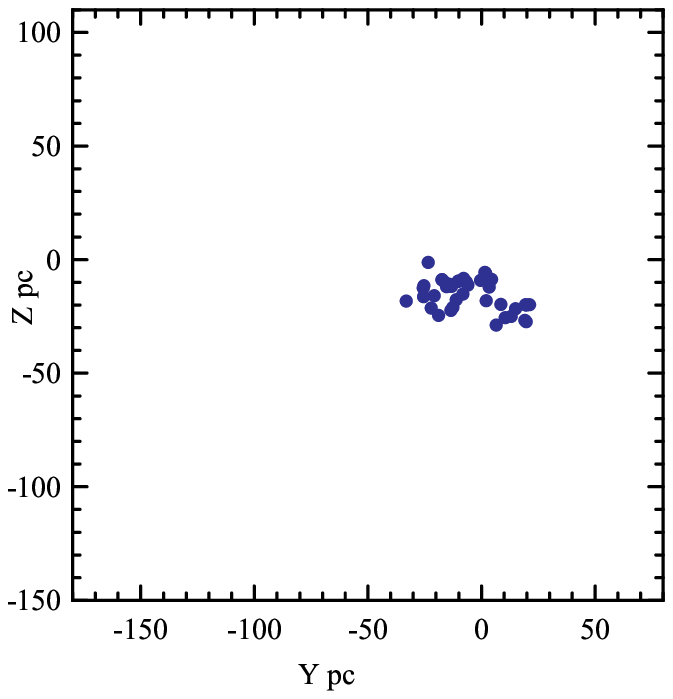}
}\\
  \end{tabular}
    \end{center}

\caption{Combinations of the sub-spaces of the UVWXYZ--space for the $\beta$~Pic
Association showing a well defined clustering in both kinematical and spatial
coordinates.}

\label{fig:bpxyz}
\end{figure}

The $\beta$~Pic Association is well defined in the SACY sample \citep{torres06}.
Using the SACY sample and all other members proposed, the convergence solution
gives 48 high probability members.
There are 30 stars from the \citet{zuckerman04} list
and 18 new proposed members. They are listed in  Table~\ref{table:bpa} and
their spatial and velocity distributions are shown in Figure~\ref{fig:bpxyz}.
Only two stars of the \citet{zuckerman04} list, namely HD~203 and HIP~79881,  are
rejected by the convergence
method and their membership probabilities are 0.75 and 0.85, respectively.
The rejection of HIP~79881 is in-line with \cite{song03} and \cite{ortega04} who
also considered this object as an outlier.
HD~203 is rejected using the radial velocity obtained in four observations made by us,
8.8$\pm$2.9~km~s$^{-1}$.
Nevertheless, its radial velocity is  less reliable as it is a fast rotator.
Using the velocity in \citet{barbier00} (6.5$\pm$3.5~km~s$^{-1}$)  it
becomes a high probability member.
(This is another example of the need for good kinematical data to properly
characterize  memberships, although good quality radial velocities are hard to obtain for hot fast rotators.)
Hence we consider it as possible member and it is included in the
Table~\ref{table:bpa}, in italics,
but not in the figures or in the statistics.

The solution in \citet{torres06} has 41 high probability members, but
subsequently a few updates have changed somewhat the solution of the
original paper, mainly by the inclusion of four new proposed members,
by revising the status of some stars, and by the addition of the stars
from \citet{moor06}.  With new radial velocity observations, HD~165189
and HD~199143 have been revised to be high probability members.
\citet{moor06} have proposed two members with debris disks, HD~15115
and HD~191089.  \citet{kalas07} detected the scattered light of an
extremely asymmetric dusty debris disk around HD~15115 that would make
it a very interesting candidate, but from the convergence method the
star is not a good member and its probability is low (p=0.7).  The
bona fide member BD+05~378 is close to HD 15115
at an angular distance of only 3.9$^\circ$
 and they share similar kinematics.  Thus, an adjustment of the
kinematical data of HD~15115 may promote it to a bona fide member.
Actually \citet{kalas07} proposed that the extreme asymmetries of its
disk are due to dynamical perturbations from BD+05~378 (itself a
single line spectroscopic binary).  As for HD~203, HD~15155 is
considered as a possible member and it is included in
Table~\ref{table:bpa}, in italics, but not in the figures or in the
statistics.

\begin{figure}[]
  \begin{center}
  \begin{tabular}{r}
 \resizebox{0.5\hsize}{!}{
\includegraphics[draft=False]{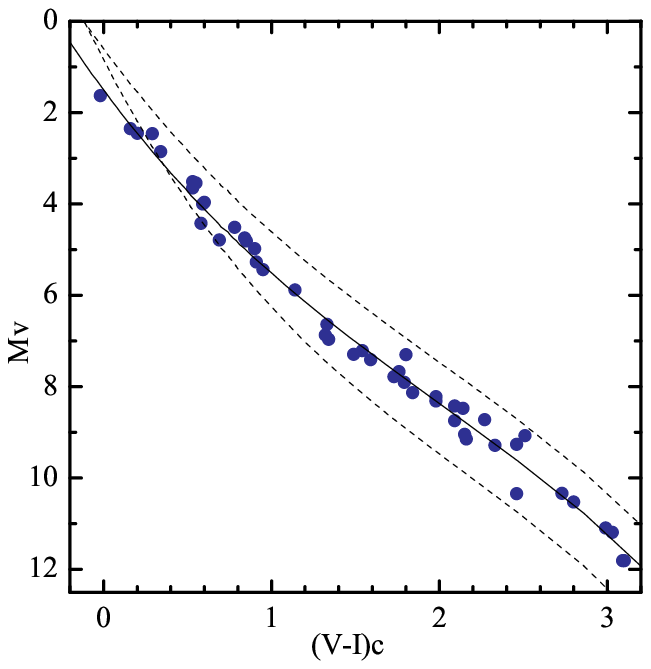}
}
\resizebox{0.5\hsize}{!} {
\includegraphics[draft=False]{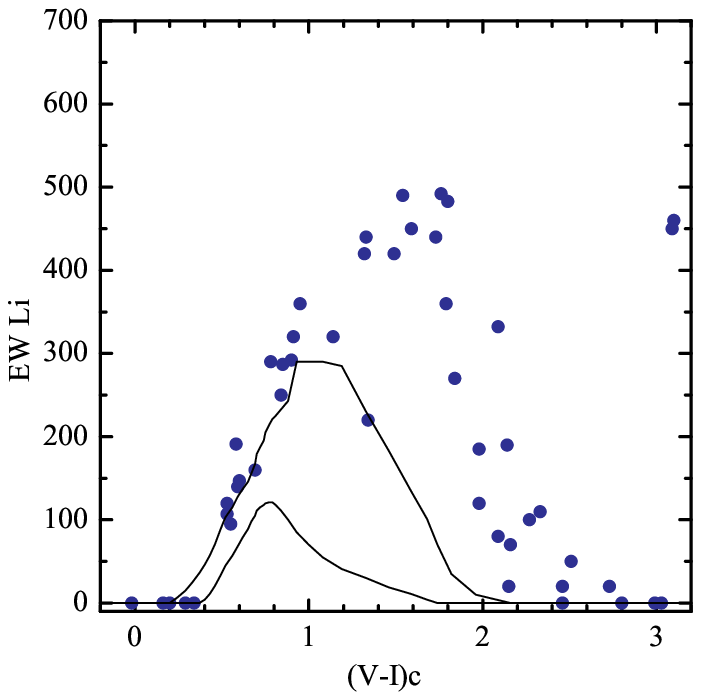}
}
   \end{tabular}
    \end{center}
\caption{{\it Left:}  The HR diagram of the members proposed for the $\beta$~Pic
Association; the over-plotted isochrones
are the {\it ad-hoc} ones given in the text for 5, 10 and 70~Myr. {\it Right:} The
distribution of the Li equivalent width
as a function of $(V-I)_C$; the curves are the upper and lower limits for the
age of the Pleiades \citep{neu97}.
}
\label{fig:bphr}
\end{figure}

The obtained age for the $\beta$~Pic Association is about 10~Myr but,
as commented before, the ages obtained with the  {\it ad-hoc} isochrones
must be taken with caution.
\cite{zuckerman01a} suggested 12$^{+8}_{-4}$~Myr, and \cite{feige06}
estimate the age of the member {GJ~3305} to be  13$^{+4}_{-3}$~Myr.
This is consistent with 11~Myr  obtained by dynamical back-tracing models
\citep{ortega02, ortega04, song03}.
These authors agree with the suggestion of \cite{mamajekasp} which places
the  birthplace of the $\beta$~Pic Association into the Sco-Cen complex.
The Li distribution (Figure~\ref{fig:bphr}) agrees well with these ages.
The two points with high Li equivalent width around $(V-I)_C$~=~3.1 in
(Figure~\ref{fig:bphr}) correspond to TX~PsA and HD~139084B.
TX~PsA forms a binary with the slightly brighter and hotter WW~PsA.
However, in contrast to the strong Li found in TX~PsA, WW~PsA has the Li fully depleted.
\citet{song02} interpreted the pair in the context of Li depletion boundary, and
concluded that it can give a strong observational constraint to age determinations using
the boundary of Li depletion.
If instead the kinematical distance is the correct one (it is at 2\,$\sigma$ of
the Hipparcos parallax error),
the stars could be 0.33~mag fainter,
and this should be considered in the discussion of the Li-age relations.
HD~139084B may help in this context.

AT~Mic and AU~Mic were already noted as young stars by \citet{eggen1968}
and their connection with the debris disk A type star $\beta$~Pic was first proposed by \citet{barrado99}.
The weak Li line of AU~Mic was  first detected by \citet{delareza81}
during an attempt to find other active  red dwarfs with Li
to explain the odd Li-rich V1005~Ori (Gl~182) \citep{bopp74}.
\citet{delareza81} tried to interpret the Li line of Gl~182 in terms of the Li production
by spallation reactions.
They discarded this possibility and proposed that V1005~Ori may be a member of a
very young kinematical group
suggested by \citet{kunkel75}, that includes also AU Mic and AT Mic.
These three stars are now proposed to belong to the $\beta$~Pic Association.

One of the new members proposed, GSC~8350-1924, is a  serendipitous  discovery
of the SACY. The TYCHO star is farther away from the ROSAT source and
we inadvertently observed the nearest star, GSC~8350-1924. The photometric observations
show that the TYCHO star actually is a hot star, the color in the catalogue
being erroneous. The proper motions used for GSC~8350-1924 are taken from UCAC2.
\citet{nuria08a} detected a companion at 0.767$\arcsec$ with similar magnitude.

Another interesting star proposed now is V1311~Ori, a wTTS known long ago (HBC~97; \citep{hbc})
but poorly studied. It has spot photometric variations with a period of 4.5~d, similar
to AU~Mic, determined by \citet{gahm95}.

The $\beta$~Pic Association has six known spectroscopic binaries
(double line spectroscopic binary stars: HIP~23418, AF Lep,
V824~Ara, HD~161460, V4046~Sgr; single line spectroscopic binary star:
BD+05~378).  One of these double line spectroscopic binary stars,
V4046~Sgr, a non-SACY star proposed by \citet{torres03a, torres03} to
be another possible member, was confirmed as high probability member
in the final analysis \citep{torres06}.  V4046~Sgr is an interesting
double line spectroscopic binary isolated cTTS with a circumbinary
disk \citep{quast00, stempels04}.  It has a possible faint optical
companion, GSC~7396-0759 (see Table~\ref{table:bpa}).  This star has
now been observed twice, and a spectral feature reported before
\citep{torres06} is not present in the new spectrum.  The two radial
velocities are very similar, so, even in the case of being a
spectroscopic binary, the systemic velocity should be near the mean
radial velocity.  We used the proper motions of V4046~Sgr but a good
proper motion determination is important to confirm the membership.

In addition to these six spectroscopic binaries, the $\beta$~Pic
Association has 13 wide visual binaries, easily recognized in
Table~\ref{table:bpa} since they have two entries, and ten close
visual binaries: HIP~23418 (sep.=1.0$\arcsec$), BD-21~1074\,BC
(sep.=1.2$\arcsec$), GSC~8350-1924 (sep.=0.8$\arcsec$), HD~161460
(sep.=0.1$\arcsec$), HD~165189 (sep.=1.7$\arcsec$), CD-64~1208
(sep.=0.2$\arcsec$), CD-26~13904 (sep.=1.1$\arcsec$), $\eta$ Tel
(sep.=4.2$\arcsec$), HD~199143 (sep.=1.1$\arcsec$) and AZ~Cap
(sep.=2.2$\arcsec$).  The very close optical companion of HD~161460,
detected by \citet{nuria08a}, is also probably the spectroscopic one.

\section{The GAYA Complex}

\begin{figure}[]
  \begin{center}
  \begin{tabular}{l}

\resizebox{1.0\hsize}{!}{
\includegraphics[draft=False]{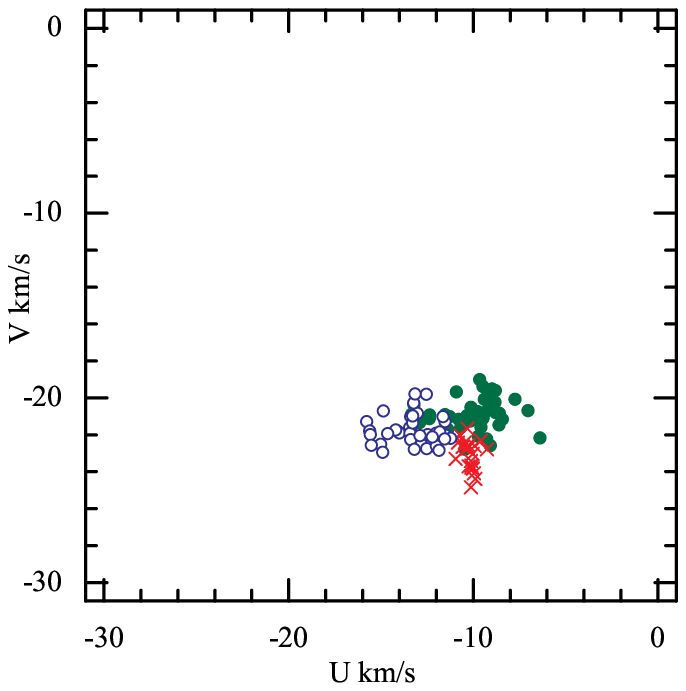}
\includegraphics[draft=False]{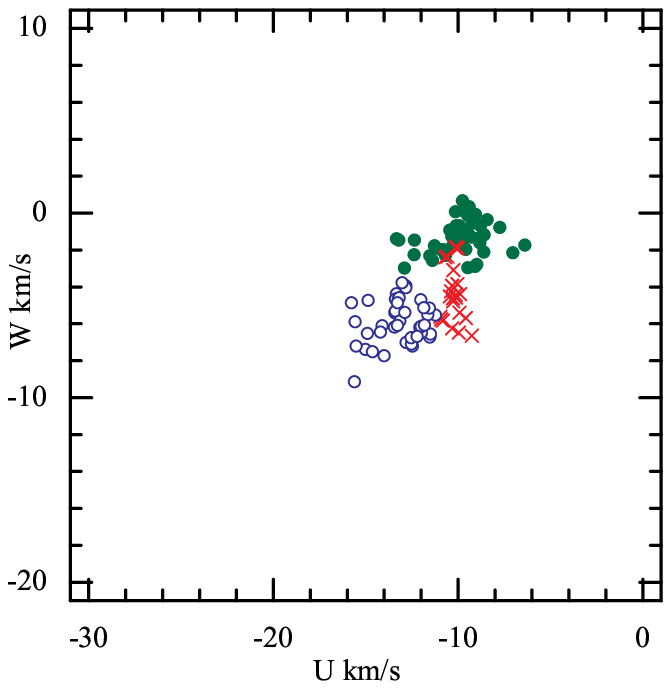}

\includegraphics[draft=False]{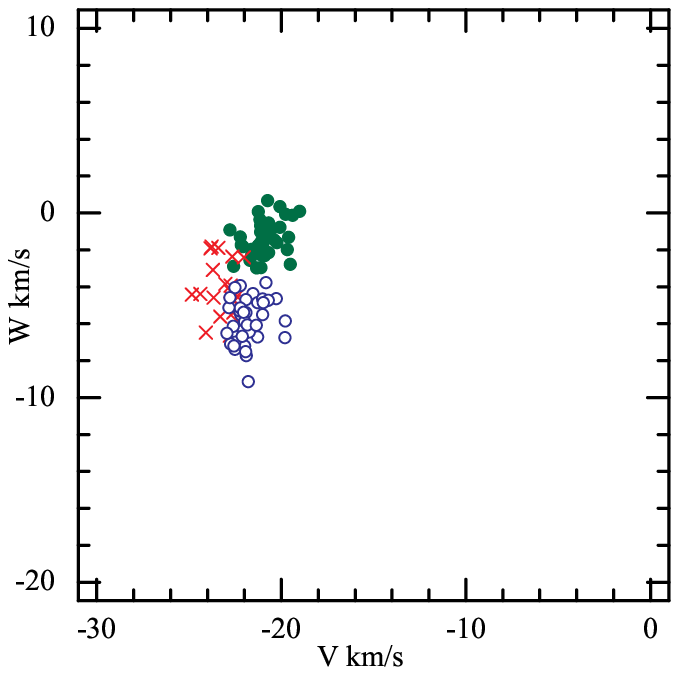}
}\\

\resizebox{1.0\hsize}{!}{
\includegraphics[draft=False]{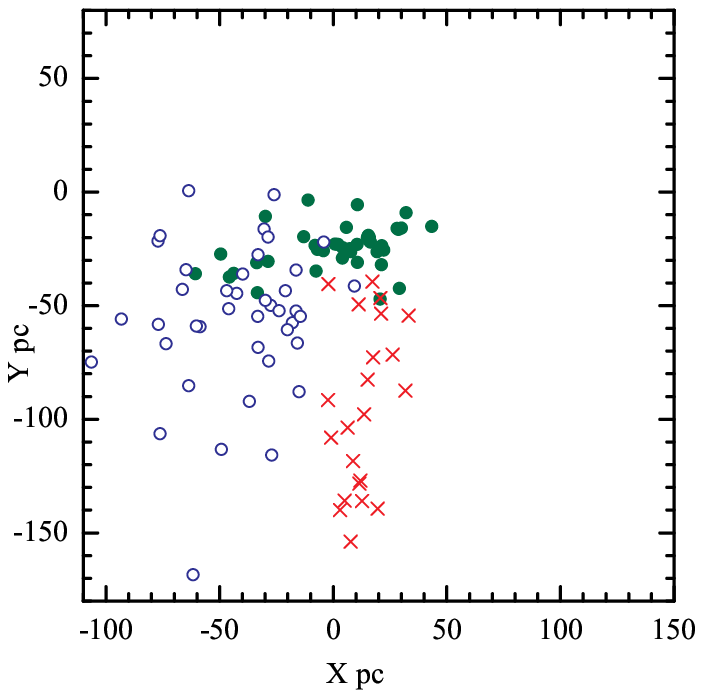}
\includegraphics[draft=False]{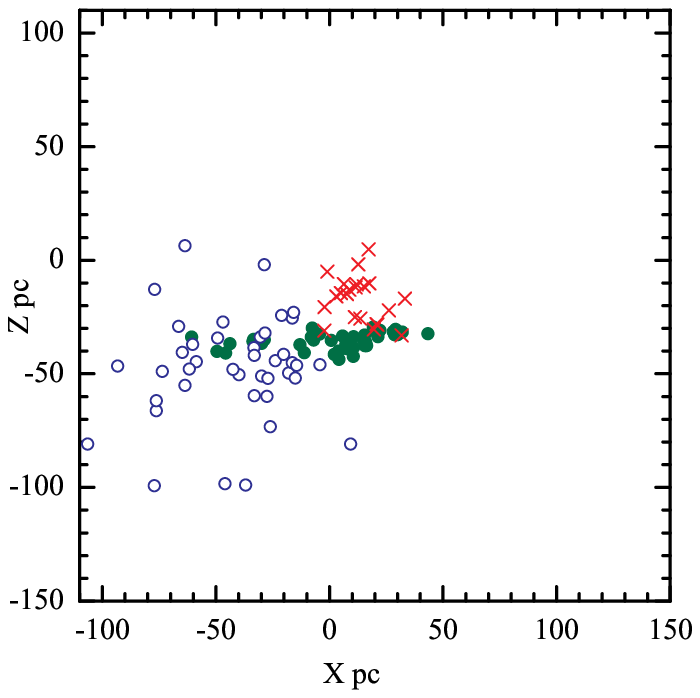}

\includegraphics[draft=False]{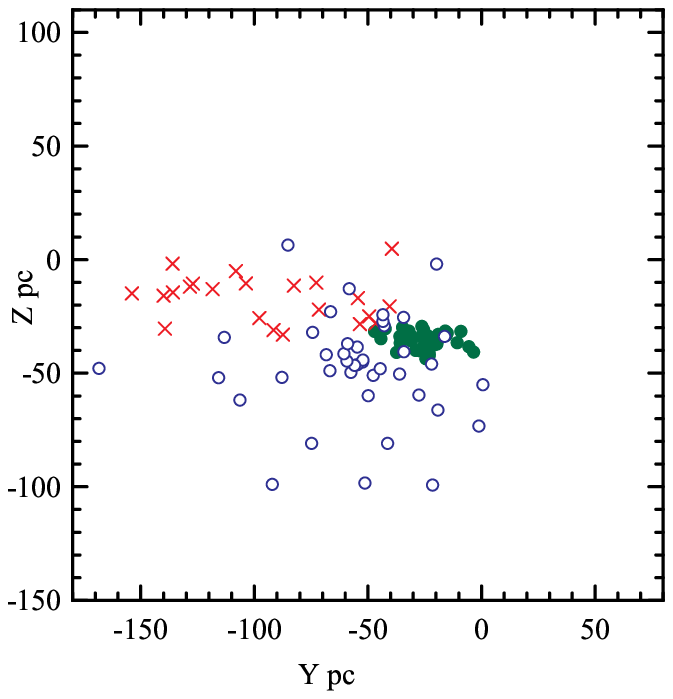}
}\\
  \end{tabular}
    \end{center}

\caption{Combinations of the sub-spaces of the UVWXYZ--space for the GAYA
complex ~--~ Tuc-Hor Association (filled circles),
Col Association (open circles)  and Car Association (crosses) ~--~ showing
their clustering in both kinematical and spatial coordinates.}

\label{fig:gaxyz}
\end{figure}
\begin{figure}[]
  \begin{center}
  \begin{tabular}{l}

\resizebox{0.5\hsize}{!}{
\includegraphics[draft=False]{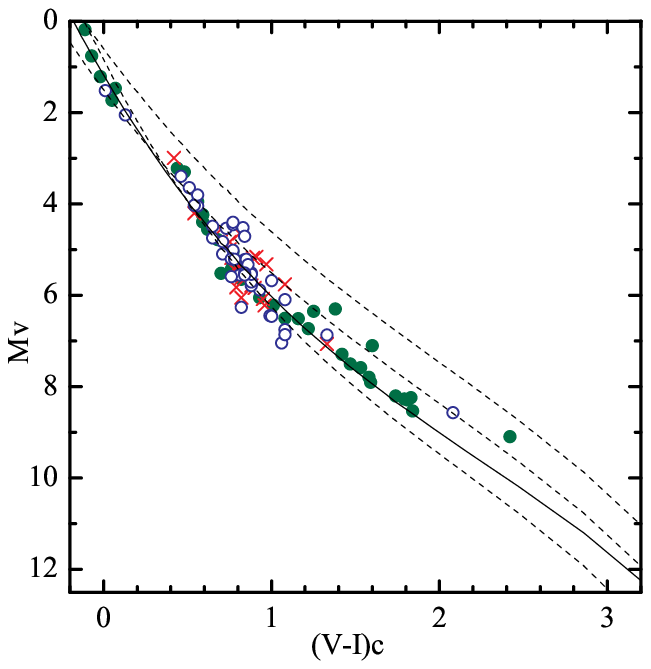}
}
\resizebox{0.51\hsize}{!}{
\includegraphics[draft=False]{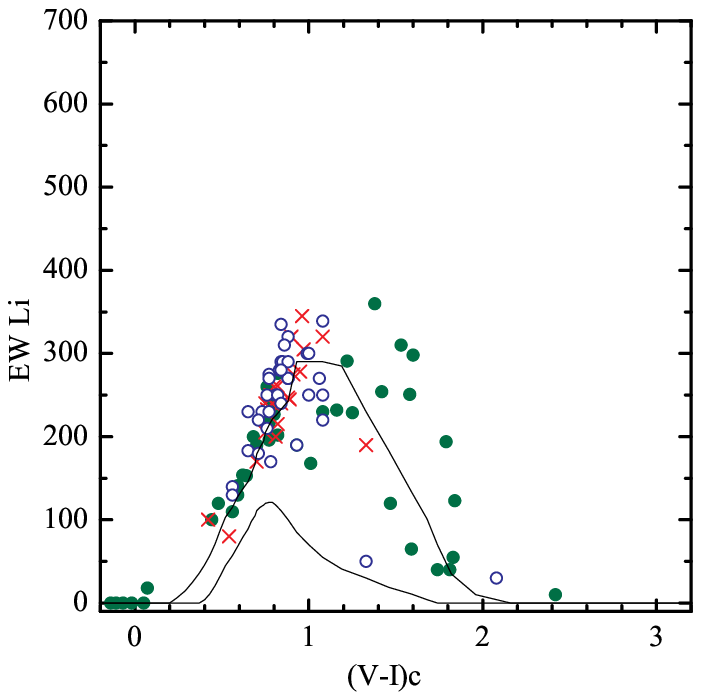}
}
   \end{tabular}
    \end{center}
\caption{{\it Left:}  The HR diagram for the associations of the GAYA complex (symbols are as in
Figure~\ref{fig:gaxyz} and the  {\it ad-hoc} isochrones are for 5, 10, 30 and 70~Myr).
{\it Right:} The distribution of Li equivalent width as a function of $(V-I)_C$; the
curves are the upper and lower limits for the age of the Pleiades \citep{neu97}.
}
\label{fig:gahr}
\end{figure}

We recall that it was the independent and almost simultaneous discovery of the
Tucana Association \citep{zuckerman00}
and the Horologium Association \citep{torres00} that motivated the SACY project.
One of its first results was to postulate a large population  of
young stars in the southern sky  with kinematical and physical properties
similar to both associations, initially
called GAYA (from Great Austral Young Association) \citep{torres01}.
Now it becomes clear that
it can be subdivided into, at least, three distinct, but similar associations:
the Tucana-Horologium Association, almost the same as the one originally defined by \citet{zuckerman01b},
the Columba Association and the Carina Association, the last two coming from a division
of the early GAYA2 \citep{torres03a, torres03, torres06}.

In Tables~\ref{table:final} and \ref{table:finalb} it can be noted that the
three associations can be distinguished
in at least one of the velocity components and that they form a
sequence in physical space.
This can be visualized in Figure~\ref{fig:gaxyz}.
Anyway these three associations are not very distinct from one another.
The Tuc-Hor  Association is clearly more compact and closer to what we expect for an association.
The other two are more difficult to define and this is reflected in the probability values.
There are many stars not included in any of these two associations;
nevertheless they have similar kinematical properties.
This situation seems similar to the large Sco-Cen Association but GAYA is older
and, therefore, more scattered.
The Col and Car associations may be like reefs in a sea of similar young stars.
Attempts to obtain solutions merging Col and Car associations (GAYA2)
gave worse results.

\citet{zuckerman04} proposed 50 stars to be members of the Tuc-Hor
Association\footnote{Actually there are 49 entries in their table, but
we discriminate both companions of the wide binary DS~Tuc.}, but only
25 of them are in the SACY sample.  The non-SACY members were also
included to be tested in the convergence method.  From these 50 stars,
31 were found to be high probability members of the new definition of
the Tuc-Hor Association; three of the Col Association; and three of
the Car Association.  Some of the 13 stars rejected could still be
classified as possible members of one of these associations of the
GAYA complex.  In fact, the quality of the proper motions and radial
velocities is crucial to establish a membership, but errors in their
kinematical data or a natural spread of the stars born in these
associations may reduce their membership probabilities.  This can be
said about any association, but it is particularly true here.
Actually four of these rejected stars are fast rotators and their
radial velocities are unreliable, but their space motions are not very
far from those of the Tuc-Hor Association and, although their
convergence values are bad, they have high values in the probability
model.  Therefore these high rotators, HD~12894 (p=0.9), HD~20385
(p=0.9), HD~208233 (p=0.8) and $\eta$ Tuc (p=0.8), can be considered
possible members.  A possible member for the Col Association is V1358
Ori, that is good in the convergence method but with very low
probability.  Another active star proposed, AT Col, has low
probability for any of the three associations and has space motions
very similar to the Car Association values, but it lies in XYZ--space
in the Col Association.  GSC~8056-0482 has badly determined proper
motions and a Lithium equivalent width of 380~m\AA, too high for the
Tuc-Hor age \citep{torres00, zucksb04}.  CD-34 2406 has the duplicity
confirmed by TYCHO and its corrected magnitude is too faint to belong
to the GAYA complex.  HIP~3556, CD-64~120, HD~53842, HD~200798 and
BS~Ind (see Section 1.3) have space motions far from those of the GAYA
associations.

\subsection{The Tuc-Hor Association}

\begin{table}[]

\caption{The high probability members proposed for the Tuc-Hor Association}
\smallskip
{
\label{table:tha}
\begin{tabular}{lcclllrl}
\tableline
\noalign{\smallskip}
Name&$\alpha_{2000}$   & $\delta_{2000}$    &\hspace{3mm}V & Sp.T.    & D  & P.&Ref.\\
&&&&&[pc]&\%&\\
\noalign{\smallskip}
\tableline
\noalign{\smallskip}

HD 105    &00 05 52.5&    -41 45 11&      \hspace{2mm}7.53 &      G0V   &40H&100&Z,S\\
HD 987    &00 13 53.0&    -74 41 18&      \hspace{2mm}8.78 &      G8V  &44H&100&Z,S\\
HD 1466    &00 18 26.1&    -63 28 39&     \hspace{2mm}7.45 &      F8V  &41H&100&Z\\
HIP 1910    &00 24 09.0&    -62 11 04&       11.49*&      M0Ve   &44&100&Z\\
CT Tuc       &00 25 14.7&    -61 30 48&       11.47 &      M0Ve  &43&100&Z\\
HD 2884    &00 31 32.7&    -62 57 30&      \hspace{2mm}4.36&      B9V&43H&90&Z\\
HD 2885      &00 31 33.5&    -62 57 56&      \hspace{2mm}4.77*&     A2V&46&100&Z\\
HD 3003      &00 32 43.9&    -63 01 53&      \hspace{2mm}5.07 &      A0V&47H&100&Z\\
HD 3221    &00 34 51.2&    -61 54 58&      \hspace{2mm}9.61 &      K4Ve  &46H&100&Z,S\\
CD-78 24     &00 42 20.3&    -77 47 40&       10.21 &      K3Ve   &50&100&S\\
HD 8558    &01 23 21.3&    -57 28 51&      \hspace{2mm}8.51 &      G7V &49H&100&Z,S\\
CC Phe       &01 28 08.7&    -52 38 19&      \hspace{2mm}9.07 &      K1V &37H&100&Z,S\\
DK Cet       &01 57 49.0&    -21 54 05 &     \hspace{2mm}8.66* &      G4V &42H&100&Z,S\\
HD 13183    &02 07 18.1&    -53 11 57&      \hspace{2mm}8.63 &      G7V  &50H&100&Z,S\\
HD 13246    &02 07 26.1&     -59 40 46&      \hspace{2mm}7.50 &      F7V &45H&100&Z,S\\
CD-60 416    &02 07 32.2&    -59 40 21&       10.68 &      K5Ve  &48&100&Z,S\\
$\phi$ Eri   &02 16 30.6&    -51 30 44&      \hspace{2mm}3.56&      B8V  &47H&100&Z\\
$\epsilon$ Hyi&02 39 35.4&   -68 16 01&      \hspace{2mm}4.12&      B9IV &47H&85&Z\\
CD-53 544    &02 41 46.8&    -52 59 52&       10.22 &      K6Ve  &42&100&Z,S\\
AF Hor       &02 41 47.3&    -52 59 31 &      12.21 &      M2Ve  &42&100&Z,S\\
CD-58 553    &02 42 33.0&    -57 39 37&       10.98 &      K5Ve   &50&100&Z,S\\
CD-35 1167   &03 19 08.7&    -35 07 00&       11.12  &         K7Ve  &44&100&S\\
CD-46 1064   &03 30 49.1&    -45 55 57 &     \hspace{2mm}9.55  &         K3Ve &44&100&S\\
CD-44 1173   &03 31 55.7&    -43 59 14&       10.90 &      K6Ve  &42&100&S\\
HD 22213     &03 34 16.4&    -12 04 07&      \hspace{2mm}8.85  &G7V&48&95&S\\
HD 22705     &03 36 53.4&    -49 57 29   &   \hspace{2mm}7.65* &      G2V    &42H&100&Z\\
BD-12 943    &04 36 47.1&    -12 09 21&  \hspace{2mm}9.86 &    K0V&  69&100&S\\
HD 29615     &04 38 43.9&    -27 02 02 &     \hspace{2mm}8.47 &      G3V  &55H&100&Z,S\\
HD 30051     &04 43 17.2&    -23 37 42   &   \hspace{2mm}7.12 &     F2IV &58H&100&Z\\
TYC8083-0455   &04 48 00.7&    -50 41 26&       11.53 &      K7Ve   &46&90&S\\
HD 32195    &04 48 05.2  &     -80 46 45&    \hspace{2mm}8.14 &F7V  & 60H&80 &Z\\
BD-20 951   &04 52 49.5  &   -19 55 02&         10.33* &      K1V(e) & 72&90&S\\
BD-19 1062  &04 59 32.0  &   -19 17 42&         10.65 &      K3V(e) & 68&90&S\\
BD-09 1108  &05 15 36.5  &   -09 30 51&    \hspace{2mm}9.79 & G5V   & 78&85 &S\\
CD-30 2310  &05 18 29.1  &   -30 01 32&         11.66 &      K4Ve   & 65&90&S\\
%T8096-0414  &05 33 25.6&     -51 17 13&    11.74 &     K7Ve   &22.6&&90\\
%CD-39 2075  &05 37 05.3&    -39 32 26&      9.52   &    K1V(e)   &16.4&&\\
%HD 42270    &05 53 29.3 &   -81 56 33&      9.14  &     K0V(e)   &18.3&&\\
$\alpha$ Pav &20 25 38.9&    -56 44 06&      \hspace{2mm}1.91  &     B2IV&56H&100&Z\\
HD 202917   &21 20 50.0&    -53 02 03&      \hspace{2mm}8.69 &      G7V  &46H&100&Z,S\\
%T9486-0927  &21 25 27.6&    -81 38 28&     11.85   &    M2Ve   &28.7&&\\
HIP 107345   &21 44 30.1&    -60 58 39&       11.61 &      M0Ve     &47&100&Z,S\\
HD 207575    &21 52 09.7&    -62 03 09 &     \hspace{2mm}7.22 &    F6V&45H&100&Z\\
HD 207964    &21 55 11.4&    -61 53 12 &     \hspace{2mm}6.56*&     F0IV   &47H&100&Z\\
\\
\smallskip
(Continued)
\end{tabular}
}
\end{table}

\begin{table}[th]
\hspace{20pt}  Table~\ref{table:tha}. ~~ (Continued)
\smallskip

{
\begin{tabular}{lcclllrl}
\tableline
\noalign{\smallskip}
Name&$\alpha_{2000}$   & $\delta_{2000}$    &\hspace{3mm}V & Sp.T.    & D  &P.&Ref.\\
&&&&&[pc]&\%&\\
\noalign{\smallskip}
\tableline
\noalign{\smallskip}

TYC9344-0293   &23 26 10.7&    -73 23 50&       11.83* &      M0Ve   &46&100&S\\
CD-86 147    &23 27 49.4&    -86 13 19&      \hspace{2mm}9.29 &      G8IV &60&85&S\\
HD 222259B   &23 39 39.1 &   -69 11 40&      \hspace{2mm}9.84 &      K3Ve  &46H&100&Z,S\\
DS Tuc       &23 39 39.5&    -69 11 45&      \hspace{2mm}8.49 &      G6V  &46H&100&Z,S\\
\noalign{\smallskip}
\tableline
\noalign{\smallskip}
\end{tabular}
}
{(*) the photometric values are corrected for duplicity.\\
The distances are from Hipparcos (H in the table) or kinematical ones,
calculated with the convergence method.\\
Z=\citet{zuckerman04}; S=in the SACY survey.}

\end{table}

The 44 high probability members of the Tuc-Hor Association are presented in Table~\ref{table:tha}.
From this list, 31 stem from the previous list of \citet{zuckerman04},
17 of them being also in the SACY sample.
We proposed 13 new members.
As can be seen in Figure~\ref{fig:gaxyz}, the Tuc-Hor Association
is more compact  than the Col or the Car associations.
Figure~\ref{fig:gahr} suggests that it may be younger than the other two.
Actually, it is the best defined association of the GAYA complex and it seems
to have a nucleus, containing at least seven stars, around $\beta$~Tuc
($\alpha$~=~00:32 $\delta$~=~--63$^\circ$ -- see Fig~\ref{fig:thpiz} and Table~\ref{table:tha})

BD-09~1108 is located in front of the Orion Complex but, as it is not correlated with
Orion CO lines,  \citet{Alcala00} suggested a possible connection with the Gould Belt.
Using TYCHO-2 proper motions we propose it as the farthest Tuc-Hor member (78\,pc).

There are nine visual double or multiple systems in the Tuc-Hor Association, four of them wide:
{\it i)} the $\beta$~Tuc multiple system, with at least five components, three in Table~\ref{table:tha} --
HD~2884 (double, sep.=2.4$\arcsec$), HD~2885 (double, sep.=0.4$\arcsec$) and HD~3003;
{\it ii)} the pair HD~13246\,/\,CD-60~416;
{\it iii)} the pair CD-53~544\,/\,AF~Hor; {\it iv)} the pair DS~Tuc\,/\,HD~222559B.
The five other systems, having separations less than 1$\arcsec$, are: HIP~1910 (sep.=0.6$\arcsec$);
DK~Cet (sep.=0.2$\arcsec$);  HD~207964 (sep.=0.4$\arcsec$); TYC~9344-0293 (sep.=0.2$\arcsec$)
and HD~22705 --
\citet{makarov07} determined its orbit, based on Hipparcos data,
and found  a period of 201 days and a semi-major axis of only 5~mas.
HD~22705 is also a possible low amplitude single line spectroscopic binary \citep{nord04},
probably the same as the astrometric one.
Another low amplitude single line spectroscopic binary is $\alpha$ Pav,
detected as early as 1907, by \citet{curtis07}.
The only double line spectroscopic binary in the Tuc-Hor Association, detected in the SACY, is BD-20~951.
Observed only once, its systemic velocity is less reliable.

The very massive star $\alpha$ Pav is leaving the main sequence, and
our polynomial {\it ad-hoc} isochrones are no more appropriate (its
color is anyway out of the limits of Equation~\ref{eq:iso30}).

\subsection{The Columba Association}

\begin{table}[]
\caption{The high probability members proposed for the Columba Association}
\smallskip
{
\label{table:cola}
\begin{tabular}{lcclllr@{\hskip7pt}l}
\tableline
\noalign{\smallskip}
Name&$\alpha_{2000}$   & $\delta_{2000}$    &\hspace{3mm}V & Sp.T.    & D  & P.&Ref.\\
&&&&&[pc]&\%&\\
\noalign{\smallskip}
\tableline
\noalign{\smallskip}

CD-52 381    &01 52 14.6&   -52 19 33&       10.89*   &K2V(e) &  \hspace{2mm}92&80 &S\\
BD-16 351    &02 01 35.6&   -16 10 01&       10.33    &K1V(e) &  \hspace{2mm}78&95&S\\
CD-44 753    &02 30 32.4&   -43 42 23&       10.42    &K5V(e) &   \hspace{2mm}52&70  &S\\
BD-11 648    &03 21 49.7&   -10 52 18&       11.32    &K0Ve   &            127& 75     &S\\
V1221 Tau &03 28 15.0&   +04 09 48&     \hspace{2mm}9.70*&K0 &\hspace{2mm}84&60&S,zi\\
HD 21955     &03 31 20.8&   -30 30 59&     \hspace{2mm}9.93*&     G7IVe &122&70&S\\
HD 21997     &03 31 53.6&   -25 36 51&      \hspace{2mm}6.38&  A3V&\hspace{2mm}73H&80&M\\
BD-04 700    &03 57 37.2&   -04 16 16&       10.61&    G8V(e)&                 103&85  &S\\
BD-15 705    &04 02 16.5&    -15 21 30&      10.17&          K3(e) &\hspace{2mm}49&75&z,S\\
HD 26980      &04 14 22.6&   -38 19 02&     \hspace{2mm}9.08&      G3V   &\hspace{2mm}83H&90&S\\
HD 27679  &04 21 10.3&    -24 32 21&     \hspace{2mm}9.43&          G2V   &\hspace{2mm}75&95&S\\
CD-43 1395  &04 21 48.7&    -43 17 33&      10.18&      G7V    &143&100&S\\
%CD-44 1533    &04 22 45.6&    -44 32 52&     10.47&      K0V &\hspace{2mm}90&30&S  \\
%CD-44 1568    &04 27 20.5&    -44 20 39&     10.91 &          K1V(e) &\hspace{2mm}92&60&S\\
CD-36 1785   &04 34 50.8&    -35 47 21 &     10.84&      K1Ve &\hspace{2mm}77&85&S\\
HD 30447     &04 46 49.5&    -26 18 09&\hspace{2mm}7.85&F3V&\hspace{2mm}78H&60&M   \\
GSC8077-1788   &04 51 53.0&    -46 47 31&      13.03&      M0Ve  &\hspace{2mm}80&85&S\\
HD 31242     &04 51 53.6&    -46 47 13&       \hspace{2mm}9.85&      G5V  &\hspace{2mm}72&85&S\\
HD 272836    &04 53 05.2&    -48 44 39&      10.78 &         K2V(e) &\hspace{2mm}76&80&S\\
TYC5900-1180 &04 58 35.8&    -15 37 31&      11.15 &         G9V&   152& 90  &S\\
BD-08 995    &04 58 48.6&   -08 43 40&       10.32&          K0V    &\hspace{2mm}83&85  &S\\
HD 32372    &05 00 51.9&    -41 01 07&       \hspace{2mm}9.50&      G5V  &\hspace{2mm}72H&95&S\\
AS Col      &05 20 38.0&    -39 45 18&        \hspace{2mm}7.34&    F6V  &\hspace{2mm}46H&70&z\\
BD-08 1115  &05 24 37.3&    -08 42 02&      \hspace{2mm}9.88&      G7V(e)&118&80 &S\\
HD 35841    &05 26 36.6&     -22 29 24&   \hspace{2mm}8.91& F3V  &\hspace{2mm}96&65&M\\
HD 274561   &05 28 55.1&    -45 34 58&    11.45 &       K1V(e)  &\hspace{2mm}78&75&S\\
HD 36329    &05 29 24.1&    -34 30 56 &       \hspace{2mm}9.22*&      G3V  &\hspace{2mm}75H&95&S\\
%HD 269620   &05 29 27.2&    -68 52 05&      ~9.56&       G6V   &9.9 &&    \\
AG Lep   &05 30 19.0&  -19 16 32& \hspace{2mm}9.62&G6V&111&95&S\\
%AT Col  &05 37 05.3&    -39 32 26&      \hspace{2mm}9.52&       K1V(e) &\hspace{2mm}50&95&z,S\\
%HD 37484&    05 37 39.6& -28 37 35&    \hspace{2mm}7.26 &      F4V&\hspace{2mm}60H& 55&M\\
%HD 2699217  &05 38 34.4&    -68 53 07&     10.28&       G7V    &11.3&&    \\
BD-08 1195   &05 38 35.0&  -08 56 40&    \hspace{2mm}9.84 &  G7V&  \hspace{2mm}83 &80 &S\\
HD 38207   & 05 43 21.0& -20 11 21&             \hspace{2mm}8.46 &F2V&\hspace{2mm}93&80&M\\
HD 38206   &  05 43 21.7& -18 33 27&    \hspace{2mm}5.73 &  A0V&\hspace{2mm}69H&95&M\\
CD-38 2198  &05 45 16.3&    -38 36 49&       10.95&       G9V    &\hspace{2mm}90&85&s\\
CD-29 2531  &05 50 21.4&    -29 15 21&      11.31&      K0V(e)    & 149&95&S\\
CD-52 1363 &05 51 01.1&    -52 38 13&       10.61&      G9IV     & 106&85&S\\
%HD 42270    &05 53 29.3&    -81 56 53 &     ~9.14&       K0V   &17.3&&    \\
HD 40216    &05 55 43.1&    -38 06 16&     \hspace{2mm}7.46 &      F7V  &\hspace{2mm}54H&80&z\\
%V1358 Ori   &06 19 08.1&    -03 26 20&      \hspace{2mm}7.91  &     F9V  &\hspace{2mm}50H&45 &z\\
%AB Pic      &06 19 13.0&    -58 03 16 &     ~9.13   &    K1V(e) &&22.0$\pm$0.8&\\
CD-40 2458  &06 26 06.9&    -41 02 54&     10.00  &         K0V  &\hspace{2mm}91&70&S\\
CD-48 2324  &06 28 06.1&    -48 26 53&     11.08 &          G9V    &135&60&S\\
TYC4810-0181&06 31 55.2&    -07 04 59&      11.80&           K3Ve   &\hspace{2mm}96&70  &S\\
%CD-37 2984  &06 39 46.9&     -37 50 10&    10.72 &          K1V    &16.3&&    \\
%HD 295290&    06 40 22.3&    -03 31 59& \hspace{2mm}9.00   &K0V(e)&\hspace{2mm}65& 55 &S\\
HD 48370 &    06 43 01.0&    -02 53 19& \hspace{2mm}7.92   &G8V&\hspace{2mm}35& 70 &S\\
%CD 49885    &06 43 46.3&    -71 58 36&      ~8.94   &    G6V     &&17.7$\pm$0.9&\\
%CD-41 2572  &06 45 37.9 &   -41 12 41&     10.82  &         K0V     &\hspace{2mm}82&45&S\\
CD-36 3202  &06 52 46.8 &   -36 36 17&     11.22  &         K2V(e)   &135&100&S\\
HD 51797    &06 56 23.5&    -46 46 55&     \hspace{2mm}9.84    &       K0V(e)  &\hspace{2mm}75&80&S\\
\\
\smallskip
(Continued)
\end{tabular}
}
\end{table}

\begin{table}[]
\hspace{20pt}  Table~\ref{table:cola}. ~~ (Continued)
\smallskip

{
\begin{tabular}{lcclllrl}
\tableline
\noalign{\smallskip}
Name&$\alpha_{2000}$   & $\delta_{2000}$    &\hspace{3mm}V & Sp.T.    & D  &P.&Ref.\\
&&&&&[pc]&\%&\\
\noalign{\smallskip}
\tableline
\noalign{\smallskip}
CD-39 3026  &07 01 51.8&    -39 22 04&     11.05 &         G9V(e)&         189&75&S\\
HD 62237    &07 42 26.6&    -16 17 00&  \hspace{2mm}9.88* &  G5V  &        104&  85&S\\
\noalign{\smallskip}
\tableline
\noalign{\smallskip}
\end{tabular}
\smallskip
}
{\\
(*) the photometric values are corrected for duplicity.
The distances are from Hipparcos (H in the table) or kinematical ones,
calculated with the convergence method.\\
S=in the SACY survey; s=observed in the SACY, but outside of the sample
definition; z=\citet{zuckerman04}; M=\citet{moor06}; zi=\citet{zickgraf05}.
(Here and in the next tables,
small letters in the last column mean a star not proposed by the authors
to belong to this association).
}
\end{table}

We found 41 members for the Col Association which are presented in Table~\ref{table:cola};
only seven are not in the SACY data sample.
Three stars (one in the SACY sample) were proposed by \citet{zuckerman04}
to be members of the Tuc-Hor Association, but they fit better in the Col Association.
There are also five stars proposed by \citet{moor06}, all having debris disks.
The authors proposed another one, HD~37484, for the Tuc-Hor Association.
In fact HD~37484 has a very good solution in the convergence method for the Col Association,
but not for the Tuc-Hor Association.
Nevertheless it has a low probability (p=0.45) and should be considered only as a possible member.

We accepted some low probability members (0.6$<$p$<$0.8) for the Col
and Car associations since the convergence method gives reliable
solutions for them.  Actually the values given by the probability
model must be interpreted in a relative sense.  For a certain cutoff
value of the probability it discriminates bona fide members from the
field stars.  This value depends on the density of the bona fide
members of the association in the UVWXYZ--space relative to the
density of the possible members in the field.  Now, the Col and Car
associations, as we noted before, are loose associations immersed in a
field of young stars with similar kinematical and physical properties,
and therefore their cutoff values may be lower.  Compared to other
well defined associations, we have chosen for both associations lower
cutoff probabilities (p=0.6), but yielding consistent membership
lists.  It is another sign of the sea of young stars in this region,
possibly belonging also to the GAYA complex, but more kinematically
dispersed.  But Col and Car Associations are also large and distant,
near the SACY limit (see Table~\ref{table:finalb}; the other three
distant associations are compact or are at special locations in the
hexa-dimensional space).  Therefore we lose almost completely their low
mass population.  A deeper survey would give a better insight into these
associations.  Another approach could be to define a "GAYA"
association, with less restrictive kinematical parameters, but losing
the details.

The age obtained for the Col Association is the same as that for the Tuc-Hor and Car
associations, but it depends largely on the badly determined low equivalent width
of the Li line of GSC~8077-1788, a M0Ve wide companion of HD~31242 (sep.=18.3$\arcsec$).
Stars near K0 indicate  an older age  (see Figure~\ref{fig:gahr}),
thus the age proposed for the  Col Association in  Table~\ref{table:finalb} should be better investigated.
Compared to all other associations exhibiting an expansion in the X direction
(Figure~\ref{fig:exp}), the Col Association is the farthest towards the Galactic anti-center.
There seems to be a correlation of age with X,
in the sense that associations farther towards the Galactic center tend to be younger.
An older age for the Col Association would follow such a trend.

CD-52~381, proposed by \citet{torres00} to belong to the Hor Association,
is a more probable member of the Col Association.
\citet{chauvin03} found a possible substellar companion,
confirmed by \citet{neuh04} and \citet{chauvin05b} as a brown dwarf (M$\sim25~M_{J}$).

BD-08~1115 (PDS~111) is a wTTS in the direction of the Orion Cloud
\citep{Torres1995}, but the convergence method gives a much shorter
distance and we propose that it belongs to the Col Association.  If
the source IRAS~05222-0844 is really associated with this star and if
the membership is correct, it would have an exceptional amount of
circumstellar material for such an old star.

BD-08~1195, very similar to PDS~111, is also in the direction of the
Orion Cloud \citep{Alcala00}, and as PDS 111 we propose it to
belong to the Col Association.

HD~36329 is a double line spectroscopic binary star, discovered in the
SACY, with similar components. HD~62237, observed once, is another
possible SB2. \citet{zickgraf05} detected V1221~Tau as a double line
spectroscopic binary, probably the same as the visual pair
(sep=0.9$''$) whose brighter component is a fast rotator
\citep{cutispoto99}. AS~Col may also be a spectroscopic binary, since
there is some spread in the published velocities, but it is a fast
rotator and a short period photometric variable \citep{cutispoto99}.

There are four visual binaries in the Col Association: besides
HD~31242, CD-52~381 and V1221 Tau mentioned above, HD~21955 is a close
visual binary (sep.=0.9$\arcsec$). V1221 Tau has a possible F8 distant
companion (sep=73$''$) measured by Hipparcos - our kinematical
distance is shorter but within the parallax errors.

\subsection{The Carina Association}

\begin{table}[!h]
\caption{The high probability members proposed for the Carina Association}
\smallskip
{
\label{table:cara}
\begin{tabular}{lcclll@{\hskip7pt}r@{\hskip7pt}l}
\tableline
\noalign{\smallskip}

Name&$\alpha_{2000}$   & $\delta_{2000}$    &\hspace{3mm}V & Sp.T.    & D  &
P.&Ref.\\
&&&&&[pc]&\%&\\
\noalign{\smallskip}
\tableline
\noalign{\smallskip}

%HD 8813     &01 23 25.9&    -76 36  42&    \hspace{2mm}8.37&       G6V  &\hspace{2mm}48H&60&S\\
HD 42270    &05 53 29.3&    -81 56 53 &    \hspace{2mm}9.14&       K0V  &\hspace{2mm}59&70&S\\
%HD 40216    &05 55 43.1&    -38 06 16&     \hspace{2mm}7.46 &      F7V  &&18.5$\pm$0.6&80\\
%V1358 Ori   &06 19 08.1&    -03 26 20&     \hspace{2mm}7.91  &    F9V  &\hspace{2mm}50& 60&z?   \\
AB Pic      &06 19 12.9&    -58 03 16 &     \hspace{2mm}9.13   &    K1V(e) &\hspace{2mm}45H&60&z,S\\
%CD-40 2458  &06 26 06.9&    -41 02 54&     10.00  &         K0V   &11.6&&100\\
%CD-48 2324  &06 28 06.1&    -48 26 53&     11.08 &          G9V    &7.8 &&90\\
%D-37 2984  &06 39 46.9&     -37 50 10&    10.72 &          K1V    &16.3&&    \\
HD 49855    &06 43 46.2&    -71 58 35&      \hspace{2mm}8.94   &    G6V     &\hspace{2mm}56H&90&z,S\\
%CD-41 2572  &06 45 37.9 &   -41 12 41&     10.82  &         K0V     &12.4&&60\\
%CD-36 3202  &06 52 46.8 &   -36 36 17&     11.22  &         K2V(e)   & 9.6&&100\\
%HD 51797    &06 56 23.5&    -46 46 55&     \hspace{2mm}9.84        &   K0V(e)  &14.3 &&90\\
HD 55279    &07 00 30.5&    -79 41 46&     10.11    &   K2V  &\hspace{2mm}64H&70&z,S\\
%CD-39 3026  &07 01 51.8&    -39 22 04&     11.05   &        G9V(e)    &7.2 &&\\
CD-57 1709&  07 21 23.7 & -57 20 37&        10.72  &      K0V     & 100   &70&S\\
CD-63 408 & 08 24 05.7& -63 34 03 &         \hspace{2mm}9.87   &     G5V &104&100&S,m\\
CD-61 2010 & 08 42 00.4& -62 18 26 &          10.95  &     K0V & 147&100&S,m\\
CD-53 1875& 08 45 52.7& -53 27 28 &          10.29       &     G2V & 145&90&S,m\\
CD-75 392 & 08 50 05.4 & -75 54 38 &        10.59  &  G9V    &101&80&S\\
CD-53 2515& 08 51 56.3& -53 55 57 &          11.06*  &     G9V & 141&100&S\\
TYC8582-3040  & 08 57 45.7& -54 08 37 &          11.71   &  K2IV(e)&159&100&S,m\\
CD-49 4008  &08 57 52.2&    -49 41 51&      10.51  &    G9V    &111&100&S,m\\
CD-54 2499  &08 59 28.8&    -54 46 49&     10.08   &    G5IV  &106&100&S,m\\
CP-55 1885& 09 00 03.3& -55 38 24 &        10.83     &     G5V & 122 &100&S\\
CD-55 2543& 09 09 29.3& -55 38 27 &        10.20         &     G8V & 133&100&S\\
CD-54 2644& 09 13 16.9& -55 29 03 &        11.36    &     G5V & 132&100&S\\
V479 Car    &09 23 35.0&    -61 11 36&     10.86   &    K1V(e)  &\hspace{2mm}85H&90&S,m\\
HD 83096 & 09 31 24.9& -73 44 49 &         \hspace{2mm}7.49*  &     F2V&\hspace{2mm}79H&90&s\\
HIP 46720B & 09 31 25.2& -73 44 51 &        10.02*&     G9V &\hspace{2mm}79H&80&S\\
CP-52 2481& 09 32 26.1& -52 37 40 &        10.86     &     G8V & 139&100&S\\
CP-62 1293  &09 43 08.8&    -63 13 04&     10.44   &    G6V  &\hspace{2mm}77&100&S,m\\
%HD 86021   & 09 54 11.0& -53 38 28 &       10.73*     &     G5V &&7.3$\pm$1.2 &60    \\
%HD 2989369  &10 13 14.6&    -52 30 54&    \hspace{2mm}9.79       &    K3Ve &19.3&&\\
CD-54 4320  &11 45 51.8&    -55 20 46&     10.24   &        K5Ve  &\hspace{2mm}44&80&S\\
HD 107722    &12 23 29.0&    -77 40 51&     \hspace{2mm}8.30   &        F6&\hspace{2mm}67&60&g\\
%CD-51 6700  &12 27 56.4&    -52 11 00&     10.00   &        K0V   &15.0&&\\
%T9486-0927  &21 25 27.6 &   -81 38 28&     11.85    &   M2Ve    &26.4&&\\

\noalign{\smallskip}
\tableline
\noalign{\smallskip}
\end{tabular}
\smallskip
}
{
{

(*) the photometric values are corrected for duplicity.\\
The distances are from Hipparcos (H in the table) or kinematical ones,
calculated with the convergence method.\\
S=in the SACY survey; s=observed in the SACY, but outside of the sample definition;
m=\citet{makarov00} proposed this star as belonging to the Carina-Vela moving
group, z=\citet{zuckerman04}; g=\citet{guenther07}.}
}
\end{table}

The convergence method yields 23 high probability members for the Car
Association,\footnote{The Car Association must not be confused with
  the  Carina-Near moving group proposed by  \citet{zuckerman06},
  supposed to be much older ($\sim$200 Myr) and therefore beyond the
  scope of this review.}
presented in Table~\ref{table:cara}.
As explained for the Col Association, we chose a low cutoff probability.
Only one member was not observed  in the SACY:
HD~107722 -- it is in front of the Cha region \citep{whittet97}.
The authors proposed a distance of 74\,pc and negligible reddening.
The data used here were taken from \citet{guenther07}.
There are also three stars that \citet{zuckerman04} proposed as members of the Tuc-Hor Association,
all in the SACY sample.

\citet{makarov00} found a sparse young moving group located in Carina and Vela and
from their list of 58 candidate members, eight are probable members of the Car Association.
Nevertheless,  their moving group is more similar to the Argus Association
and it will be discussed in Section~7.
The age of the Car Association is similar to the one of the Tuc-Hor Association.

AB~Pic was proposed by \citet{song03} as a member of the Tuc-Hor Association
and we now change it to the Car Association.
It has a low mass companion detected by \citet{chauvin05c},
at the  brown dwarf/planet boundary,
and it may be one of the few planets detected by direct imaging (see Section 9).

There are two double line spectroscopic binary stars in the
{Carina Association}: CD-53~2515,
discovered in SACY and observed on 17 nights with an orbital period of
24.06 days,
and the faint visual companion of HD~83096
(sep.=1.9$\arcsec$), the only visual binary of this association,
observed only once spectroscopically.

\section{The TW~Hya Association}

Although the TW~Hya Association was the first of the nearby associations
discovered, it is one of the hardest to establish memberships for by
the convergence method, due to the lack of good kinematical data.
The reasons for this lack are two-fold.
First, as many of the members proposed are faint they have no accurate
proper motions, like those of the Hipparcos mission.
And second, there is a large number of binary stars in  the TW~Hya Association.
Close visual binaries make the determination of proper motions using
photographic plates less reliable.
If the orbit has a relatively short period it is necessary to use the systemic motion,
which is also not easy to obtain.
Moreover, there are suspected long period single line spectroscopic
binaries, systems for which it is hard to obtain  systemic velocities.

One example is TWA~3 (Hen~3-600), a visual binary separated by 1.5$\arcsec$:
its barycenter may not coincide with the photocenter on a plate.
There could be orbital motion not taken into account in the published proper motions.
The separation of 1.5$\arcsec$ does not allow to obtain separate spectra
with FEROS or similar fiber spectrographs.
For slit spectrographs the effort to obtain separate spectra may arise in bad slit
centering, resulting in velocity shifts.
We may therefore expect that the radial velocity of TWA~3 is unreliable.
Moreover (or as a consequence...), one of the stars of the pair  is suspected to be a long period
single line spectroscopic binary, maybe even both!
Actually \citet{Ray06} claim that TWA~3A is a double line spectroscopic binary
star, but as they give no details, we can not correct the photometry for duplicity and
TWA~3 also requires  more precise
proper motion determinations (the USNO proper motions used here are of lower quality).
In our solution both stars are over-luminous, which may be another indication of these
possible missing companions.

\begin{table}[!t]
\caption{The high probability members proposed for the TW~Hya Association}
\smallskip
{
\label{table:twa}
\begin{tabular}{lcclllrl}
\tableline
\noalign{\smallskip}
Name&$\alpha_{2000}$   & $\delta_{2000}$    &\hspace{3mm}V & Sp.T.    & D  & P.&Ref.\\
&&&&&[pc]&\%&\\
\noalign{\smallskip}
\tableline
\noalign{\smallskip}
%TWA 21       &10 13 14.6     &-52 30 54      &\hspace{2mm}9.79  &      K3Ve  &19.2&             &70\\
TWA 7       &10 42 30.1     &-33 40 16      &11.65    &M2Ve  &28     &100&W,S\\
TWA 1       &11 01 51.9     &-34 42 17      &11.07       &K6Ve  &53  &100&G,S\\
TWA 2       &11 09 13.8     &-30 01 40      &11.43*   &M2Ve  &41     &90&G,S\\
TWA 3B      &11 10 27.8    &-37 31 53      &13.07*   &M4Ve  &37       &90&G\\
TWA 3A      &11 10 27.9    &-37 31 52      &12.57*   &M4Ve  &37       &90&G\\
TWA 13A     &11 21 17.2     &-34 46 46      &11.46*   &M1Ve  &55      &100&St\\
TWA 13B     &11 21 17.4     &-34 46 50      &11.96*    &M1Ve  &55     &100&St\\
TWA 4       &11 22 05.3     &-24 46 40      &\hspace{2mm}9.42*&K5V   &44&100&G,S\\
TWA 5A      &11 31 55.3     &-34 36 27      &12.12*  &M2Ve  &45        &90&G,S\\
TWA 5B      &11 31 55.4     &-34 36 29      &20.40     &M8Ve  &45      &90&W\\
TWA 8A      &11 32 41.2     &-26 51 56      &12.23     &M3Ve &  39     &100&W\\
TWA 8B      &11 32 41.2     &-26 52 09      &15.3      &M5Ve & 40      &100&W\\
TWA 26      &11 39 51.1     &-31 59 21      &20.5      &M8Ve  &41      &100&Gi\\
TWA 9B      &11 48 23.7     &-37 28 48	    &14.00     &M1Ve  &68      &90&W,S\\
TWA 9A      &11 48 24.2     &-37 28 49	    &11.26     &K5Ve  &68      &100&W,S\\
TWA 27      &12 07 33.4     &-39 32 54      &20.2      &M8Ve  &53      &100&Gi\\
TWA 25      &12 15 30.7     &-29 48 43      &11.44     &M1Ve  &51      &100&So,S\\
TWA 20      &12 31 38.1     &-45 58 59      &13.3      &M3Ve  &73      &100&R\\
TWA 16      &12 34 56.4     &-45 38 07      &12.8*     &M1Ve  &70     &100&Zb\\
TWA 10      &12 35 04.2     &-41 36 39      &12.96     &M2Ve  &52     &100&W\\
TWA 11B     &12 36 00.6     &-39 52 16      &12.80    &M2Ve  &    67H&90&Zb,S\\
TWA 11A     &12 36 01.1     &-39 52 10      &\hspace{2mm}5.78 &A0V&67H &100&Zb\\
\noalign{\smallskip}
\tableline
\noalign{\smallskip}
\end{tabular}
\smallskip
}
{\hspace{8mm}(*) the photometric values are corrected for duplicity.\\
The distances are from Hipparcos (H in the table) or kinematical ones,
calculated with the convergence method.\\
S=in the SACY survey; G=\citet{gregorio92}; Gi=\citet{gizis02};
R=\citet{reid03}; So=\citet{song03}; St=\citet{sterzik99}; W=\citet{webb99};
Zb=\citet{zuckerman01c}.}
\end{table}

Lists of potential members of the TW~Hya Association appear in
\citet{reid03}, \citet{torresg03}, \citet{zuckerman04},
\citet{mamajek05} and \citet{barrado06}.  Stars in the TW~Hya
Association have traditionally been designated with association names.
There are 28 stars with TWA designations, eight of them being visual
binaries with separation greater than 1$\arcsec$ that allow us to
study each component separately.  From these 36 stars, 11 are in the
SACY sample and we obtained spectra for 17 stars.  We tried to
introduce all proposed stars to be tested by the convergence method.
For the sake of getting the most homogeneous data we used SACY data whenever
possible.  The other data were taken from \citet{mamajek05}, who got
proper motions for some stars and compiled radial velocities for all
stars.  There are four stars that lack kinematical data and could not
be used in this way: TWA~15B, TWA~22, TWA~23 and TWA~28
\cite[SSPM~J1102-3431: a young brown dwarf,][]{sholz05}.

Applying the convergence method to the 32 remaining stars we found 22
to be high probability members (seven in the SACY sample).  They are
listed in Table~\ref{table:twa}.  From the ten stars rejected, six are
also rejected by \citet{mamajek05}: TWA~12 (p=0.6, and farther in Y
direction); TWA~17 (p=0, distant and with bad resultant spatial
velocity); TWA~18, TWA~19A and B, TWA~24 (all with p=0 and belonging
to the Sco-Cen Association).  This possible membership of the Sco-Cen
Association for some of the TW~Hya Association candidate stars was
proposed earlier by \citet{mamajekasp}.  The other four rejected stars
are: TWA~15 (p=0, the proper motions are unreliable but, anyway, the
photometric data impose an unacceptably large distance); TWA~21
(p=0.6, is farther in Z direction); TWA~6 (p=0.7, it is in the SACY
sample); TWA~14 (p=0.3, is somewhat farther in Y direction).  These
last two stars show spreads in their published radial velocities and
could be single line spectroscopic binaries.  With the determination
of the systemic velocities they could eventually become bona fide
members of the TW~Hya Association.  TWA~6 and TWA~14 should be
considered as possible TW~Hya Association members.

TW~Hya itself (TWA~1) is a cTTS \citep{RucinskiKrautter1983}, it is
very well studied and possesses an optical circumstellar disk
\citep{krist00}.  There is an extensive bibliography both about TW~Hya
and its disk.
% (see SIMBAD).
It is also one of the oldest cTTS known.  As already mentioned in
Section~1.1, in the PDS program we had searched for other young stars
around it using IRAS source or emission line stars in the \citet{SW72}
catalog, resulting in the discovery of four stars \citep{delareza89,
gregorio92}, later confirmed by \citet{Kastner97}.  They were
designated with the names TWA~2 to TWA~5.

TWA~2 (CD-29~8887) was serendipitously discovered \citep{delareza89}
at the beginning of the PDS survey (the nearby IRAS source is a
galaxy) and no disk has been detected for this star.  It is a visual
binary (sep.=0.6$\arcsec$).

TWA~3 (Hen~3-600, see above) was the first star observed in the PDS survey.
Although the binary system has roughly equal mass components, only the primary
(possibly a double line spectroscopic binary)  has a disk \citep{ray99}.

\begin{figure}[]
  \begin{center}
  \begin{tabular}{c}
  \resizebox{1.0\hsize}{!}{
\includegraphics[draft=False]{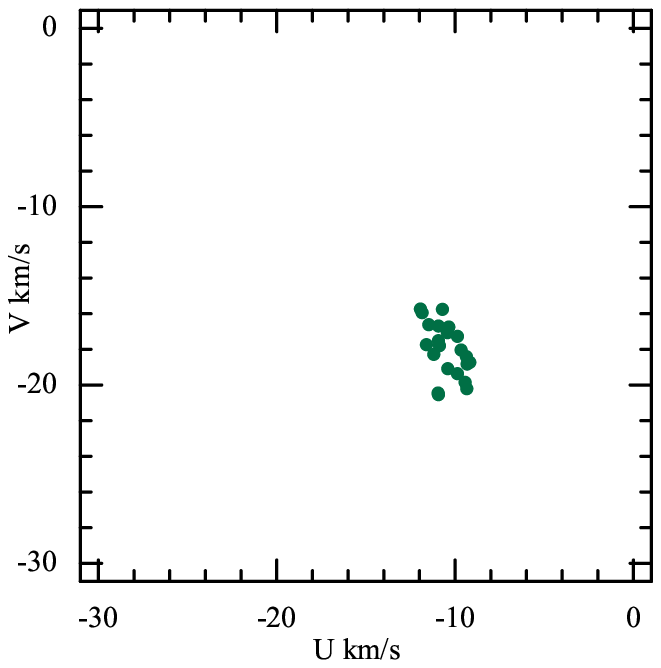}
\includegraphics[draft=False]{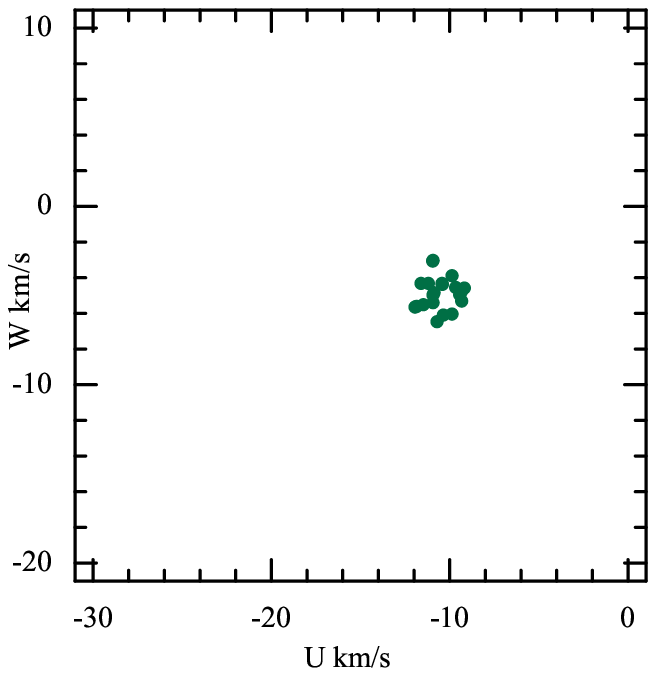}

\includegraphics[draft=False]{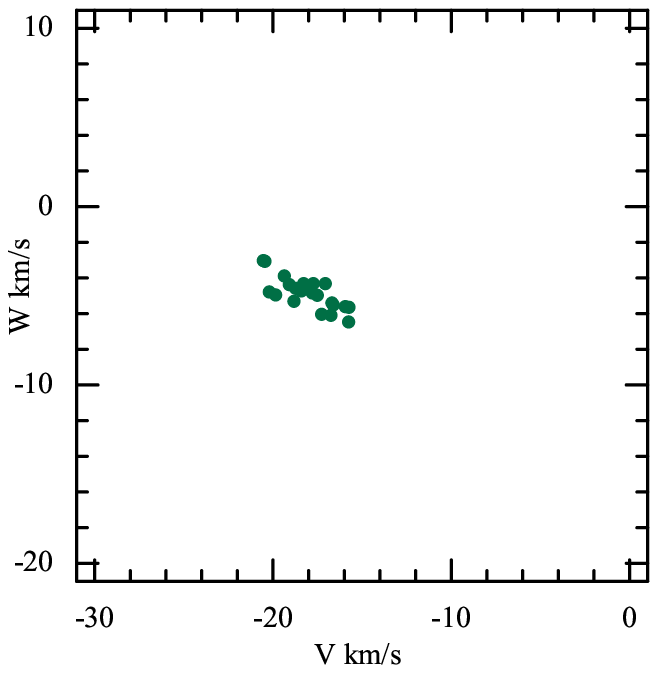}
}\\
\resizebox{1.0\hsize}{!}{
\includegraphics[draft=False]{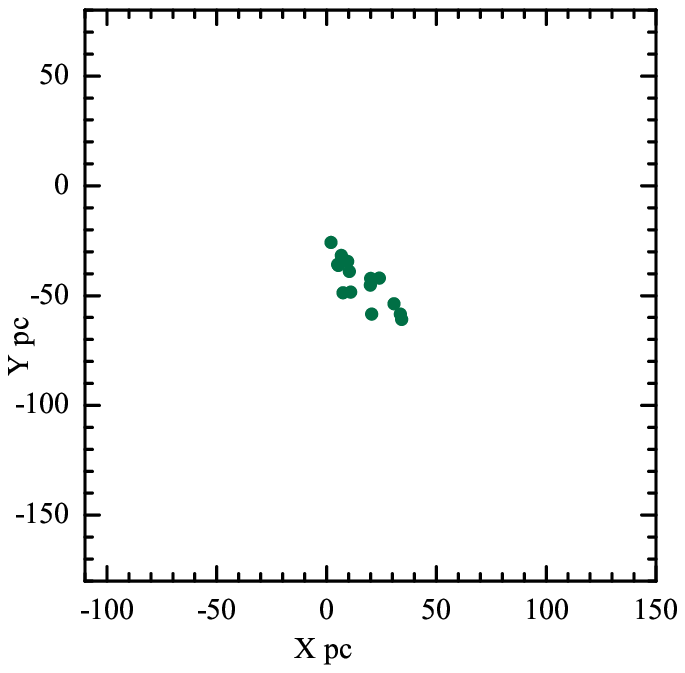}
\includegraphics[draft=False]{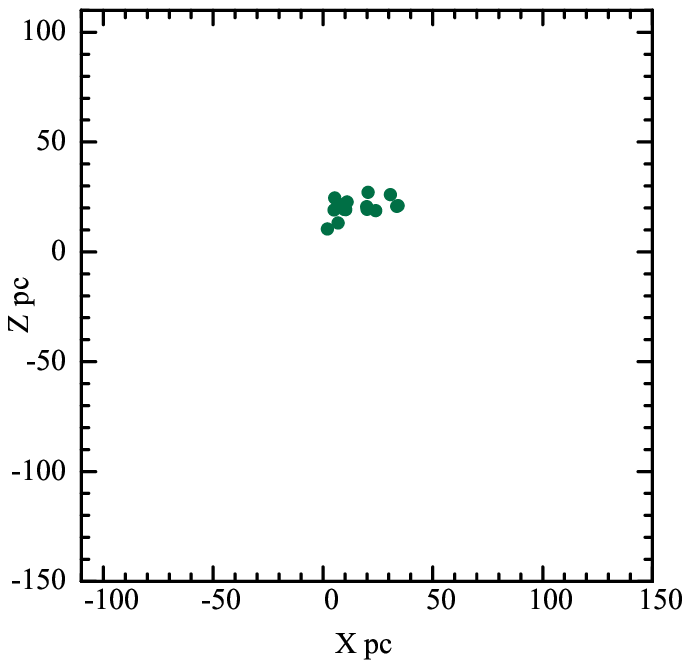}

\includegraphics[draft=False]{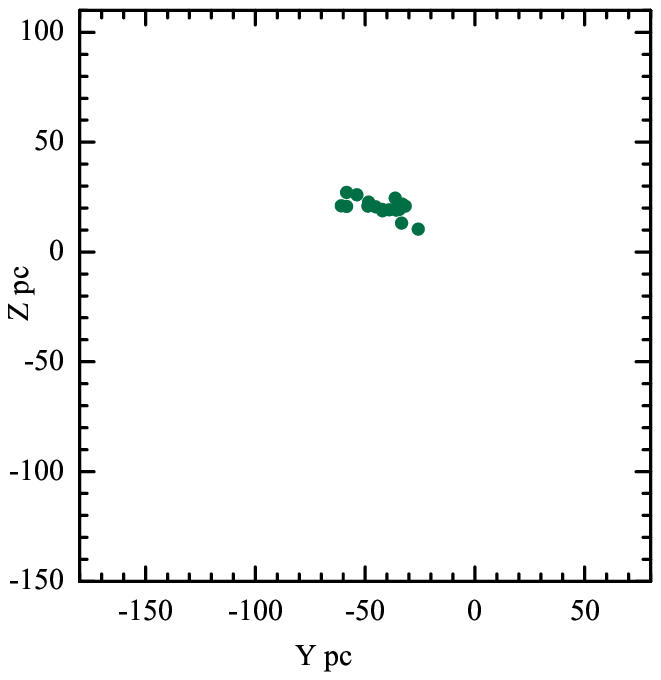}
}\\
  \end{tabular}
    \end{center}
\caption{Combinations of the sub-spaces of the UVWXYZ--space for the TW~Hya
Association showing a well defined clustering
in both kinematical and spatial coordinates.
}
\label{fig:twxyz}
\end{figure}
\begin{figure}[]
  \begin{center}
  \begin{tabular}{c}
 \resizebox{1.0\hsize}{!}{
\includegraphics[draft=False]{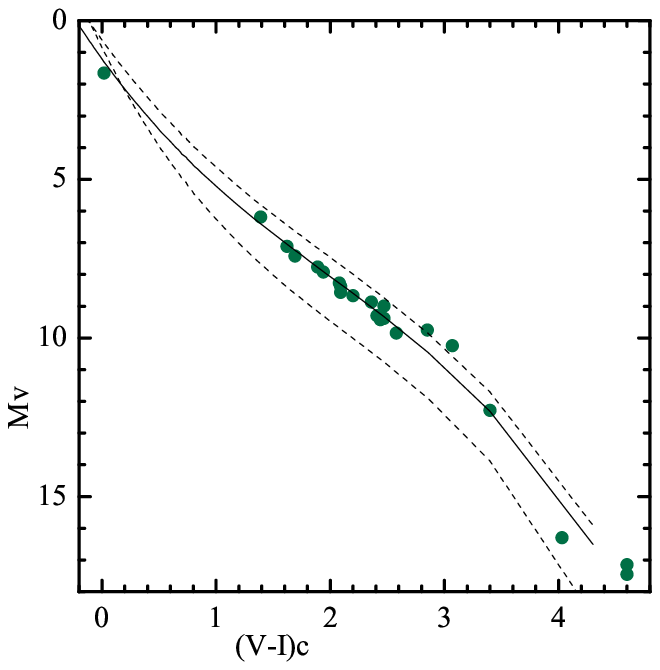}
\includegraphics[draft=False]{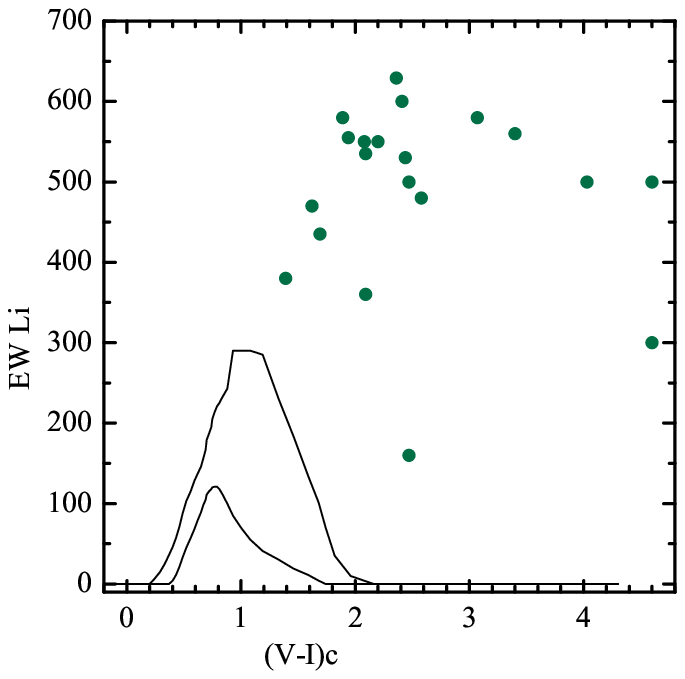}
}
   \end{tabular}
    \end{center}
\caption{{\it Left:}  The HR diagram of the members proposed for the TW~Hya
Association; the over-plotted isochrones
are the ones for 5, 8, and 70~Myr.
{\it Right:} The distribution of Li equivalent width as a function of $(V-I)_C$; the
curves are the upper and lower limits for the age of the Pleiades
\citep{neu97}.
}
\label{fig:twhr}
\end{figure}

TWA~4 (HD~98800; TV~Crt) was first proposed to be a young star by
\citet{gregorio92}.  It is a visual binary star (sep.=0.8$\arcsec$),
and \citet{torres95} established that both visual components are
themselves spectroscopic binaries, the primary being a single line
spectroscopic binary (period=262~d) and the secondary a double line
spectroscopic binary (period=315~d).  A preliminary orbit of the
visual system is presented by \citet{tokovinin99}.  The subsystem
TWA~4B has a visual and physical orbit determined by \citet{bolden05}
using interferometry with the Keck telescope and images from the Hubble
Space telescope.  TWA~4B is surrounded by a disk \citep{koerner00}.

TWA~5 (CD-33~7795) harbors
the brown dwarf companion  TWA~5B (sep.=2$\arcsec$), \citep{lowrance99, webb99}.
TWA~5A is also resolved as a close visual binary with nearly equal components by
\citet{mac00}, and an orbit solution with a 5.9~yr period is presented by \citet{kono07}.
The situation is even more complicated as some authors detected radial velocity
variations in TWA~5A, indicating another object in the system  \citep[see][]{torresg03}.
The lines sometimes suggest a double line spectroscopic binary
\citep[see][]{reid03} and could be from the close visual components.
The apparent high rotation may come from the blended lines of both components.
Nevertheless, the mean lines of the 10 SACY spectra are consistent with a
sinusoidal variation with a period
very close to 1~day.

TWA~7 (GSC~7190-2111) was proposed as a TW~Hya Association member by \citet{webb99}
(together with TWA~6 to TWA~10).
A suspected disk \citep{zuckerman01d,low05} was confirmed by \citet{matthews07}.

TWA~8 (GSC~6659-1080) was discovered as an active star by
 \citet{fleming98}, and was proposed as a TW~Hya Association member by
 \citet{webb99}. They also detected a
 faint companion at a distance of 13$\arcsec$ (TWA~8B).

TWA~9 (HIP~57589), another one of the members proposed by \citet{webb99}, was
independently discovered by \citet{jensen98}
and it is also a visual binary (sep.=5.8$\arcsec$).

TWA~10 (GSC~7766-0743), proposed as a TW~Hya Association member by
\citet{webb99}, has H$\alpha$ variations, probably from a chromospheric flare
event, not due to accretions \citep{Ray06}.

TWA~11 (HR~4796) is the only early-type star (spectral type A0) in  the TW~Hya
Association -- all other members are later than K5!
The presence of a disk around it  was proposed by \citet{jura93} and later confirmed
through infrared imaging by \citet{koerner98} and \citet{ray98}.
\citet{webb99} proposed that this star belongs to  the TW~Hya Association.
\citet{zuckerman01c} searched for young stars near TWA~11 and found six more TTS.
They proposed all these stars to belong to  the TW~Hya Association (TWA~14-19).
In our analysis, only TWA~16 is a bona fide member.
\citet{jura93} showed that TWA~11A/B (sep.=7.8$\arcsec$) form a physical
system with the A0 primary (they rejected the ``C" star) and  they suggested
that TWA~11B may be a young star, later confirmed by \citet{stauffer95}.

TWA~13 (CD-34~7390) is a visual binary (sep.=5.1$\arcsec$).
Both components have TTS signatures and are proposed as
TW~Hya Association members by \citet{sterzik99}.

TWA~16 is  a member proposed by \citet{zuckerman01c}.
They argue it is a close visual binary (sep.=0.7$\arcsec$), but this needs confirmation.

TWA~20 (GSC~8231-2642)  is rejected by \citet{song03} due to its weak Li (EW=160~m\AA).
Indeed, this value is somewhat discrepant, as  can be seen in Figure~\ref{fig:twhr}.
However, it may depend critically on its photometry (which may be highly uncertain),
and Li may easily be depleted in this star.
\citet{mamajek05} also agrees with its membership.
However, TWA~20 is also claimed to be a spectroscopic binary \citep{Ray06},
so its radial velocity might be uncertain.

TWA~21-25 were proposed by \citet{song03}.
We rejected TWA~21 and TWA~24 and there are no complete kinematical data to test TWA~22 and TWA~23.
TWA~25 (GSC~7760-0283) is a bona fide member.

TWA~26 and TWA~27 are brown dwarfs proposed as  TW~Hya Association members by
\citet{gizis02}.

TWA~27 is a visual binary with a companion of Jupiter mass \citep{chauvin04,chauvin05a}.
The system was analyzed by \citet{mohanty07} who found that the planetary mass
object would be underluminous.
They suggested an edge-on disk around it to explain this discrepancy \cite[but see] []{marley07}.
Its age and distance match very well with those proposed here.
It has also an accretion disk \citep{riaz06, riaz07}.
New trigonometric distances have been recently determined by three groups:
\citet{gizis07} found 54$\pm$3\,pc;
\citet{biller07},  59$\pm$7\,pc; \citet{ducourant07}, 52.4$\pm$1.1.
These determinations agree very well with our kinematical distance (53\,pc; see
Table~\ref{table:twa}).

For a discussion of rotation in the TW~Hya Association see \citet{lawson05}.
Using the period distribution they arrived at a similar conclusion about the
TW~Hya Association memberships presented here.
One of the more enigmatic properties of  the TW~Hya Association is the absence
of F, G and early-type K stars.
As they would be very bright, this deficit can not be explained by observational bias.

\section{The $\epsilon$~Cha Association }

A kinematic group of young stars in the Cha region was proposed earlier by \citet{frink98}.
The group was discussed by \citet{terranegra99}, who defined 12 members (four in the SACY sample)
and reviewed by \citet{mamajek00}, when studying the $\eta$~Cha cluster,
calling it the  $\epsilon$~Cha group.
Using the first papers, they suggested nine members, only one in the SACY sample.
\citet{feige03} selected X-ray sources, using the
Chandra X-ray Observatory, near one of the proposed stars, DX~Cha, and found
some interesting faint young stars that could be related to the star $\epsilon$~Cha.
However, no kinematical data are available for the application of the convergence method.
Using IR data, \citet{luhman04} proposed three more candidates that also lack  kinematical data.
\citet{zuckerman04}, reviewing the literature, suggest the existence of  a similar association in this
region, with 17 members, which they call the "Cha-Near" region.
It is considerably larger in space dimension than the proposed $\epsilon$~Cha group.
Among their list, there are only four stars in the SACY sample and
six proposed members have no complete kinematical data.
Their list excludes the region close to the star  $\epsilon$~Cha,
but four of the "Cha-Near" members  are in the group proposed by \citet{mamajek00}.
Thus the situation is unclear -- {\it are these three proposed groups the same? }

\citet{mamajek00} used the available kinematical data of the members of the  $\eta$~Cha cluster,
obtaining a heliocentric velocity of  (--11.8, --19.1, --10.5)~km~s$^{-1}$ and a
distance of 97.3$\pm$3.0~pc, close to the velocities and distances we had found
for the $\epsilon$~Cha Association in previous analysis \citep{torres03a, torres03}.
This poses another question:
{\it what connection exists between the $\epsilon$~Cha group and the $\eta$~Cha cluster?}

\begin{table}[!h]
\caption{
The high probability members proposed for the $\epsilon$~Cha Association}
\smallskip
{
\label{table:cha}
\begin{tabular}{l@{\hskip7pt}c@{\hskip7pt}c@{\hskip7pt}lll@{\hskip7pt}r@{\hskip7pt}l}
\tableline
\noalign{\smallskip}
Name&$\alpha_{2000}$   & $\delta_{2000}$    &\hspace{3mm}V & Sp.T.    & D  &
P.&Ref.\\
&&&&&[pc]&\%&\\
\noalign{\smallskip}
\tableline
\noalign{\smallskip}
\multicolumn{8}{c} {The $\eta$~Cha cluster kinematical members}\\
\noalign{\smallskip}
\tableline
\noalign{\smallskip}
EG Cha   &08 36 56.2      &-78 56 46  &11.36*&K4Ve&\hspace{2mm}99&100&S,m\\
$\eta$~Cha&08 41 19.5&   -78 57 48  &\hspace{2mm}5.46&B8V&\hspace{2mm}97H&90&m\\
RS Cha   &08 43 12.2&   -79 04 12   &\hspace{2mm}6.80*&A7+A8&\hspace{2mm}98H&100&m\\
EQ Cha   &08 47 56.9&   -78 54 54   &13.92* & M3e&122&100&m,c\\
\tableline
\noalign{\smallskip}
\multicolumn{8}{c} {Field members}\\
\noalign{\smallskip}
\tableline
\noalign{\smallskip}
HD 82879 &09 28 21.1&   -78 15 35   &\hspace{2mm}8.99& F6V&120&80&c\\
CP-68 1388& 10 57 49.3&  -69 14 00&  10.39&K1V(e)&112&70&S\\
DZ Cha      &11 49 31.9    &-78 51 01&12.9V &M0Ve  &110 &100&L,g,c\\
T Cha      &11 57 13.5    &-79 21 32&12.0V &K0Ve  &109 &100&M,T,Z\\
GSC9415-2676&11 58 26.9&   -77 54 45&14.29& M3e  &123&100&L,T\\
EE Cha     &11 58 35.4&   -77 49 31&\hspace{2mm}6.73 & A7V&105H &100&M,Z\\
$\epsilon$~Cha&11 59 37.6&   -78 13 19&\hspace{2mm}5.34* & B9V&111H &100&M\\
HIP 58490     &11 59 42.3&  -76 01 26&11.31&     K4Ve&\hspace{2mm}93H&90&S,M,T,Z\\
DX Cha      &12 00 05.1&   -78 11 35&\hspace{2mm}6.73* &A8Ve&116H &100&M,T\\
HD 104237D  &12 00 08.3&   -78 11 40& 14.28 &  M3Ve&116 &100&F\\
HD 104237E  &12 00 09.3&   -78 11 42& 12.08 & K4Ve&116 &100&F\\
HD 104467   &12 01 39.1&    -78 59 17& \hspace{2mm}8.56& G3V(e) &104&100&S,T,Z\\
GSC9420-0948 &12 02 03.8&   -78 53 01& 12.48& M0e    &120&100&c\\
GSC9416-1029 &12 04 36.2&   -77 31 35& 13.81*& M2e    &123&100&Z,T\\
HD 105923   &12 11 38.1&    -71 10 36& \hspace{2mm}9.16 &G8V &115&80&S\\
%TYC8986-0497&12 16 30.1&    -67 11 48& 11.56&1.44&K4IVe& 7.9&&80\\
GSC9239-1495&12 19 43.5&    -74 03 57&  13.08&M0e  &105&100&Z,T\\
GSC9239-1572&12 20 21.9&    -74 07 39&  12.85*&K7e  &110&100&Z,T\\
CD-74 712  &12 39 21.2&    -75 02 39 & 10.30&K3e  &103&90&S,T,Z\\
CD-69 1055 &12 58 25.6&    -70 28 49 & \hspace{2mm}9.95&K0Ve &101&80&S\\
MP Mus     &13 22 07.6&    -69 38 12 & 10.35&K1Ve&103&80&S,g\\

\tableline
\noalign{\smallskip}
\end{tabular}
}
{
{

(*) the photometric values are corrected for duplicity.\\
The distances are from Hipparcos (H in the table) or kinematical ones,
calculated with the convergence method.\\
S=in the SACY survey; F=\citet{feige03}; L=\citet{luhman08};
M=\citet{mamajek00}; T=\citet{terranegra99};
Z=\citet{zuckerman04}, member of the "Cha-Near";
c=\citet{covino97}, wTTS in Cha region;
g=\citet{gregorio92}, (PDS);
m=\citet{mamajek99}, member of the $\eta$~Cha Cluster.}
}
\end{table}

Unfortunately, to explore these questions, we have few stars in the SACY sample
in any of these proposed groups,
and only a fraction of their proposed members have kinematical data.
Thus we included samples of young stars from other surveys in this region --
\citet{covino97}, \citet{melo03}, \citet{james06}, \citet{guenther07} and some PDS data.
We also added  six stars of the $\eta$~Cha cluster that have kinematical data:
EG~Cha is in SACY; for EM~Cha we took the data from \citet{covino97}; for EO~Cha and EQ~Cha
from \citet{guenther07}; for $\eta$~Cha from SIMBAD and for RS~Cha  from \citet{alecian05}.

\begin{figure}[!h]
  \begin{center}
  \begin{tabular}{r}
  \resizebox{1.0\hsize}{!}{
\includegraphics[draft=False]{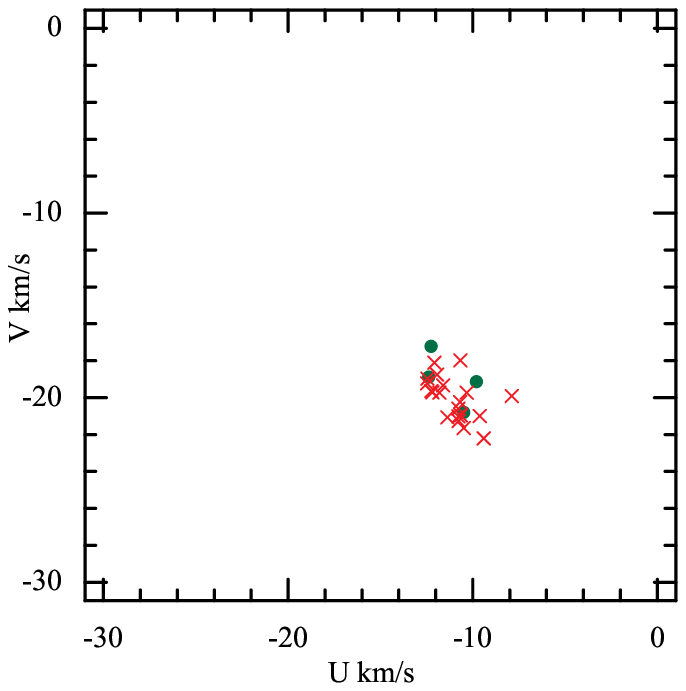}
\includegraphics[draft=False]{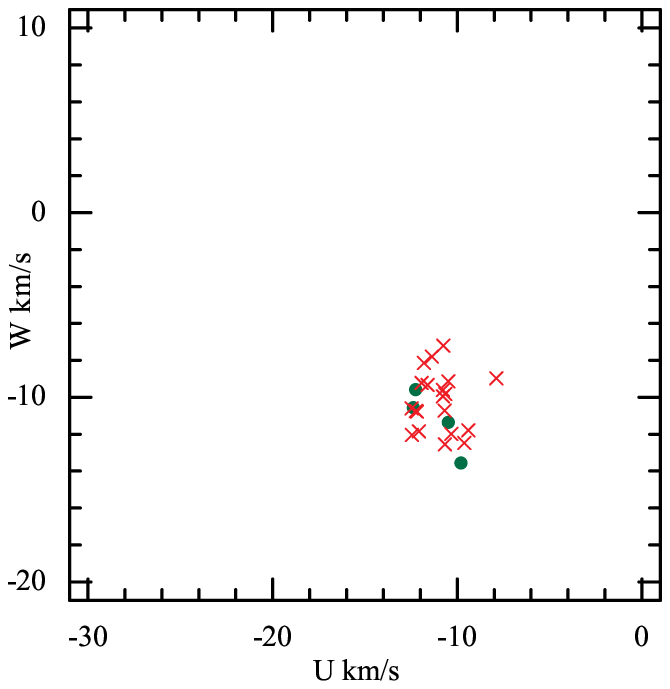}

\includegraphics[draft=False]{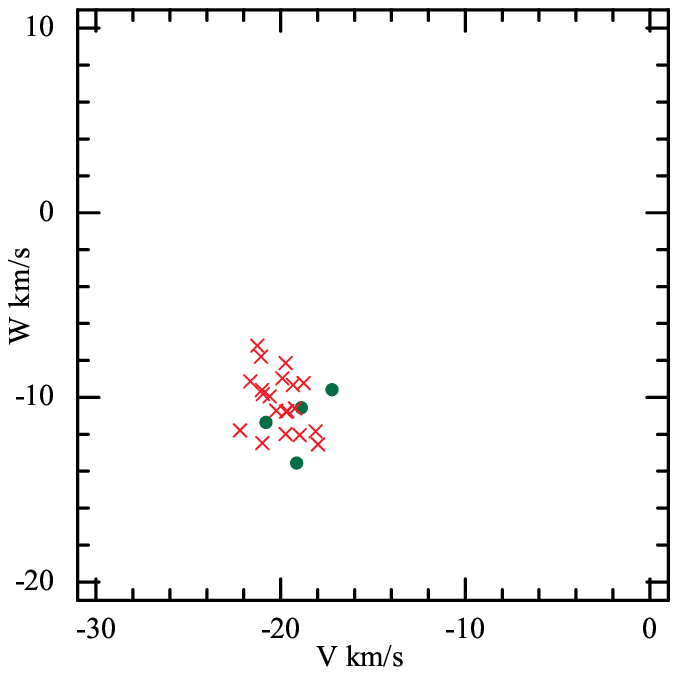}
}\\
  \resizebox{1.0\hsize}{!}{
\includegraphics[draft=False]{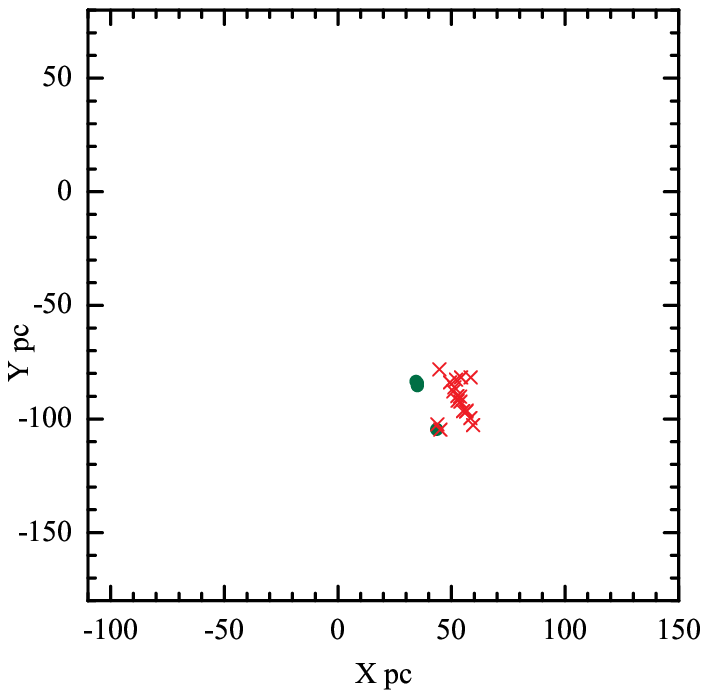}
\includegraphics[draft=False]{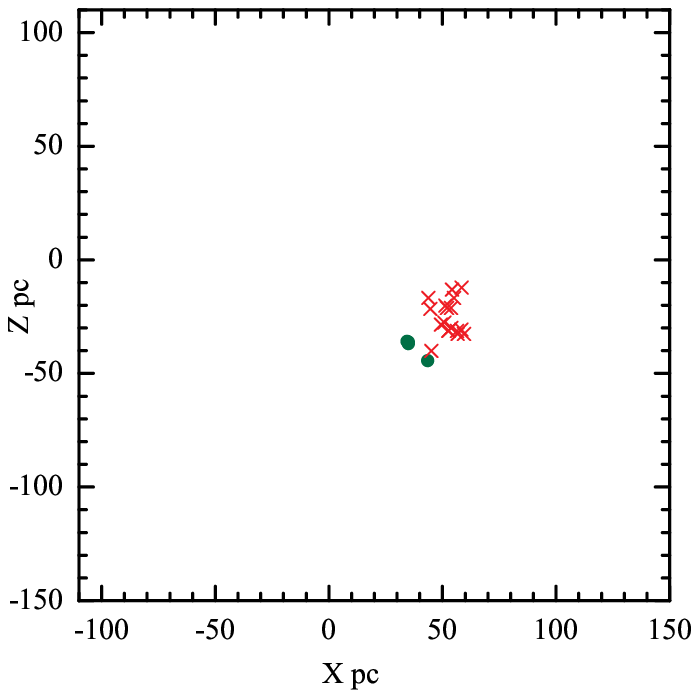}

\includegraphics[draft=False]{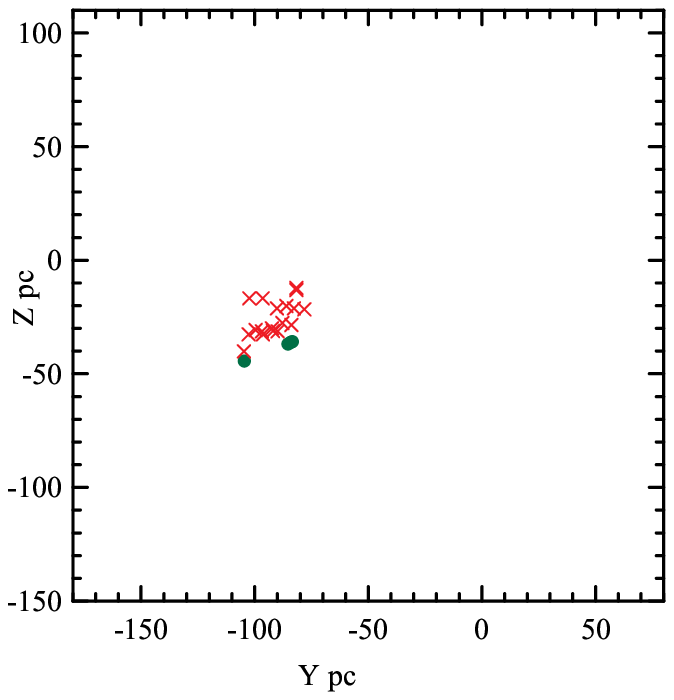}
}\\
  \end{tabular}
    \end{center}
    \caption{Combinations of the sub-spaces of the UVWXYZ--space for the
$\epsilon$~Cha Association showing a well defined clustering in both kinematical
and spatial coordinates. Crosses represent the $\epsilon$~Cha Association field
members and filled circles, the $\eta$~Cha members.}
\label{fig:echxyz}
\end{figure}

\begin{figure}[!h]
   \begin{center}
  \begin{tabular}{c}
\resizebox{1.0\hsize}{!}{
\includegraphics[draft=False]{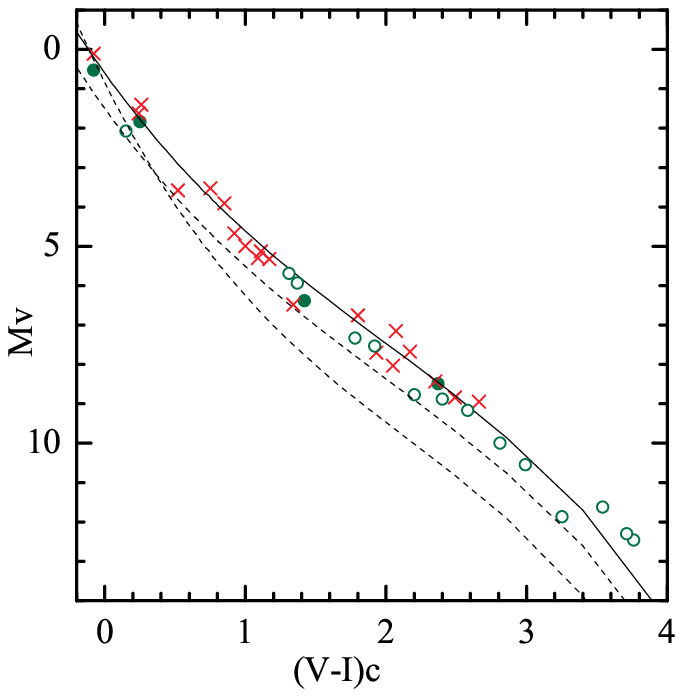}
\includegraphics[draft=False]{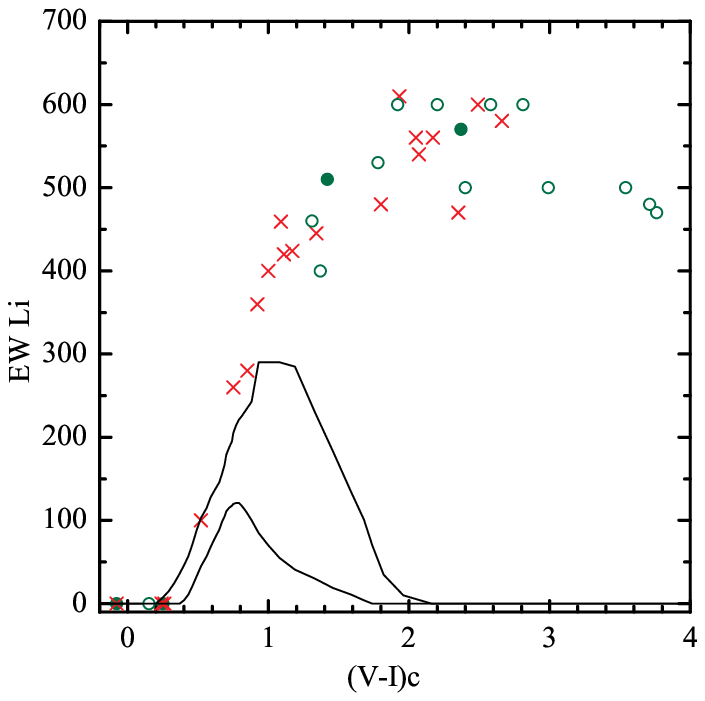}
}
     \end{tabular}
   \end{center}
\caption{{\it Left:} The HR diagram of the members proposed for the $\epsilon$~Cha
Association;
the over-plotted isochrones are the ones for 5, 8, and 70~Myr.
Crosses represent the $\epsilon$~Cha Association
field members, filled circles the $\eta$~Cha members kinematically proposed as
belonging to
the $\epsilon$~Cha Association, and  open circles the $\eta$~Cha members without
complete
kinematical data (their distance was supposed to be 108\,pc).
{\it Right:} The distribution of Li equivalent width as a function of $(V-I)_C$; the
curves are the upper and lower limits for the age of the Pleiades
\citep{neu97}.
}
\label{fig:echr}
\end{figure}

The application of the convergence method yields 24 {\it kinematical} members proposed for
the $\epsilon$~Cha Association (Table~\ref{table:cha} and
Figures~\ref{fig:echxyz} and ~\ref{fig:echr}), near the 5~Myr isochrone,
and at a mean distance of 108~pc, eight of them in the SACY sample.
Due to the low extension in X direction (Figure~\ref{fig:echxyz}) we used no expansion.
This solution is distinct from the preliminary one by  \citet{torres03a, torres03},
being farther away and larger.
It has eight new members proposed, four of them belonging to the SACY sample, besides
those of previous suggested associations.
As this "new"  association  includes  $\epsilon$~Cha itself, its brightest member,
we keep with the old designation of the association.
This association is nearer and older than the more compact background
ChaI star forming region (see the chapter by Luhman in this book).

The present solution includes nine of the 12 stars of the kinematical
group proposed by \citet{terranegra99} (only two not in the
"Cha-near").  Two stars of their list have no published radial
velocity and only one star is rejected (GSC~9415-1685).  It also
includes five stars of the $\epsilon$~Cha group \citep{mamajek00}
(three of them are also in the "Cha-Near" list), and two faint
companions of HD~104237 (DX~Cha), proposed by \citet{feige03}.  It is
larger than "Cha-Near", including eight of its members proposed.
Three of the "Cha-Near" stars with kinematical data were rejected:
HD~99827 is a member proposed for the AB~Dor Association (see
Section~8); again, GSC~9415-1685, using proper motions from UCAC2 and
radial velocity from \citet{guenther07}, is very far from the
convergence solution; and the wTTS DW~Cha.  This last star has a close
visual companion (sep.=0.07$\arcsec$) detected by \citet{koler01} and
a wide one, GSC~9415-2676, at 16.1$\arcsec$.  The solution of the
convergence method for GSC~9415-2676, using the UCAC2 proper motions
and the radial velocity from \citet{covino97}, defines it as a high
probability member, with a larger distance than the Hipparcos parallax
for DW~Cha (but still at 2.6$\sigma$).  Perhaps the close visual
companion, not detected by Hipparcos, perturbs the parallax and proper
motions (and even the radial velocities) of DW~Cha.  If we use the
UCAC2 proper motions and no trigonometric parallax, the star would be
an $\epsilon$~Cha Association member and form a physical pair with
GSC~9415-2676.  Even using the Hipparcos data it has a 70$\%$
membership probability, but with very bad kinematics.  Then, DW~Cha
must be considered a possible member of the $\epsilon$~Cha Association
and its astrometric data deserve re-examination.  After this analysis
was finished, \citet{luhman08} published a new list of members, two of
them are in Table~\ref{table:cha} and one is GSC~9415-2676.  Except
for DW~Cha, the other members proposed have no good-quality
kinematical data to be tested by the convergence method.

Finally, from the six members of the $\eta$~Cha cluster having kinematical data,
four are high probability members of this solution for the $\epsilon$~Cha Association.
The two not included in the solution are EM~Cha, a double line spectroscopic
binary  \citep{lyo03} without systemic velocity,
and EO~Cha, with large discrepancy between
the proper motions from UCAC2 and from \citet{ducourant05}.
This kinematical study indicates that the $\eta$~Cha cluster may belong to the
$\epsilon$~Cha Association.
Since this conclusion is based on only four of the 18 members proposed for
the $\eta$~Cha cluster, this must be investigated further, with more extensive kinematical data.


When we try to obtain a photometric parallax for the other members of the cluster,
using the 5~Myr isochrone, we find a mean distance of 120~pc.
In fact both these 14 $\eta$~Cha cluster members, using the 108~pc mean distance, and
the 24 members proposed for the $\epsilon$~Cha Association are slightly
displaced below the 5~Myr isochrone  (see Figure~\ref{fig:echr}).
A compromise could be to consider the $\epsilon$~Cha Association having an age around 6-7~Myr.
In fact, \citet{jilinski05}, using their dynamical approach,  argue that
the $\epsilon$~Cha Association and the $\eta$~Cha open cluster were formed together 6.7~Myr ago.
It is interesting to note that the ages obtained with the convergence method for the $\epsilon$~Cha,
the TW~Hya and the $\beta$~Pic associations  (all of them possibly formed in the Sco-Cen
Association), that is, $\sim$6~Myr, 8~Myr
and 10~Myr (Table~\ref{table:finalb}), all agree with dynamical ages
obtained for the same associations: $\epsilon$~Cha Association, with 6.7~Myr \citep{jilinski05};
TW~Hya Association with 8.3~Myr \citep{reza06}  and $\beta$~Pic Association,
with 11.2~Myr \citep{ortega02, ortega04}.
Clearly distinct ages for these three groups appear to be present.

The new definition of the $\epsilon$~Cha Association, given in
Table~\ref{table:cha}, with its young age,
explains better the evolutionary status of
MP~Mus (PDS~66), commonly classified as an ``old'' cTTS.
Its old age  arises mainly from the suggestion
that it is a member of the LCC \citep{mamajek02}, that would have an age of about 15~Myr
(but see the discussion of Preibisch \& Mamajek in their chapter in this book).
Our convergence method gives a higher likelihood for MP~Mus to be an
$\epsilon$~Cha Association member than a LCC one.
\citet{silverstone06} find an optically thick disk, and \citet{argiroffi07} study its X-ray emission,
but the conclusion of these papers may change with the new proposed age.

Actually MP~Mus would not be an exception in the $\epsilon$~Cha
Association.  Two other members of the $\eta$~Cha cluster are also
cTTS: ET~Cha \citep{lawson02} and RECX~16 \citep{song04}.  T~Cha, a
very interesting TTS, was classified by \citet{alcala93} as a wTTS,
for the strength of the H$\alpha$ emission.  In the PDS we found its
H$\alpha$ highly variable, ranging from absorption to 25~\AA\,
emission.  Thus, in its active phase it would also be classified as a
cTTS.  Its photometric variations are discussed in \citet{batalha98}.
It could be associated with the cloud FS~195 (\citet{covino97}) and,
in that case, it would be the only star in young associations near a
cloud.  T~Cha definitely deserves a fresh look.  If we accept T~Cha as
a cTTS, then the $\epsilon$~Cha Association has four cTTS.  This is
one more indication that the $\epsilon$~Cha Association is younger
than the TW~Hya and the $\beta$~Pic associations, which have only one
cTTS each (TW~Hya and V4046~Sgr).

In Table~\ref{table:cha} there are three double line spectroscopic
binary stars: the eclipsing binary RS~Cha, DX~Cha and GSC~9416-1029.
RS~Cha has also signatures of $\delta$~Scuti type pulsations
\citep{alecian05}.  The authors revisited the solution for the system
and found that the orbital period is not constant.  \citet{alecian07}
give an age of 9.5~Myr for the system, older than our kinematical one.
The HAEBE star DX~Cha (HD~104237A) has also signatures of
$\delta$~Scuti type pulsations \citep{kurtz99}.  \citet{bohm04}
studied the spectroscopic orbit and the $\delta$~Scuti pulsations.
This very interesting system, with at least two more TTS components
(presented in Table~\ref{table:cha} as HD~104237D and E), has been
studied in detail by \citet{grady04} using many different techniques.
GSC~9416-1029 has recently been detected as a double line
spectroscopic binary star by \citet{doppmann07} with a period of
5.35~days and an age of 5~Myr.  EQ~Cha, $\eta$~Cha, T~Cha and
HD~104467 are suspected to be spectroscopic binaries. EG~Cha, EQ~Cha,
$\epsilon$~Cha, DX~Cha and GSC~9239-1572 have close visual companions.
Besides RS~Cha and DX~Cha, another $\delta$~Scuti star in the
$\epsilon$~Cha Association is EE~Cha \citep{kurtz99}.

The possible connection between the $\epsilon$~Cha Association and the
nearby $\eta$~Cha open cluster must be investigated further.  It is
very important to obtain radial velocities for the stars having only
proper motions and to search for candidates in the region between the
$\eta$~Cha cluster and the main concentration of the $\epsilon$~Cha
association.  For example, knowing now the spatial motion of this
association, the candidates can be found in deeper proper motion
surveys.

\subsection{The $\eta$~Cha Open Cluster}

\begin{figure}[]
   \begin{center}
\plotone{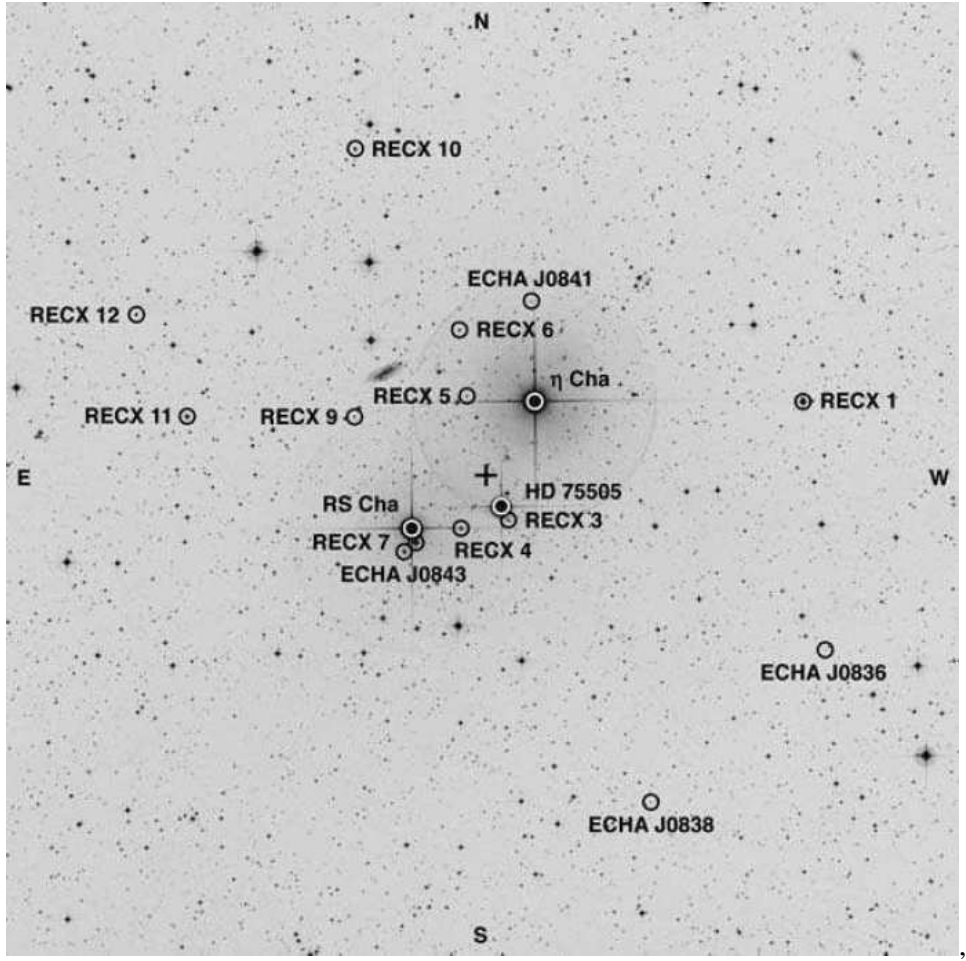},
    \end{center}
\caption{ 45 x 45 arcmin POSS-II red-band image of the $\eta$ Cha cluster region,
centred (plus) at $\alpha_{2000}$ = 8h42m $\&\,\, \delta_{2000}$ = --79$\deg$01$\arcmin$.
(RECX~16 is outside of the field.) From \citet{lyo04}.}
\label{fig:eta}
\end{figure}

\begin{table}[]
\caption{Members of the $\eta$~Cha cluster, not included in
Table~\ref{table:cha}.%{\bf based on literature}
}
\smallskip
{
\label{table:eta}
\begin{tabular}{llcclll}
\tableline
\noalign{\smallskip}
Name&other&$\alpha_{2000}$   & $\delta_{2000}$    &\hspace{3mm}V & Sp.T.&Ref.
\\
%&&&&&&[mas]&[mas]~~~~~~&\\
\noalign{\smallskip}
\tableline
\noalign{\smallskip}
RECX 18&         &08 36 10.6    &-79 08 18  &17.66&M5e&So\\
RECX 17    &     &08 38 51.5&    -79 16 14  &16.82&M5e&So  \\
RECX 14&ES Cha   &08 41  30.6&   -78 53 07  &17.07&M5e&L \\
RECX 3& EH Cha   &08 41 37.2&   -79 03 31   &14.37& M3e&M\\
RECX 13&HD 75505 &08 41 44.7&   -79  02 53  &\hspace{2mm}7.27& A1V&M \\
RECX 4& EI Cha   &08 42 23.7&   -79 04 04   &12.73& M1e&M\\
RECX 5& EK Cha   &08 42 27.3&   -78 57 48   &15.20& M4e&M\\
RECX 6& EL Cha   &08 42 39.0&   -78 54 44   &14.08& M3e&M\\
RECX 7& EM Cha   &08 43 07.7&   -79 04 52   &10.89*& K6e&M\\
RECX 15& ET Cha  &08 43 18.4&   -79 05 21   &13.97&M3e&L \\
RECX 16     &    &08 44 09.1&   -78 33 46   &17.5  &M5e&So\\
RECX 9& EN Cha   &08 44 16.6&   -78 59 09   &15.75* & M4e&M \\
RECX 10& EO Cha   &08 44 32.2&   -78 46 32   &12.53 & M0e&M \\
RECX 11& EP Cha   &08 47 01.8&   -78 59 35   &11.13 & K6e &M   \\

\noalign{\smallskip}
\tableline
\noalign{\smallskip}
\end{tabular}
}

{(*) the photometric values are corrected for duplicity.\\
L=\citet{lawson02}; M=\citet{mamajek99}; So=\citet{song04}.\\
RECX 13 is not a X-ray source but this designation is in SIMBAD.}
\end{table}

On the basis of a spatial clustering of X-ray sources, \citet{mamajek99} found
an open, young, and poor cluster in the vicinity of $\eta$~Cha.
They proposed 13 members, 12 of them being X-ray sources.
%{\bf(the exception is HD~75505)}.
Their main properties are investigated in \citet{mamajek00}, where the
authors first show a possible connection with the $\epsilon$~Cha
Association.  \citet{lawson01} present a photometric study of ten
low-mass candidate members and find that all are variables, with
periods ascribed to rotational modulation by starspots.  They estimate
an age of 4-9~Myr.  \citet{lawson02}, \citet{song04}, and \citet{lyo04} propose
five more members and the cluster now consists of 18 primary stars
(see Figure~\ref{fig:eta}), four of which, with good kinematical data,
we found as high probability members of the $\epsilon$~Cha
Association.  The 14 members of the $\eta$~Cha cluster, not included
in Table~\ref{table:cha}, are presented, for completeness, in
Table~\ref{table:eta}, with the \citet{luhmans04} identification
numbers.  Deeper searches to find fainter members did not produce
results so far \citep{luhman04, lyo06}.

Besides the binaries mentioned for the $\epsilon$~Cha Association, EN~Cha is a
close visual double having similar components (sep.=0.2$\arcsec$)
\citep{brandeker06} and EM~Cha is a double line spectroscopic binary star with a
period of 2.6~d, equal to  its photometric period \citep{lyo03}.

\section{The Octans Association}

The Oct Association was already postulated in SACY \citep{torres03a,
torres03}.  In their first solution they found only six members, none
in the Hipparcos.  Of course, this fact limits the accuracy of the
distances and ages obtained by the convergence method.  We applied the
convergence method assuming an age of 10~Myr based on the lithium
distribution, using a high p value in Equation~\ref{eq:f}, and no
expansion (the more conservative option).  Under these assumptions,
the solution has 15 members which are shown in Table~\ref{table:oca}.
It includes all six members previously proposed (the last six entries in
Table~\ref{table:oca}) which gave the association its name.

\begin{table}[]
\caption{The high probability members proposed for the Octans Association}
\smallskip
%\begin{center}
{
\label{table:oca}
\begin{tabular}{lccl
%r
llrl}
\tableline
\noalign{\smallskip}
Name&$\alpha_{2000}$   & $\delta_{2000}$    &\hspace{3mm}V &
%V-I &
Sp.T.    & D  & P.&Ref.\\
&&&&&
%&
[pc]&\%&\\
\noalign{\smallskip}
\tableline
\noalign{\smallskip}

CD-58 860   &04 11 55.6    &-58 01 47&10.01    &G6V     & \hspace{2mm}82&90&S\\
CD-43 1451  &04 30 27.3    &-42 48 47&10.75    &G9V(e)  & 120&80&S\\
CD-72 248   &05 06 50.6    &-72 21 12&10.91    &K0IV & 143&100&S\\
HD 274576    &05 28 51.4    &-46 28 18&10.57    &G6V & 116&100&S\\
CD-47 1999  &05 43 32.1    &-47 41 11&10.19    &G0V & 167&100&S\\
TYC7066-1037&05 58 11.8    &-35 00 49&11.24   &G9V & 149&100&S\\
CD-66 395   &06 25 12.4    &-66 29 10&10.92     &K0IV &147&100&S\\
CD-30 3394A &06 40 04.9    &-30 33 03&\hspace{2mm}9.84&F6V&141&70&s\\
CD-30 3394B &06 40 05.7    &-30 33 09&  10.24&    F9V&161&80&S\\
%GSC9415-1685 &11 50 28.9    &-77 04 38&11.98  &K2e  &\hspace{2mm}83&100&c,z\\
HD 155177   &17 42 09.0&   -86 08 05& \hspace{2mm}8.88& F5V &159&100&s\\
TYC9300-0529  &18 49 45.1&   -71 56 58& 11.59 & K0V & 196&100&S\\
TYC9300-0891  &18 49 48.7&    -71 57 10& 11.43*& K0V(e) & 175&100&S\\
CP-79 1037  &19 47 03.9&    -78 57 43&  11.16&G8V & 164&100&S\\
CP-82 784   &19 53 56.7&    -82 40 42&  10.87&K1V & 152&100&S\\
CD-87 121   &23 58 17.7&    -86 26 24 &  \hspace{2mm}9.98&G8V &122&100&S\\

\noalign{\smallskip}
\tableline
\noalign{\smallskip}
\end{tabular}
}
%\end{center}
{(*) the photometric values are corrected for duplicity.\\
The distances are calculated with the convergence method.\\
S=in the SACY survey; s=observed in the SACY, but outside of the sample
definition.}
\end{table}

\begin{figure}[]
  \begin{center}
  \begin{tabular}{c}

\resizebox{1.0\hsize}{!}{
\includegraphics[draft=False]{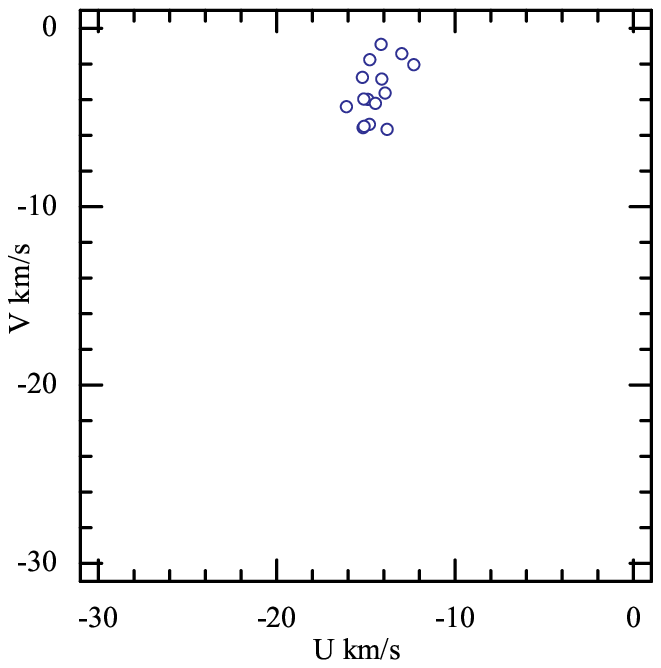}
\includegraphics[draft=False]{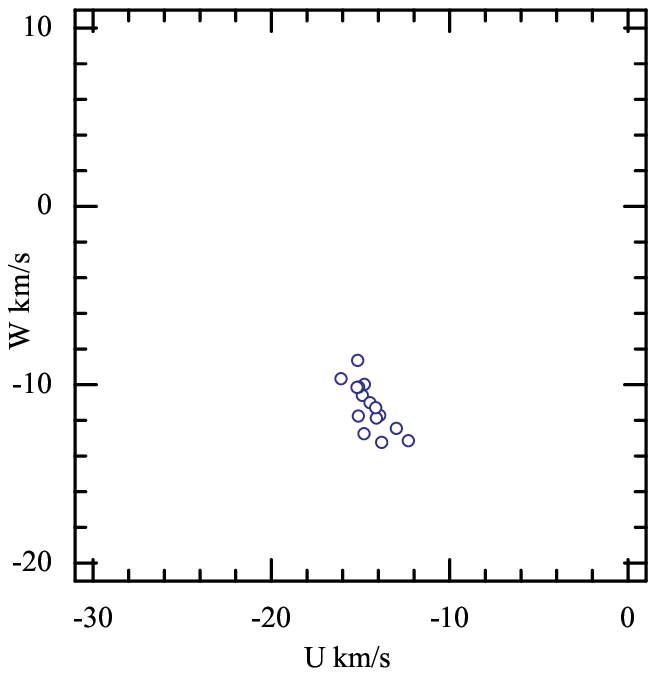}

\includegraphics[draft=False]{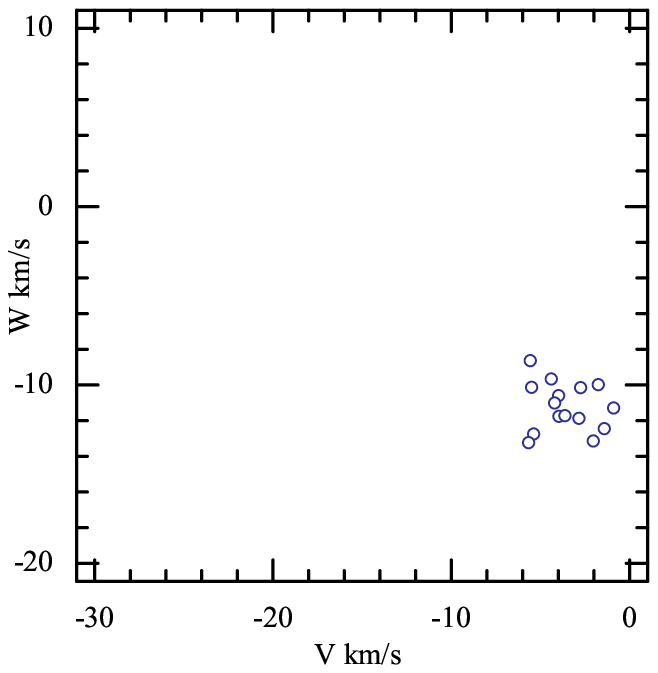}
}\\

\resizebox{1.0\hsize}{!}{
\includegraphics[draft=False]{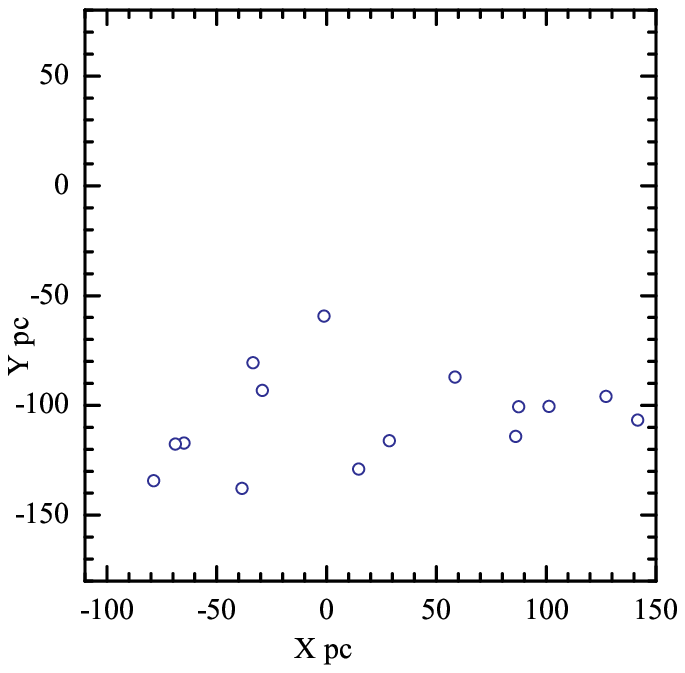}
\includegraphics[draft=False]{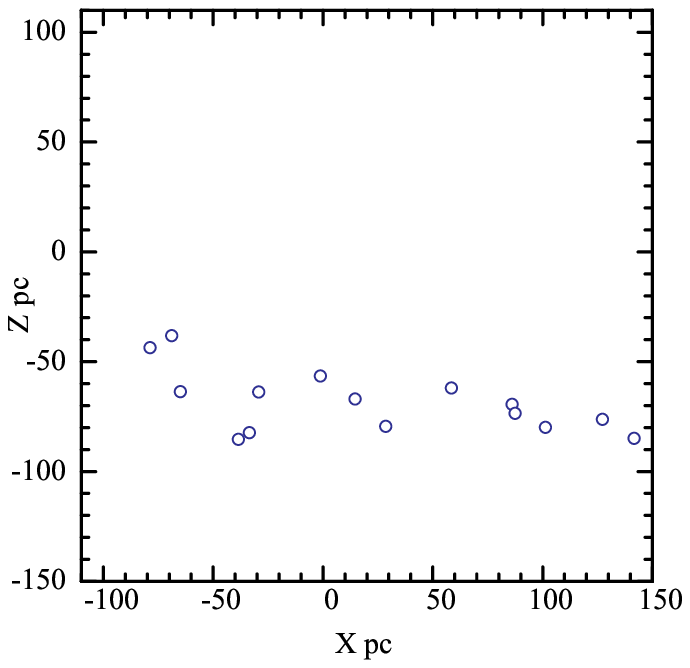}

\includegraphics[draft=False]{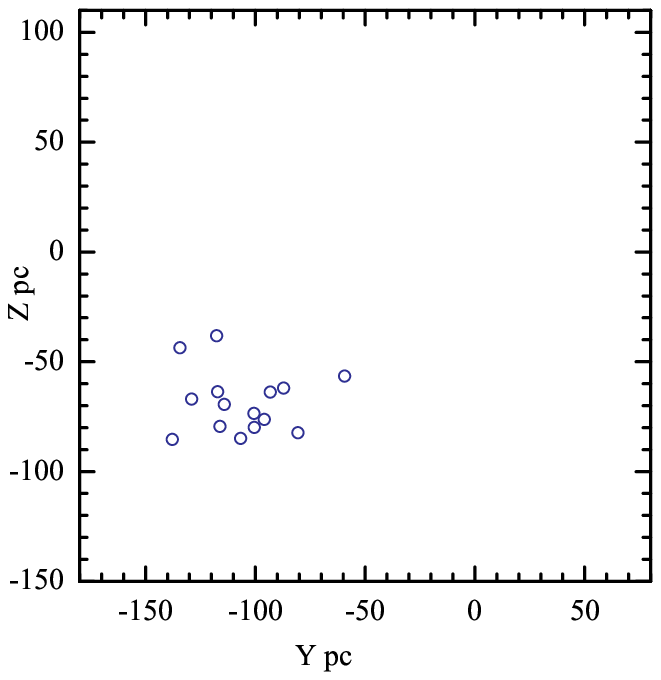}
}\\
  \end{tabular}
    \end{center}

\caption{Combinations of the sub-spaces of the UVWXYZ--space for the Octans
Association showing a well defined clustering
in both kinematical and spatial coordinates.
}

\label{fig:ocxyz}
\end{figure}

\begin{figure}[]
   \begin{center}
  \begin{tabular}{c}
 \resizebox{1.0\hsize}{!}{
\includegraphics[draft=False]{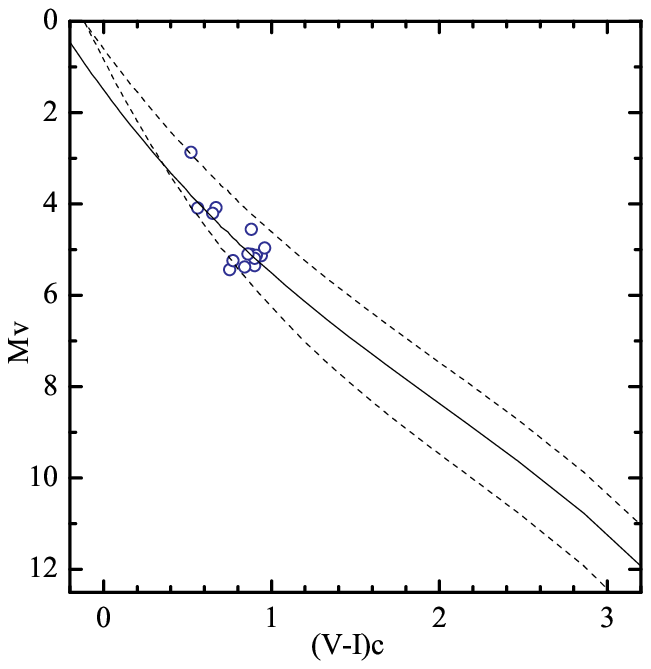}
\includegraphics[draft=False]{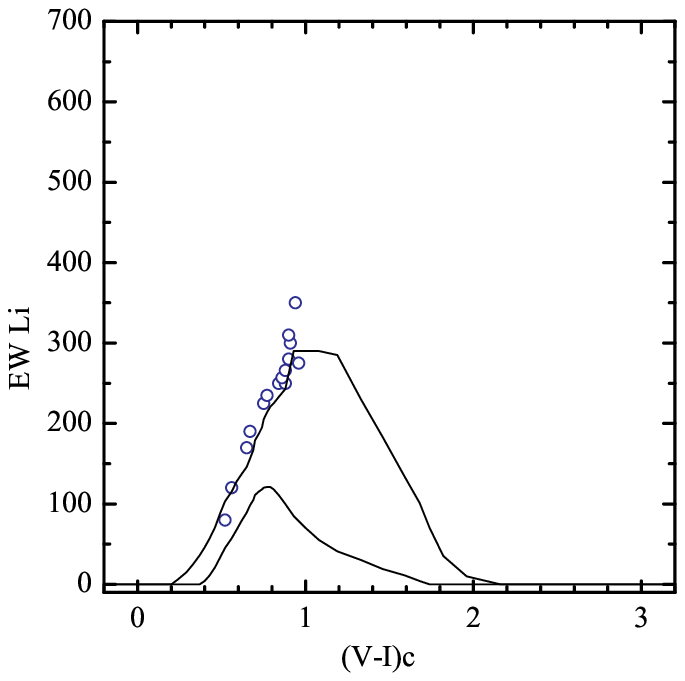}
}
    \end{tabular}
    \end{center}
\caption{{\it Left:} The HR diagram of the members proposed for the  Oct Association;
the over-plotted isochrones
are the ones for 5, 10, and 70~Myr.
{\it Right:} The distribution of Li equivalent width as a function of $(V-I)_C$; the
curves are the upper and lower limits for the age of the Pleiades
\citep{neu97}.
}
\label{fig:ochr}

\end{figure}

The mean distance, 141\,pc, is near the observational limit of  SACY and
thus explains the absence of redder stars.
The members proposed include two middle F stars:
CD-30~3394, observed as its visual companion is a SACY star
and HD~155177, observed in another similar program.
CD-72~248 and CD-66~395 are very fast rotators (vsin(i)~$\sim$200\,km~s$^{-1}$).
They were observed only once, therefore their radial velocities have large errors.
In fact, all members proposed, except for the close visual binary TYC~9300-0891,
have vsin(i)~$\geq$~20\,km~s$^{-1}$, another indication that the association must be young.
The companion of TYC~9300-0891 is at 0.9$\arcsec$, according to TYCHO-2,
 and they are  20$\arcsec$ from TYC~9300-0529.
Another wide visual binary in this association is the pair CD-30~3394, at a separation of 12.4$\arcsec$.
There is also a faint star 10.9$\arcsec$ from CD-87~121, a possible companion.

Based on these 15 stars, the Oct Association is compact in Y and Z directions,
but in X direction it appears much larger than expected for the age assumed in the convergence method.
As a compromise we propose in  Table~\ref{table:finalb} an age of about 20~Myr.
Its V velocity is peculiar in relation to all other nearby young associations
and it has the lowest Z.
These peculiarities and its position in the sky are the main reasons to believe that this association is
real despite  the small number of members and their distribution in X direction.
Anyway, that is why we prefer not to use a solution with expansion, although
using it we obtain a similar solution, but even more extended in X direction.
A good solution and better age determination must await fainter members and some trigonometric distances.

\section{The Argus Association}

\begin{table}[!ht]
\caption{The high probability members of the Argus Association
\centerline{(a)\hspace{2mm} IC~2391 members}
}
\smallskip
%\begin{center}
{
\label{table:ic}
\begin{tabular}{lcclllrl}
\tableline
\noalign{\smallskip}
Name&$\alpha_{2000}$   & $\delta_{2000}$    &\hspace{3mm}V & Sp.T.    & D  & \%
&Ref.\\
&&&&&[pc]&\%&\\
\noalign{\smallskip}
\tableline
\noalign{\smallskip}
%\multicolumn{9}{c} {\bf {IC 2391 members}}\\
%\noalign{\smallskip}
%\hline
%\noalign{\smallskip}
PMM 7422       &08 28 45.6&    -52 05 27&       10.49&          G6   &120&100&p,m\\
PMM 7956      &08 29 51.9&     -51 40 40  &      11.62&         Ke   &141&100&p\\
PMM 1560      &08 29 52.4&     -53 22 00 &      10.66&          G3   &135&95&p\\
PMM 6974      &08 34 18.1&     -52 15 58&       12.26&               &147&100&p\\
PMM 4280       &08 34 20.5&    -52 50 05&       10.34&          G5   &145&95&p\\
PMM 6978      &08 35 01.2&     -52 14 01&       12.07&               &128&100&p\\
PMM 2456       &08 35 43.7&    -53 21 20&       12.20&          K3e  &135&100&p\\
PMM 351       &08 36 24.2&     -54 01 06&       10.18&           G0  &128&100&p\\
PMM 3359      &08 36 55.0&     -53 08 34&       11.51&               &137&100&p\\
PMM 5376      &08 37 02.3&     -52 46 59&       14.30&          Me   &132&100&p\\
PMM 4324       &08 37 47.0&    -52 52 12&\hspace{2mm}9.66 &   F5V    &137&100&p\\
PMM 665        &08 37 51.6&    -53 45 46&       11.35&          G8   &139&100&p\\
PMM 4336       &08 37 55.6&    -52 57 11&       11.58&     G9        &139&100&p\\
PMM 4362      &08 38 22.9&     -52 56 48&       10.97&               &147&100&p\\
PMM 4413      &08 38 55.7&     -52 57 52&       11.06*&     G2       &141&100&p\\
PMM 686       &08 39 22.6&     -53 55 06&       12.63&           Ke  &137&100&p\\
PMM 4467      &08 39 53.0&     -52 57 57&       11.86V&      K0(e)   &137&100&p\\
PMM 1083       &08 40 06.2&    -53 38 07&       10.45V&      G0      &139&100&p,m\\
PMM 8415      &08 40 16.3&     -52 56 29&       11.84&      G9(e)    &141&100&p\\
PMM 1759      &08 40 18.3&     -53 30 29&       13.54&      K3e      &135&100&p\\
PMM 1142       &08 40 49.1&    -53 37 45&       11.08&      G6       &147&100&p\\
PMM 1174       &08 41 22.7&    -53 38 09&\hspace{2mm}9.54&     F3V   &139&100&p\\
PMM 1820      &08 41 25.9&     -53 22 41&        12.56V&    K3e      &154&90&p\\
PMM 4636       &08 41 57.8&    -52 52 14&       13.57&      K7e      &135&100&p\\
PMM 3695      & 08 42 18.6&    -53 01 57&       13.96&      M2e      &137&100&p\\
PMM 756        &08 43 00.4&    -53 54 08&       11.16&         G9    &149&100&p\\
PMM 5811       &08 43 17.9&    -52 36 11&\hspace{2mm}9.16&     F2V   &145&100&p\\
PMM 2888       &08 43 52.3&    -53 14 00&\hspace{2mm}9.76&     F5    &135&100&p\\
PMM 2012       &08 43 59.0&    -53 33 44&       11.67&      K0(e)    &135&100&p\\
PMM 4809       &08 44 05.2&    -52 53 17&       10.85V&      G3(e)   &154&100&p\\
PMM 1373      &08 44 10.2&     -53 43 34&       12.25&               &141&100&p\\
PMM 5884       &08 44 26.2&    -52 42 32&       11.46V&      G9(e)   &139&100&p\\
PMM 4902      &08 45 26.9&     -52 52 02&       12.76V&      K3e     &133&100&p\\
PMM 6811       &08 45 39.1&    -52 26 00&\hspace{2mm}9.91V&  F8Ve    &135&100&p,m\\
PMM 2182       &08 45 48.0&    -53 25 51&       10.22&         G2(e) &152&100&p\\
\noalign{\smallskip}
\tableline
\noalign{\smallskip}
\end{tabular}
\smallskip
}
%\end{center}
{

(*) the photometric values are corrected for duplicity.\\
Names as in \citet{platais07}.
The distances are calculated with the convergence method.\\
p=\citet{platais07};
m=\citet{makarov00}, member of the Car-Vela moving group.\\
 }
\end{table}

\begin{table}[!ht]
\caption{ The high probability members of the Argus Association
\centerline{(b)\hspace{2mm} Field members}}
\smallskip
%\begin{center}
{
\label{table:arga}
\begin{tabular}{lcclllrl}
\tableline
\noalign{\smallskip}
Name&$\alpha_{2000}$   & $\delta_{2000}$    &\hspace{3mm}V & Sp.T.    & D  & \%
&Ref.\\
&&&&&[pc]&\%&\\
\noalign{\smallskip}
\tableline
\noalign{\smallskip}
BW Phe       &00 56 55.5&    -51 52 32&     \hspace{2mm}9.62*&K3Ve&\hspace{2mm}56&90&S\\
CD-49 1902    &05 49 44.8&    -49 18 26&       11.37&      G7V    &133&90&S\\
CD-56 1438    &06 11 53.0&    -56 19 05&       11.11&       K0V   &101&95&S\\
CD-28 3434    &06 49 45.4&    -28 59 17&       10.38&      G7V    &109&90&S\\
CD-42 2906    &07 01 53.4&    -42 27 56&      10.61&K1V&\hspace{2mm}96&100&S\\
CD-48 2972    &07 28 22.0&    -49 08 38&     \hspace{2mm}9.84&G8V&\hspace{2mm}85&100&S\\
CD-48 3199    &07 47 26.0&    -49 02 51&       10.61&G7V&\hspace{2mm}95&100&S,m\\
CD-43 3604    &07 48 49.6&    -43 27 06&       10.88 &K4Ve&\hspace{2mm}79&100&S\\
TYC8561-0970   &07 53 55.5&    -57 10 07&       11.50&          K0V &133&100&S,m\\
HD 67945      &08 09 38.6&    -20 13 50&      \hspace{2mm}8.08&F0V&\hspace{2mm}64&95&s\\
CD-58 2194    &08 39 11.6&     -58 34 28&       10.11&      G5V    &106&100&S,m\\
CD-57 2315    &08 50 08.1&    -57 45 59&       10.21& K3Ve&\hspace{2mm}94&100&S,m\\
TYC8594-0058  &09 02 03.9&    -58 08 50&       11.30&      G8V    &141&100&S,m\\
CD-62 1197    &09 13 30.3&    -62 59 09&       10.46&      K0V(e) &111&100&S,m\\
TYC7695-0335  &09 28 54.1&    -41 01 19&       11.70&        K3V  &143&100&S\\
HD 84075     &09 36 17.8&    -78 20 42&\hspace{2mm}8.59&G1&\hspace{2mm}63H&100&g\\
%BD-20 2977   &09 39 51.5&    -21 34 17&       10.22&          G9V    &\hspace{2mm}92&85&S\\
%TYC8945-1078 &09 42 35.3&    -62 28 35&       12.02&          M0V   &33.1&&0.75\\
TYC9217-0641 &09 42 47.4&    -72 39 50 &      12.40&          K1V    &132&100&S\\
CD-39 5883    &09 47 19.9&    -40 03 10&      10.89  &          K0V  &100&95&S\\
HD  85151A   &09 48 43.3&    -44 54 08&\hspace{2mm}9.61 &     G7V&\hspace{2mm}69&95&S\\
HD  85151B   &09 48 43.5&    -44 54 09&     10.21 &     G9V    &\hspace{2mm}69&100&S\\
CD-65 817    &09 49 09.0&    -65 40 21&       10.33*&   G5V         &141&100&S,m\\
HD 309851    &09 55 58.3&    -67 21 22&\hspace{2mm}9.90&      G1V   &103&100&S,m\\
HD 310316    &10 49 56.1&    -69 51 22&       10.82*&    G8V        &128&100&S\\
CP-69 1432   &10 53 51.5&    -70 02 16&     10.66  &      G2V       &164&100&S\\
%HD 103742     &11 56 42.3&    -32 16 05&     \hspace{2mm}7.63&G3V   &\hspace{2mm}32&&S\\
%HD 103743    &11 56 43.8&    -32 16 03&    \hspace{2mm}7.83  &       G4V  &\hspace{2mm}32&&S\\
%CD-42 7422    &12 06 32.9&    -42 47 51 &      10.66&      K0V   &115&85&S\\
%AF Cru      &12 19 07.1&    -63 09 54&     10.02*&           K0V   &211&0.90\\
CD-74 673   &12 20 34.4  &   -75 39 29&    10.72&       K3Ve   &\hspace{2mm}50&90&S,g\\
%V347 Hya   &13 34 57.4&    -29 55 24&    \hspace{2mm}9.26*& G3V   &\hspace{2mm}83&75&S\\
CD-75 652   &13 49 12.9&    -75 49 48&\hspace{2mm}9.67 &       G1V&\hspace{2mm}81&95&s,g\\
HD 129496   &14 46 21.4&    -67 46 16&\hspace{2mm}8.78  &     F7V&\hspace{2mm}85&95&s\\
NY Aps      &15 12 23.4&    -75 15 15 &\hspace{2mm}9.42&       G9V&\hspace{2mm}50H&90&S\\
%CD-52 7875  &16 59 55.2&    -52 46 20&     10.75 &      G9V   & 8.8&0.90\\
%HD 160257   &17 39 55.6&    -23 03 41 &      8.61 &     G2V    &&12.3$\pm$&1.00\\
CD-52 9381  &20 07 23.8&    -51 47 27&      10.59    &   K6Ve&\hspace{2mm}29&95&S\\

%}
\noalign{\smallskip}
\hline
\noalign{\smallskip}
\end{tabular}
\smallskip
}
%\end{center}

{(*) the photometric values are corrected for duplicity.\\
The distances are from Hipparcos (H in the table) or kinematical ones,
calculated with the convergence method.\\
S=in the SACY survey \citet{torres06}; s=not in the SACY sample definition, but
observed in this program;
g=\citet{guenther07}, wTTS in Cha region;
m=\citet{makarov00}, member of the Car-Vela moving group.
 }

\end{table}

\begin{figure}[!ht]
\begin{center}
\begin{tabular}{c}
\resizebox{1.0\hsize}{!}{
\includegraphics[draft=False]{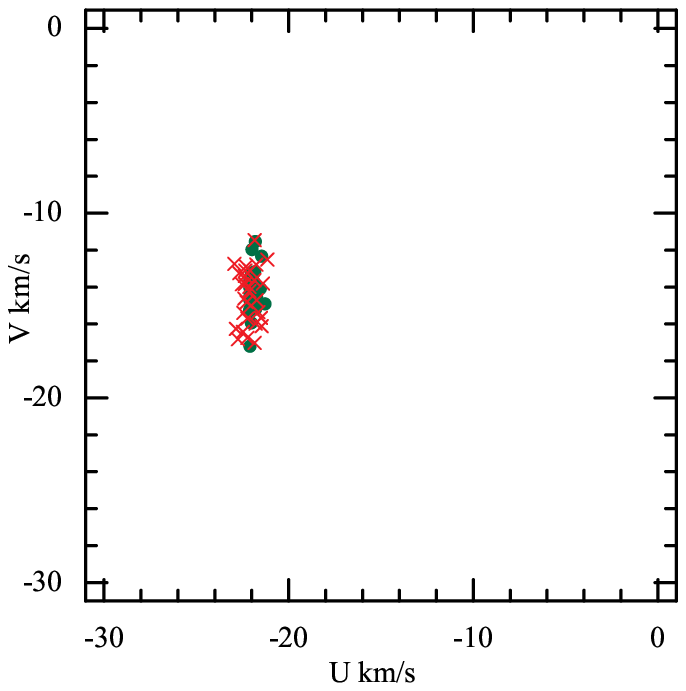}
\includegraphics[draft=False]{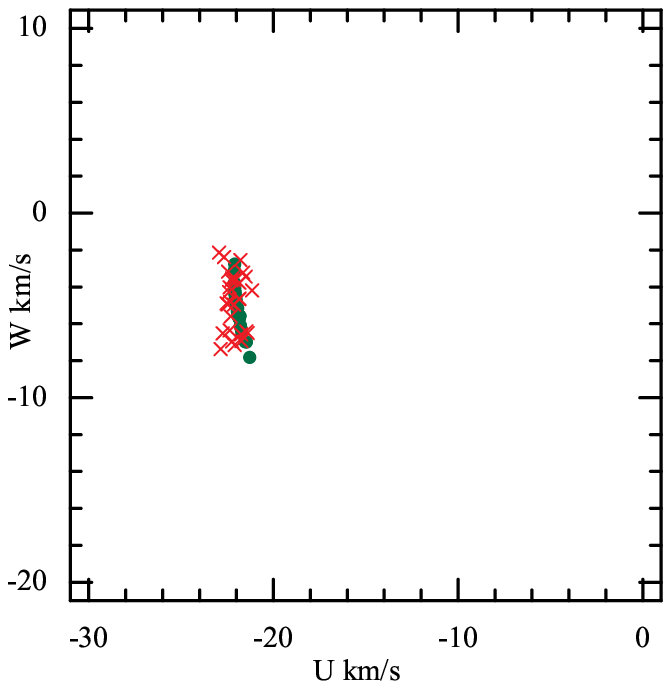}

\includegraphics[draft=False]{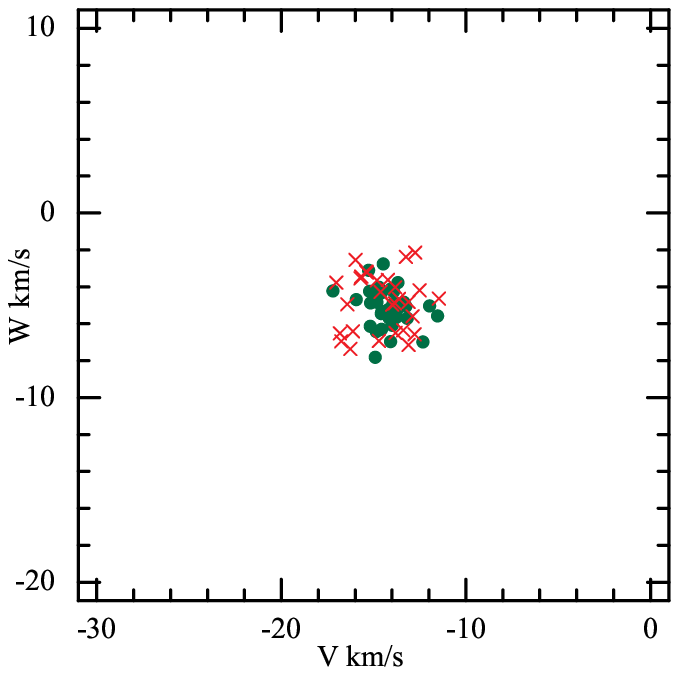}
}\\
\resizebox{1.0\hsize}{!}{
\includegraphics[draft=False]{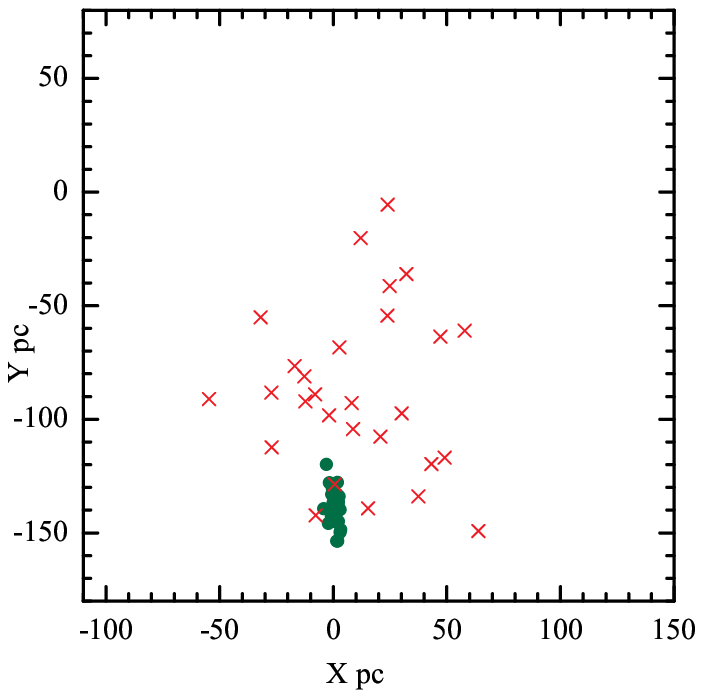}
\includegraphics[draft=False]{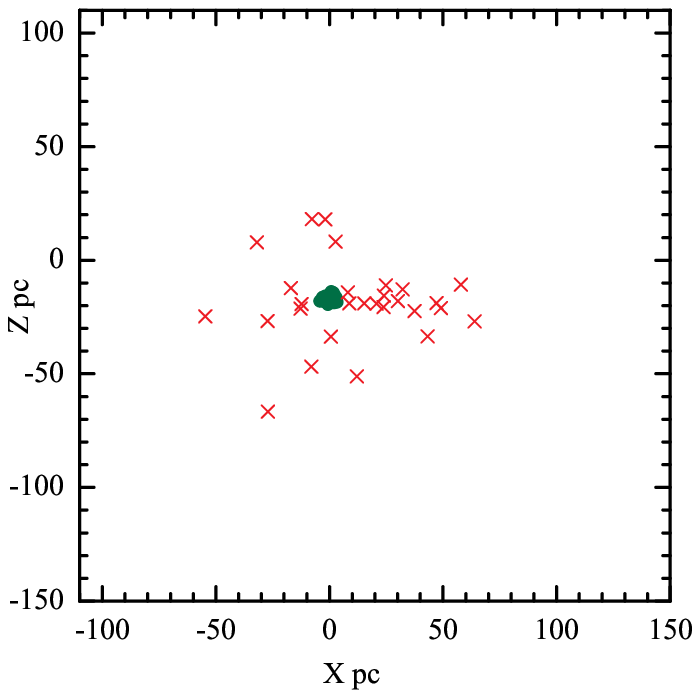}

\includegraphics[draft=False]{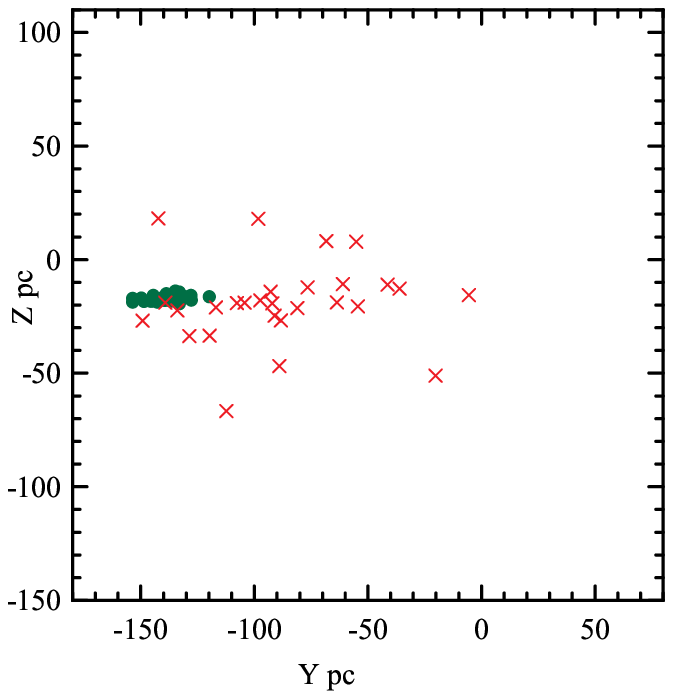}
}
  \end{tabular}
    \end{center}
\caption{Combinations of the sub-spaces of the UVWXYZ--space for the Argus
Association showing a well defined clustering in both kinematical and spatial
coordinates. Crosses represent  the field members and filled circles, the
IC~2391 members. }
\label{fig:arxyz}
\end{figure}
\begin{figure}[!h]
\begin{center}
 \begin{tabular}{c}
 \resizebox{0.5\hsize}{!}{
\includegraphics[draft=False]{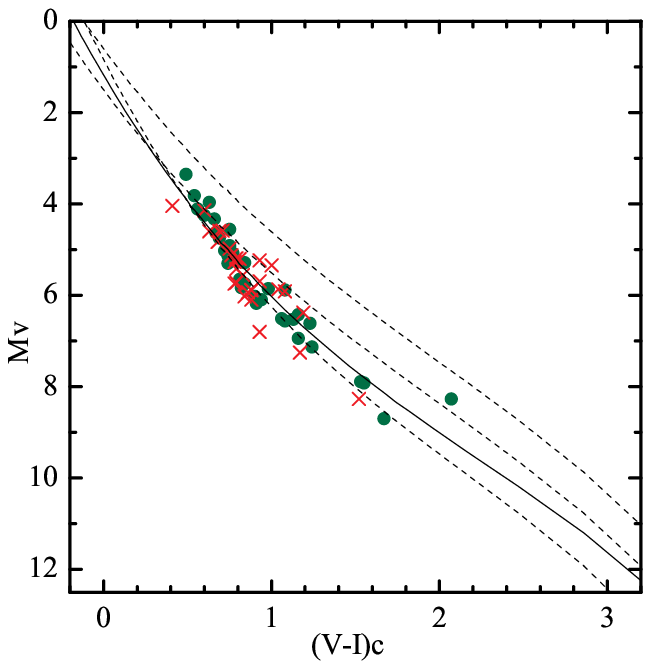}
}
\resizebox{0.5\hsize}{!}{
\includegraphics[draft=False]{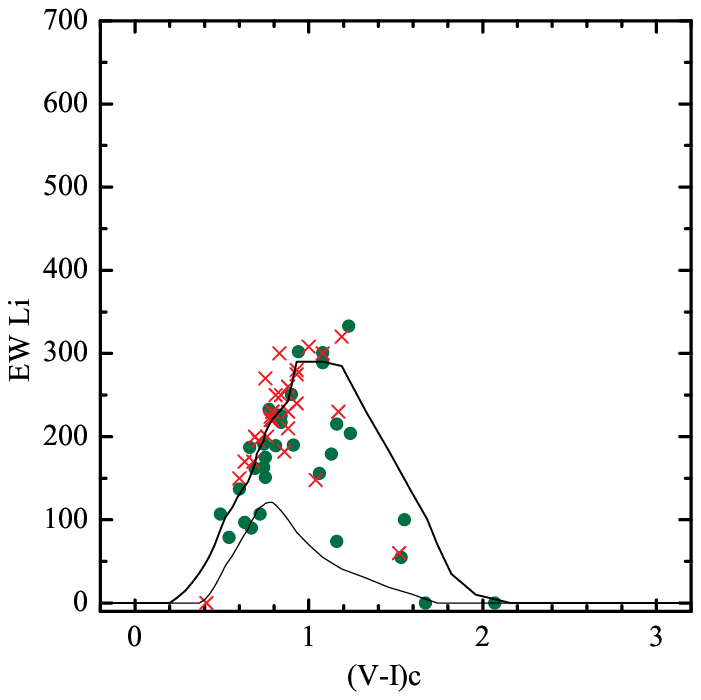}
}
 \end{tabular}
 \end{center}
\caption{{\it Left:} The HR diagram of the members proposed for the Argus Association;
the over-plotted isochrones
are the ones for 5, 10, 30 and 70~Myr. (Symbols are as in Figure~\ref{fig:arxyz}).
{\it Right:} The distribution of Li equivalent width as a function of $(V-I)_C$; the
curves are the upper and lower limits for the age of the Pleiades
\citep{neu97}. }
\label{fig:arhr}
\end{figure}

The Argus Association was easily discovered in the SACY survey due to its
special U velocity \citep{torres03}.
\citet{makarov00}, analyzing a global proper motion convergence map, found a sparse young
moving group located in Carina and Vela, having considerable geometrical depth.
From their list of 58 candidate members, 30 are in SACY, but eight of them are
proposed as members of  the Car Association (Table~\ref{table:cara}).
Only 11 stars of their list are considered as members of the
new Argus Association (Tables~\ref{table:ic} and ~\ref{table:arga}).
This challenges their proposed moving group.
Our new definition of the Argus Association is distinct
and much larger than the moving group mentioned.

\citet{makarov00} also proposed the open cluster IC~2391 as part of this moving group.
IC~2391 is reviewed by Pettersson in the chapter on Puppis and Vela in this book.
In order to test the possibility that IC~2391 may be connected with the Argus Association,
we used proper motions and  radial velocities of the possible cluster members
measured by \citet{platais07}.
There are 41 stars with all kinematical data, excluding the spectroscopic
binaries (20) except PMM~4413,
for which they published an orbital solution.
To convert their relative proper motions to the TYCHO-2 system we used 17 stars in common.
The convergence method actually shows quite a good match between IC~2391 and the
Argus Association from \citet{torres03}.

In  Table~\ref{table:arga} we present an updated solution with the 29 field
members, and in Table~\ref{table:ic} the 35 IC~2391 members
that have similar kinematical and physical data. Actually, in
Figure~\ref{fig:arhr} it is hard to note
any physical distinction between both kinds of members.
Two of the IC~2391 members proposed (PMM~1820 and PMM~5376) are not considered
bona fide ones by \citet{platais07},
and only one of their bona fide members (PMM~5829) is among the six IC~2391
stars we rejected.
The mean distance of IC~2391 obtained by this convergence method is 139$\pm$7\,pc.

In Figure~\ref{fig:arxyz} the concentration of IC~2391 can be noted at X $\sim$0
and Y $\sim$--140.
The obtained distribution in Y axis (a "finger-of-god" effect) is an artifact of
the convergence method.
As IC~2391 is along the Y--direction on the sky, the only possible spread coming
from the kinematical errors is in the distance (i.e., along the Y--axis).
Were IC~2391 located in a different place on the sky, the spread would
also be  in the other space axes and the distribution in space would be more spherical.
As the distance of IC~2391 is somewhat extreme for the Argus Association,
we can suppose that the association must be larger and that we have detected only its nearby members.
To avoid confusion with the  IC~2391 Supercluster proposed by \citet{eggen91},
we remained with our designation of "Argus Association".
In any case, if the Supercluster is real, it eludes detection in the SACY
and we can not check it as its members proposed are mainly early-type stars
that are not in the SACY database.

Very few of the 29 SACY field members have been observed elsewhere.
BW~Phe was proposed by \citet{zuckerman01b} as a possible member of the
Tucana Association.  CD-39~5883 has been proposed by \citet{reid03}
to be a possible member of the TW Hya Association.  Three stars are in
front of the Cha complex, and were observed by \citet{covino97} and
\citet {guenther07}: HD~84075, CD-74~673 and CD-75~652.

\citet{guenther07} found CD-74~673 as a long period (613.9d)
single line spectroscopic binary star, and until now it is the only  spectroscopic
binary among the field stars of the Argus Association.

There are four known visual binaries in this association: BW~Phe
(sep.= 0.2$\arcsec$), HD~85151 (sep.=2.3$\arcsec$), CD-65~817
(sep.=2.0$\arcsec$), and HD~310316 (sep.=0.6$\arcsec$).  The star CD-48~2972
has a possible bright companion at 81.7$\arcsec$, HD~59659, a F7V star
with V~=~8.80.  Both their proper motions are similar, and Hipparcos
gives a distance of 90\,pc for HD~59659.

\section{The AB~Dor Association}

\begin{table}[]
\caption{The high probability members proposed for the AB~Dor Association}
\smallskip
%\begin{center}
{
\label{table:aba}
\begin{tabular}{l@{\hskip7pt}ccl
%r
l@{\hskip7pt}lr@{\hskip7pt}l}
\tableline
\noalign{\smallskip}
Name&$\alpha_{2000}$   & $\delta_{2000}$    &\hspace{3mm}V &
%V-I &
Sp.T.    & D  & P.&Ref.\\
&&&&&
%&
[pc]&\%&\\
\noalign{\smallskip}
\tableline
\noalign{\smallskip}

PW And     &00 18 20.9    &+30 57 22&      \hspace{2mm}9.14&          K2V(e) &\hspace{2mm}27&100&Z\\
HD 4277    &00 45 50.9    &+54 58 40       &\hspace{2mm}7.80*    &F8V  &\hspace{2mm}49H& 95&Z\\
HD 6569    &01 06 26.2    &-14 17 47       &\hspace{2mm}9.50&    K1V &\hspace{2mm}50H&100&Z,S\\
BD-12 243  &01 20 32.2    &-11 28 03       &\hspace{2mm}8.43    &G9V   &\hspace{2mm}35H&100&Z,S\\
CD-46 644  &02 10 55.4    &-46 03 59       &11.24  &K3IVe &\hspace{2mm}70& 90&S\\
HD 13482   &02 12 15.4    &+23 57 30       &\hspace{2mm}7.94         &K1V &\hspace{2mm}32H& 100&Z\\
HIP 12635  &02 42 21.0    &+38 37 21       &10.28 &       K2V   &\hspace{2mm}49&100&Z\\
HD 16760   &02 42 21.3    &+38 37 07       &\hspace{2mm}8.77         &G2V  &\hspace{2mm}48& 100&Z\\
HD 17332B  &02 47 27.2    &+19 22 21       &\hspace{2mm}8.11*    &G6V   &\hspace{2mm}33H&100&Z\\
HD 17332A  &02 47 27.4    &+19 22 19       &\hspace{2mm}7.32*    &G1V   &\hspace{2mm}33H& 100&Z\\
IS Eri     &03 09 42.3    &-09 34 47       &\hspace{2mm} 8.48    &G0V   &\hspace{2mm}40H&100 &L\\
HIP 14807  &03 11 12.3    &+22 25 23       &10.47 &       K6   &\hspace{2mm}45&100&Z\\
HIP 14809  &03 11 13.8    &+22 24 57       &\hspace{2mm}8.51     &G5V  &\hspace{2mm}50H& 100&Z\\
V577 Per   &03 33 13.5    &+46 15 27       &\hspace{2mm}8.26     &G5V  &\hspace{2mm}34H& 100&Z\\
HD 21845B  &03 33 14.0    &+46 15 19       &11.20          &M0Ve  &\hspace{2mm}34H& 100&Z\\
HIP 17695  &03 47 23.3    &-01 58 20       &11.51    &M3Ve    &\hspace{2mm}16&100&Z,S,z\\
HD 24681   &03 55 20.4    &-01 43 45       &\hspace{2mm}9.05&G8V&\hspace{2mm}53&100&S,z\\
HD 25457   &04 02 36.7    &-00 16 08       &\hspace{2mm}5.38     &F6V  &\hspace{2mm}19H&100&Z\\
HD 25953   &04 06 41.5    &+01 41 02&       \hspace{2mm}7.83    &F5   &\hspace{2mm}55H& 100&Z\\
TYC5899-0026&04 52 24.4   &-16 49 22&      11.61       &M3Ve          &\hspace{2mm}16& 100&S\\
CD-56 1032B &04 53 30.5   &-55 51 32&      12.10      &M3Ve &\hspace{2mm}11&100&S\\
CD-56 1032A &04 53 31.2   &-55 51 37&     11.16     &M3Ve &\hspace{2mm}11&100&S\\
HD 31652    &04 57 22.3   &-09 08 00&      \hspace{2mm}9.98& G8V&\hspace{2mm}88&95&S\\
CD-40 1701 &05 02 30.4    &-39 59 13&      10.57         &K4V   &\hspace{2mm}42& 100&S\\
HD 32981    &05 06 27.7   &-15 49 30&     \hspace{2mm}9.13&F9V&   \hspace{2mm}81& 100&S\\
HD 293857   &05 11 09.7   &-04 10 54&     \hspace{2mm}9.26&G8V&   \hspace{2mm}78&100&S\\
HD 33999   &05 12 35.8    &-34 28 48&     \hspace{2mm}9.37      &F8V  &106&100&s\\
HD 35650   &05 24 30.2    &-38 58 11&     \hspace{2mm}9.08    &K6V& \hspace{2mm}18H&100   &Z,S\\
AB~DorB    & 05 28 44.4   &-65 26 47&       13.2*      &M4Ve &\hspace{2mm}15H&100&Z,S\\
AB~Dor     &05 28 44.8    &-65 26 56&       \hspace{2mm}6.88    &K0Ve &\hspace{2mm}15H&100&Z,S\\
UX Col    &05 28 56.5     &-33 28 16&      10.46      &K3Ve   & \hspace{2mm}57&100&S\\
CD-34 2331 &05 35 04.1    &-34 17 52&      11.84         &K3Ve   &\hspace{2mm}78& 100&S\\
HIP 26369  &05 36 55.1    &-47 57 48&       \hspace{2mm}9.81     &K6Ve  &\hspace{2mm}24H & 100&Z,S\\
UY Pic    &05 36 56.9     &-47 57 53&       \hspace{2mm}7.84     &K0V  &\hspace{2mm}24H & 100&Z,S\\
WX Col    &05 37 12.9     &-42 42 56&     \hspace{2mm}9.55*     &G7V  &\hspace{2mm}75H &100&S\\
HIP 26401B &05 37 13.2    &-42 42 57&      10.65*    &K1V   &\hspace{2mm}75H&100&S\\
Par 2752   &05 38 56.6    &-06 24 41&      10.91     &G8V   &116            &100&S\\
CP-19 878  &05 39 23.2    &-19 33 29&       10.71     &K1V   &\hspace{2mm}71 &100&S\\
TYC7605-1429 &05 41 14.4  &-41 17 59&      12.29         &K4IVe  &128&  100&S\\
CD-26 2425 & 05 44 13.4   &-26 06 15 &      10.88        &K2Ve  &\hspace{2mm}70& 100&S\\
TZ Col   &05 52 16.0      &-28 39 25&      \hspace{2mm}9.04     &G3V  &\hspace{2mm}88H& 90&S\\
TY Col     &05 57 50.8    &-38 04 03&       \hspace{2mm}9.56     &G6V(e)&\hspace{2mm}68&  100&S\\
\\
(Continued)
\end{tabular}
}
%\end{center}
\end{table}

\begin{table}[]
%\caption{The high probability members proposed for the AB~Dor Association}
\hspace{20pt}  Table~\ref{table:aba}. ~~ (Continued)
\smallskip
%\begin{center}
{

\begin{tabular}{l@{\hskip7pt}ccl
%r
l@{\hskip7pt}lr@{\hskip7pt}l}
\tableline
\noalign{\smallskip}
Name&$\alpha_{2000}$   & $\delta_{2000}$    &\hspace{3mm}V &
%V-I &
Sp.T.    & D  & P.&Ref.\\
&&&&&
%&
[pc]&\%&\\
\noalign{\smallskip}
\tableline
\noalign{\smallskip}
BD-13 1328 &06 02 21.9    &-13 55 33&       10.58&       K4V(e)&  \hspace{2mm}39& 95&S\\
CD-34 2676 &06 08 33.9    &-34 02 55&      10.17         &G9Ve  &\hspace{2mm}72&  100&S\\
CD-35 2722 &06 09 19.2    &-35 49 31&      10.98         &M1Ve  &\hspace{2mm}24&  100&S\\
HD 45270   &06 22 30.9    &-60 13 07&      \hspace{2mm}6.53     &G1V  &\hspace{2mm}23H&  100&Z,S\\
GSC8894-0426&06 25 56.1   &-60 03 27&        12.7&         M3Ve &\hspace{2mm}23H&100&Z\\
AK Pic     &06 38 00.4    &-61 32 00&       \hspace{2mm}6.27*    &G2V  &\hspace{2mm}22H&100&Z,S\\
CD-61 1439 &06 39 50.0    &-61 28 42&       \hspace{2mm}9.71  &K7V(e)&\hspace{2mm}22H& 100&Z,S\\
TYC7627-2190 &06 41 18.5  &-38 20 36&      11.08  &       K2Ve  &\hspace{2mm}78&  100&S\\
GSC8544-1037 &06 47 53.4  &-57 13 32&     11.8     &       K4V     &143&95&s\\
CD-57 1654 &07 10 50.6    &-57 36 46&      10.44  &       G2V   &103 & 100&S\\
BD+20 1790 &07 23 43.6    &+20 24 59&      \hspace{2mm}9.93   &       K5Ve &\hspace{2mm}26&100&L,S\\
HD 59169   &07 26 17.7    &-49 40 51&      10.22  &       G7V   &118H &  90&S\\
V372 Pup   &07 28 51.5    &-30 14 47&      10.13*    &M1Ve   &\hspace{2mm}13 & 95&Z,S\\
CD-84 80   &07 30 59.5    &-84 19 28&      \hspace{2mm}9.96         &G9V  &\hspace{2mm}71&  100&S\\
HD 64982   &07 45 35.5    &-79 40 09&       \hspace{2mm}8.96* &   G0V  &\hspace{2mm}83H&  95&S\\
BD-07 2388 &08 13 51.0    &-07 38 25&       \hspace{2mm}9.38&   K1V(e) &\hspace{2mm}93&  100&S\\
CD-45 5772 &10 07 25.2    &-46 21 50&        10.89  &    K4V  & \hspace{2mm}70&95&S\\
BD+01 2447  &10 28 55.6   &+00 50 28&       \hspace{2mm}9.65   &M2V &\hspace{4mm}6.6&100&L,z\\
HD 99827   &11 25 17.7    &-84 57 16&      \hspace{2mm}7.70*    &F5  &\hspace{2mm}83H&95&c\\
PX Vir    &13 03 49.7     &-05 09 43&       \hspace{2mm}7.69  &   K1V  &\hspace{2mm}22H&  100&Z,S\\
HD 139751  &15 40 28.4    &-18 41 46&      10.44* &   K5Ve &\hspace{2mm}37&100&Z,S\\
HIP 81084  &16 33 41.6    &-09 33 12&      11.30  &   M0e   &\hspace{2mm}30& 100&Z,S\\
HD 152555  &16 54 08.1    &-04 20 25&       \hspace{2mm}7.82  &   G0   &\hspace{2mm}48H&  100&Z\\
HD 317617  &17 28 55.6    &-32 43 57&         10.45&K3V                 &\hspace{2mm}56& 85&S\\
HD 159911  &17 37 46.5    &-13 14 47&       10.10& K4Ve& \hspace{2mm}45 &100&S\\
HD 160934  &17 38 39.6    &+61 14 16&       10.45*  &   K7Ve  &\hspace{2mm}33&100&Z\\
HD 176367  &19 01 06.0    &-28 42 50&       \hspace{2mm}8.48&G1V&\hspace{2mm}63H&95&s\\
HD 178085  &19 10 57.9    &-60 16 20&       \hspace{2mm}8.34  &   G1V  &\hspace{2mm}57H&  100&S\\
TYC0486-4943& 19 33 03.8  &+03 45 40&       11.29&  K3V & \hspace{2mm}71 &100&S\\
HD 189285 &  19 59 24.1   &-04 32 06&          \hspace{2mm}9.43&G7V     &\hspace{2mm}95& 90& S\\
BD-03 4778 &  20 04 49.4  &-02 39 20&       10.02&  K1V &  \hspace{2mm}70&100&S\\
HD 199058  &  20 54 21.1  &+09 02 24&   \hspace{2mm}8.61&G5V& \hspace{2mm}75&100&S\\
TYC1090-0543& 20 54 28.0  &+09 06 07&       11.29& K4Ve&\hspace{2mm}75&100&S\\
HD 201919   &21 13 05.3   &-17 29 13&      10.64  &   K6Ve  &\hspace{2mm}39& 100&S\\
LO Peg     &21 31 01.7    &+23 20 07&       \hspace{2mm}9.19  &   K5Ve &\hspace{2mm}25H&  100&Z\\
HD 207278  &21 48 48.5    &-39 29 10&       \hspace{2mm}9.66  &   G7V  &\hspace{2mm}84H&  100&S\\
HIP 107948&21 52 10.4     &+05 37 36&       12.11      &M2Ve&\hspace{2mm}30 &100&S,z\\
HIP 110526A &22 23 29.1   &+32 27 34&      11.45*  &   M3e   &\hspace{2mm}15& 100&Z\\
HIP 110526B &22 23 29.1   &+32 27 32&      11.55*  &   M3e   &\hspace{2mm}15& 100&Z\\
HD 217343  &23 00 19.3    &-26 09 14&       \hspace{2mm}7.49 & G5V&\hspace{2mm}32H&100&Z,S\\
HD 217379&   23 00 28.0   &-26 18 43&          10.45* & K7V   & \hspace{2mm}33&100&Z,S\\
HIP 114066  &23 06 04.8   &+63 55 34&      10.87  &   M1e   &\hspace{2mm}25H& 100&Z\\
\\
(Continued)
\end{tabular}
}
%\end{center}
\end{table}

\begin{table}[]
%\caption{The high probability members proposed for the AB~Dor Association}
\hspace{20pt}  Table~\ref{table:aba}. ~~ (Continued)
\smallskip
%\begin{center}
{

\begin{tabular}{lccl
%r
llrl}
\tableline
\noalign{\smallskip}
Name&$\alpha_{2000}$   & $\delta_{2000}$    &\hspace{3mm}V &
%V-I &
Sp.T.    & D  & P.&Ref.\\
&&&&&
%&
[pc]&\%&\\
\noalign{\smallskip}
\tableline
\noalign{\smallskip}

HD 218860A &23 11 52.1    &-45 08 11&       \hspace{2mm}8.75  &   G8V  &\hspace{2mm}51H&  100&Z,S\\
HD 218860B &23 11 53.6    &-45 08 00&       13.8   &       M3Ve &\hspace{2mm}51H &  100&S\\
HIP 115162  &23 19 39.5   &+42 15 10&       \hspace{2mm}8.94  &   G4   &\hspace{2mm}49H&100 &Z\\
HD 222575  &23 41 54.3    &-35 58 40&       \hspace{2mm}9.39  &   G8V  &\hspace{2mm}62H&  100&S\\
HD 224228  &23 56 10.7    &-39 03 08&       \hspace{2mm}8.22&     K2V  &\hspace{2mm}22H&  100&Z,S\\
\noalign{\smallskip}
\tableline
\noalign{\smallskip}
\end{tabular}
%\smallskip
%\end{center}
}

(*) the photometric values are corrected for duplicity.\\
The distances are from Hipparcos (H in the table) or kinematical ones,
calculated with the convergence method.\\
S=in the SACY survey; s=observed in the SACY, but outside of the sample
definition;
L=\citet{lopez06}; Z=\citet{zucksb04}; c=\citet{covino97}, z=\citet{zickgraf05}.
%\end{center}
\end{table}

The AB~Dor Association is a relatively old association ($\sim$70~Myr),
independently postulated by \citet{zucksb04}, and in the SACY project
\citep{torres03a, torres03} with the designation of AnA.  From the 37
members proposed by \citet{zucksb04}, 18 are in the SACY sample.
\citet{lopez06} proposed 11 other possible members\footnote{Although
the authors say they added 13 stars, actually two of them are in the
list of \citet{zucksb04}}, five of them within SACY.  The convergence
method was applied including the non-SACY members that were proposed.
In Table~\ref{table:aba} we present the 89 high probability members.
The solution has a velocity dispersion a little higher than the other
associations but this may be expected since it is the oldest one (see
Tables~\ref{table:final} and \ref{table:finalb}).  All members
proposed by \citet{zucksb04} are also in this solution.  In
Table~\ref{table:aba} there are actually 40 stars from their list
since we included separately the components of three visual binaries,
not discriminated by them.  From the 11 members proposed by
\citet{lopez06} we classified only three as high
probability members (one in the SACY).  We added 46 new members, 42
from the SACY sample.

\citet{luhman05} have argued  that the  AB~Dor Association is a remnant of the large-scale star
formation event that has also formed the Pleiades, and being somewhat older (75-150~Myr).
As their arguments are based on the \citet{zucksb04} list of members it will be important
to revisit the question of the origin of the AB~Dor Association on the basis of
the more extensive list presented here.
Nonetheless, our newly derived space motions are even more similar to those of the Pleiades cluster.
This common origin of the  AB~Dor Association and the Pleiades has been recently reinforced by \citet{ortega07},
using Galactic dynamics calculations.
This common origin would have occurred 119$\pm$20~Myr ago
at a height of about 250~pc below the Galactic plane.

AB~Dor itself is a well-studied active star, forming a system of four stars \citep{close05}.
We have taken the  proper motions and the parallax from the recent re-analysis
of the system by  \citet{guirado06}, using Hipparcos and VLBI data.
\citet{janson07} obtain an age of about 50 to 100~Myr for the system.
\citet{close07}  infer an age of about 75~Myr by analyzing new photometric
and spectroscopic measurements of AB~DorC.

\begin{figure}[]
  \begin{center}
  \begin{tabular}{c}

\resizebox{1.0\hsize}{!}{
\includegraphics[draft=False]{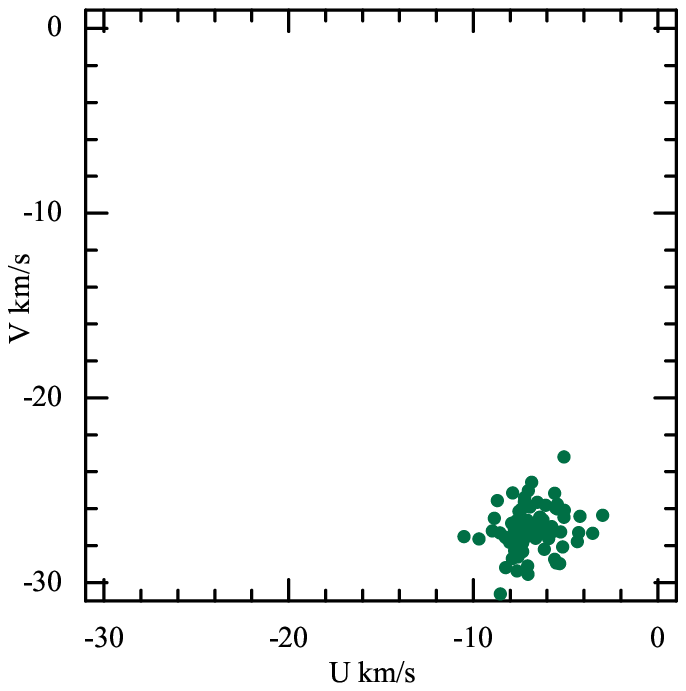}
\includegraphics[draft=False]{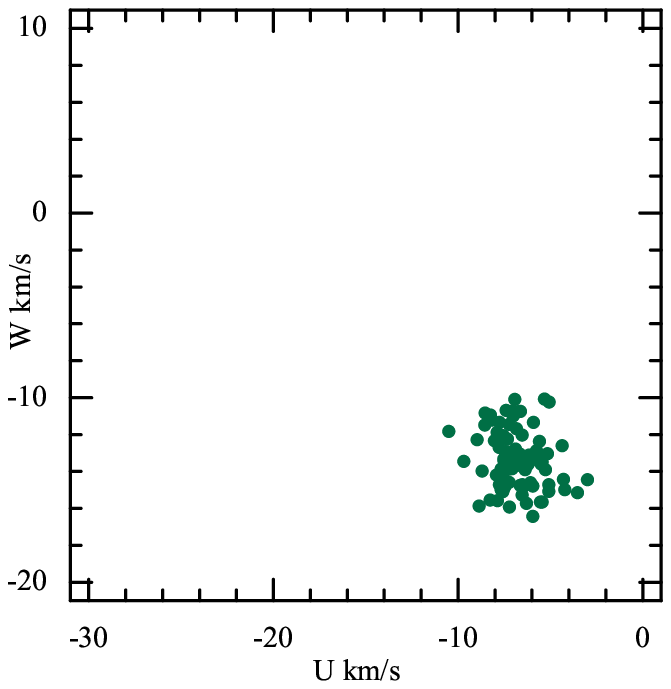}

\includegraphics[draft=False]{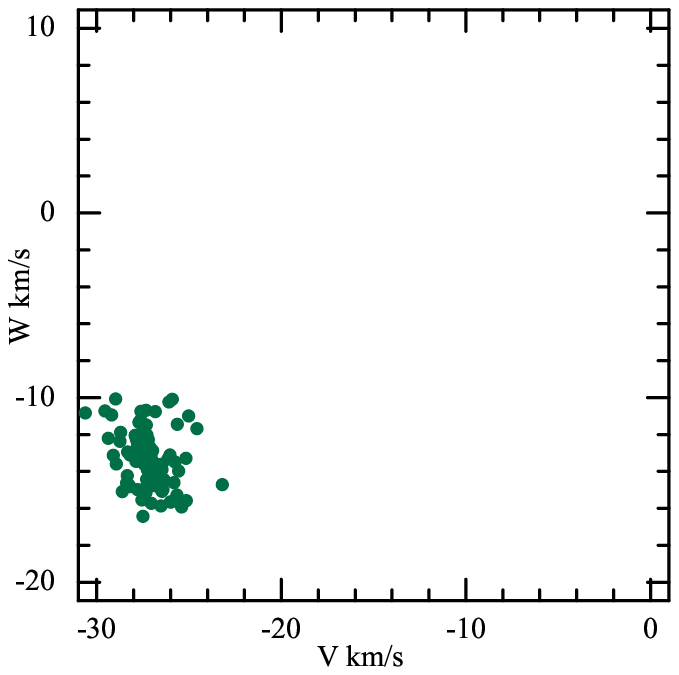}
}\\

\resizebox{1.0\hsize}{!}{
\includegraphics[draft=False]{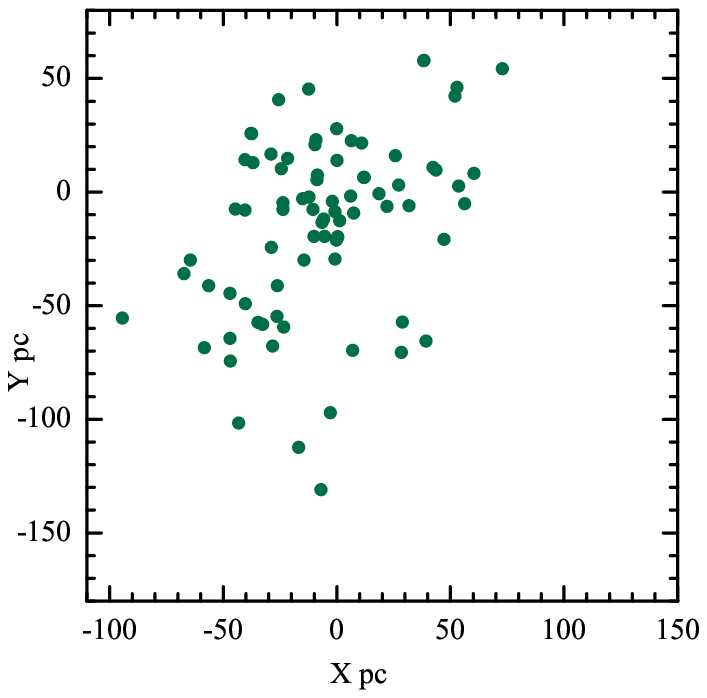}
\includegraphics[draft=False]{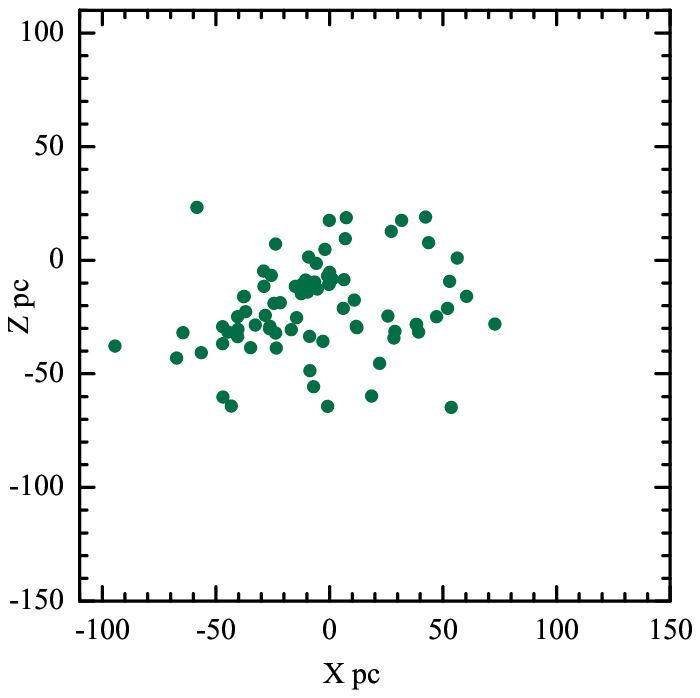}

\includegraphics[draft=False]{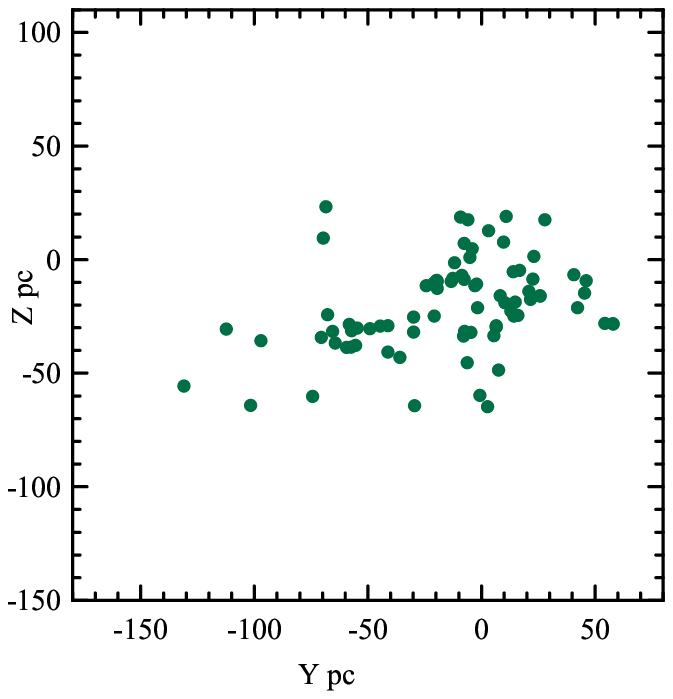}
}\\
  \end{tabular}
    \end{center}

\caption{Combinations of the sub-spaces of the UVWXYZ--space for the AB~Dor
Association showing a well defined clustering
in both kinematical and spatial coordinates.}

\label{fig:abxyz}
\end{figure}
\begin{figure}[]
\centering
\includegraphics[draft=False,width=0.48\textwidth]{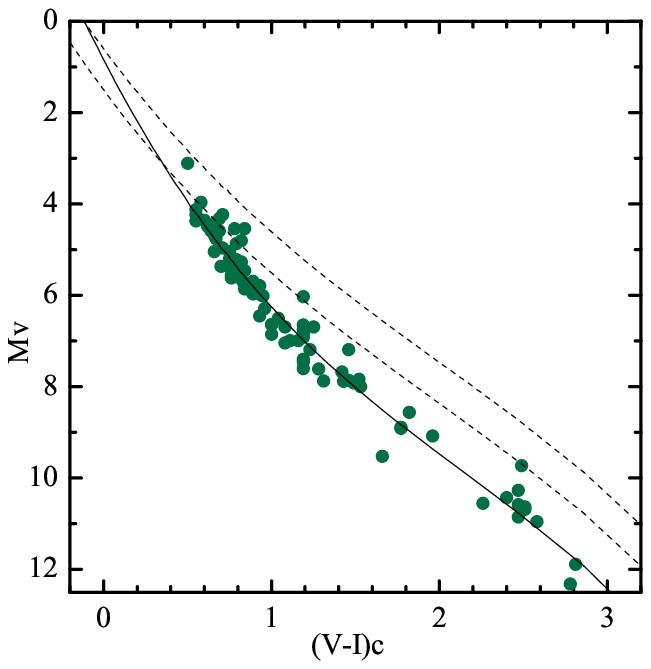}
\includegraphics[draft=False,width=0.48\textwidth]{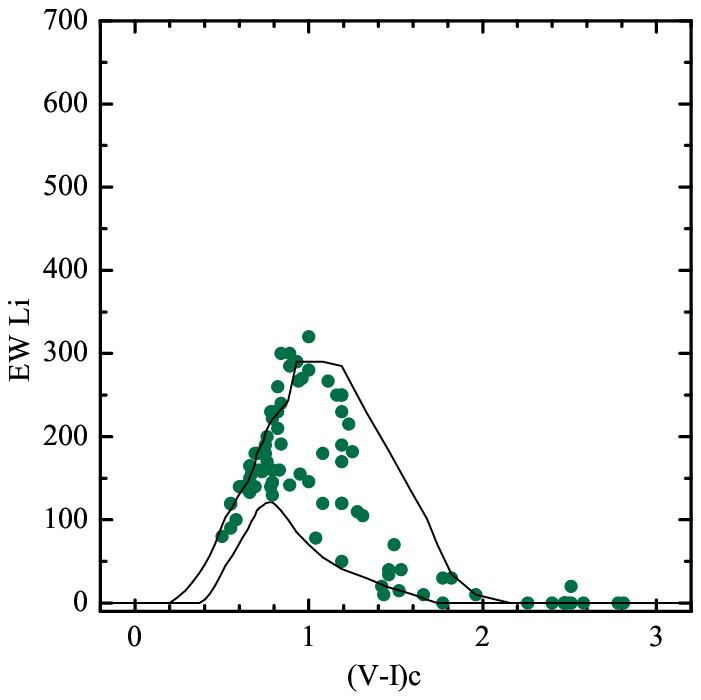}
\caption{{\it Left:} The HR diagram of the members proposed for the AB~Dor
Association; the over-plotted isochrones
are the ones for 5, 10, and 70~Myr.
{\it Right:} The distribution of Li equivalent width as a function of $(V-I)_C$; the
curves are the upper and lower limits for the age of the Pleiades \citep{neu97}.
}
\label{fig:abhr}

\end{figure}

Not only the star AB~Dor, but also many other stars in this association have been proposed as RS~CVn or
BY Dra variables. This can be noted by the names in Table~\ref{table:aba}.
As, in general, they are fast rotators, the spread of the
radial velocity values may indicate a spectroscopic binary nature, but few were confirmed.
As they are young fast rotators, they have strong X-ray emission and spots,
having  therefore been confused with the true RS~CVn variables.
In fact, all are spotted variable stars and they are important for rotation and corona studies.
In the AB~Dor Association there are 15 of this kind of variables:
PW~And, IS~Eri, V577~Per, AB~Dor,  UX~Col, UY~Pic, WX~Col, TZ~Col, TY~Col, AK~Pic, BD+20~1790,  V372~Pup,
PX~Vir, HD~160934, LO~Peg.
%PW~And (P=1.76, vsini=22.6, R=0.78), IS~Eri (P=5.41, vsini=7.2, R=0.77), V577~Per (P=1.45, vsini=7, R=0.2),
%AB~Dor (P=0.51, vsini=53, R=0.53), UX~Col (P=2.29, vsini=41.5, R=1.78) UY~Pic (P=4.52, vsini=9, R=0.80),
%WX~Col (P= ?, vsini=5.5), TZ~Col (P=2.27, vsini=20, R=0.89), TY~Col (P=3.82, vsini=55.0, R=4.1),
%AK~Pic (P=2.60, vsini=17.6, R=0.90), BD+20 1790 (P=2.78, vsini=13, R=0.71),
%V372~Pup (P=1.64, vsini=20.5, R=0.66), PX~Vir (P=6.53d, vsini=6, R=0.77),
%HD~160934 (P=1.84, vsini=17, R=0.61), LO~Peg (P=0.42, vsini=65.8, R=0.54).

HD~33999 is a close visual binary (sep.=0.6$\arcsec$), not resolved by our spectrograph.  The spectrum exhibits three features in the
cross-correlation function.  We interpret the stronger feature as
coming from the hotter component, and the other two lines
as coming from the fainter star, forming a double line spectroscopic
binary.

UX Col had its photometric variations observed by \citet{cutispoto03}
who determined a period of 2.29~days.  \citet{torres06} obtained a
vsin(i)~= 41.5\,km~s$^{-1}$, similar to the one found by
\citet{tagliaferri94}.  This would imply a minimum radius of
1.9~R$_\odot$ which is substantial for a K3V star at the age of the
AB~Dor Association. This could be resolved if one assumes
that the true rotational period is actually a shorter alias period.

TY Col was observed by \citet{cutispoto01} who
found a period of 3.82~days.  \citet{torres06} obtained a
vsin(i)~=~55.0\,km~s$^{-1}$, similar to the one found by
\citet{tagliaferri94}.  As for UX~Col, this would imply a minimum
radius of 4.1~$R_{\odot}$, incompatible with the age of 70\,Myr .
Again, the solution could be a short rotational period.  Our spectral
classification is similar to that of \citet{cutispoto98} who discusses
these difficulties.

GSC~8894-0426 was proposed by \citet{zuckerman04} as a member of the
AB~Dor Association, but it has no astrometric data.  It is separated
by 27.2$\arcmin$ from HD~45270 (itself a wide binary, the companion
being an active late-K star \citep{cutispoto02}).  If they are at the
same distance, this would imply a separation of 0.2~pc.  Using the
HD~45270 Hipparcos astrometric data, GSC~8894-0426 becomes a bona fide
member.

AK~Pic (HD~48189) has a companion at 0.8$\arcsec$.
This separation has changed in time, but it has no orbit determined yet.
Proper motions of Hipparcos and TYCHO-2 disagree in RA and
Hipparcos gives distinct RA proper motions for the components, thus we used the TYCHO-2 values.
The radial velocities measured by different authors show a spread greater than the errors
indicating a single line spectroscopic binary star.
We used a compromise among the published values (30~km~s$^{-1}$).

HD~99877 lies in front of the Cha region and the difference between
the radial velocity measured by \citet{covino97} and by \citet{nord04}
is larger than the observational errors, indicating a possible spectroscopic
binary.

PX~Vir may also be a single line spectroscopic binary, given the
spread of the radial velocity in the literature.  In fact Hipparcos
found an astrometric orbit, with P=231~d, and a semi-major axis of
11~mas, thus explaining this spread in radial velocities.  We used a
compromise value of 0~km~s$^{-1}$.  It is also a spotted variable star
with P=6.5~days \citep{gaidos00, strass00}.

HD~160934 is another single line spectroscopic binary with very long
period \citep{galvez06}.  It was optically resolved at 0.2$\arcsec$ by
\citet{hormuth07} who suggest a period of $\sim$8.5~years from both
spectroscopic and visual data.  This system has also a dM4e companion
at 19.1$\arcsec$.

HD~217379 is a visual binary (sep.=1.8$\arcsec$), not separated in our spectrum.
The spectrum shows a triple line system, formed by one K5V and two K7V stars.
The brighter visual component seems to be the double line spectroscopic K7V binary.
We used the velocity of the K5V component as the systemic velocity, supposing
that the velocity of the visual spectroscopic single secondary is near the systemic one.

HD~218860 had its red companion recognized for the first time in the
SACY.  The stars have similar radial velocities, and for the
red star we used the proper motions of the primary.

The  AB~Dor Association has 25 known visual binaries (two are triples and one is quadruple),
presented in Table~\ref{table:vbab}.
There are also in the AB~Dor Association two triple line spectroscopic multiples
(HD~33999  and HD~217379),
three single line spectroscopic binaries (AK~Pic, PX~Vir and HD~160934) and
three stars  are possible spectroscopic binaries (CD-26~2425, TZ~Col and HD~99827).

\begin{table}[]
\caption{The visual binaries in AB Dor Association}
\smallskip
\begin{center}
{
\label{table:vbab}
\begin{tabular}{lrllll}
\tableline
\noalign{\smallskip}
\hspace{2mm}Name&sep.   &\hspace{6mm}Name    &sep. &  \hspace{6mm}Name& sep.\\
 &$\arcsec$\,\,\,&       &\hspace{2mm}$\arcsec$\,\,\,&   &\hspace{2mm}$\arcsec$\\
\noalign{\smallskip}
\tableline
\noalign{\smallskip}
HD 4277& 3.8 &   \hspace{4mm}AB Dor AC&    \hspace{2mm}0.2& \hspace{4mm}HD 99827&  \hspace{2mm}3.5\\
CD-46 644&21.7&  \hspace{4mm}AB Dor BaBb& \hspace{2mm}0.06& \hspace{4mm}PX Vir  & \hspace{2mm}0.01\\
HD 13482& 1.8&   \hspace{4mm}UY Pic&      10.3&             \hspace{4mm}HD 139751&\hspace{2mm}0.9\\
HD 16760& 14.6&  \hspace{4mm}WX Col&       \hspace{2mm}3.9& \hspace{4mm}HD 160934 AB&\hspace{2mm}0.2\\
HD 17332&  3.6&  \hspace{4mm}HD 45270  &   16.2&            \hspace{4mm}HD 160934 AC&19.1\\
HIP 14809&33.2&  \hspace{4mm}AK Pic  &     \hspace{2mm}0.8& \hspace{4mm}HD 176367&11.2\\
V577 Per& 9.5&   \hspace{4mm}HD 59169  &    \hspace{2mm}1.2& \hspace{4mm}HIP 110526& \hspace{2mm}1.8\\
CD-56 1032& 7.8& \hspace{4mm}V372 Pup AB &  \hspace{2mm}0.2& \hspace{4mm}HD 217379&\hspace{2mm}1.8\\
HD 33999&   0.7& \hspace{4mm}V372 Pup AC&   \hspace{2mm}6.6& \hspace{4mm}HD 218860&19.6\\
AB Dor AB& 9.0&  \hspace{4mm}HD 64982&     \hspace{2mm}5.7&&\\
\noalign{\smallskip}
\tableline
\noalign{\smallskip}
\end{tabular}
%\smallskip
}
\end{center}
\end{table}

\section{Disks and Sub-stellar Objects}

The discovery of 51~Peg~b by \cite{mayorqueloz95} is without doubt
among the scientific highlights of the last century. It triggered an
incredible number of observational and theoretical studies aiming to
find new worlds and understand their physics.  In relation to the
question of planetary system formation, it appeared already that, in
the light of the discovery of 51~Peg~b, the relevant timescales (disk
evolution, planet formation, and migration) are all roughly of the
same order of magnitude, i.e., $\sim10$~Myr \citep[e.g.][]
{pollack96,idalin04,alibert05,haisch01}.  These timescales predicted
by the models depend on the physical properties of the circumstellar
disks, such as density profile, composition, grain size, etc.
\citep[e.g.][]{hub04}, which in turn are themselves not well known.
Clearly, a better understanding of disk properties and how they evolve
in time is of paramount importance to planetary formation theories.

The link of the young, nearby  associations to planetary system research is
evident. Most of these
associations are within an age range where disks are quickly evolving from a
rather massive, accreting,
gas-rich disk as found around TTS towards more quiescent, cold, debris  disk
where grain coagulation and re-processing, vertical settling, and planetesimal
formation are thought to have
occurred and large gaps may have opened as a result of the formation of planets.
The literature about this topic is vast, and the
{\em Spitzer Space Observatory} launched in 2003 continues to contribute in this
field, e.g. via the key program ``from disks to planets''
\citep[for a review see, e.g.,][]{werner06}.
In the following we focus on recent results  that concern the young nearby
associations discussed here.
For  a broader view on disk evolution we refer to recent reviews of
\cite{Hillenbrand05} and \cite{Hartmann05}.

A number of near (JHK) to mid-infrared (LMN) studies have been carried out during the  last
years in order to probe disk frequencies and dispersal times for stars with different ages.
Based on the extrapolation of their relation between frequency of  stars showing
JHKL--excess and age,
\cite{Haisch01} conclude that no excess is expected beyond 6~Myr. In  other words,
the inner region ($<$1~AU) of primordial hot and gas-rich circumstellar  disks
are thought to have dissipated after $\sim$6~Myr.

%Indeed
\cite{ray99} \citep[for a summary, see also][]{ray01} conducted a
mid-IR survey of the TW~Hya Association members.  They find that most
of the TW~Hya Association stars have little or no disk emission at
10~$\mu$m.  Among those showing some 10~$\mu$m emission, gaps in disks
appear to be established, suggesting that critical disk evolution has
taken place already at the age of the TW~Hya Association (see below).
\cite{mamajek04} present an N-band survey of 14 young stars in the
$\sim$30~Myr old Tuc-Hor Association to search for evidence of warm,
circumstellar dust disks.  They find that none of the stars have a
statistically significant N-band excess compared to the predicted
stellar photospheric flux, corroborating the notion that at this age,
warm thick disks have already dissipated.  In contrast to the TW~Hya
Association or the Tuc-Hor Association where few or no disks were
detected, the $\eta$~Cha cluster\footnote{Although throughout this
section we refer to $\eta$~Cha cluster, our analysis presented in
Section~5 suggests that $\eta$~Cha cluster belongs to what we call the
$\epsilon$~Cha Association.}  seems to be an interesting exception to
the disk frequency-age relation found by \cite{Haisch01}.
\cite{lyo03} found that 60\% of the observed stars have near infrared
excess.  According to \cite{lawson01b}, the age of the $\eta$~Cha
cluster is about 9~Myr.  The direct implication is that stars in the
$\eta$~Cha cluster have retained their primordial disks longer than
the observed trend.  Clearly, the age of the $\eta$~Cha cluster plays
a key role in the discussion. \cite {Haisch05} have recently revisited
the issue.  Their disk frequency is much smaller than that reported by
\cite{lyo03}, 28\%$\pm$13 versus 60\%$\pm$13\%.  This revised disk
frequency along with the age estimate of \cite{luhmans04} which gives
6~Myr (similar to what we found in Section~5) instead of 9~Myr puts
the $\eta$~Cha cluster back into the linear relation of
\cite{Haisch01}.  \cite{Haisch05}, on the other hand, do not exclude
that the $\eta$~Cha cluster is indeed 9~Myr old. In this case, after a
rapid decline during which the dust in the inner disk is dissipated or
accumulates into larger bodies, the disk fraction in clusters would
decrease more slowly, with a small number of stars ($ \sim$10\%)
retaining their disks for times comparable to the cluster age.
\cite{Mageath05} presented {\em Spitzer} observations which show that
one member of the $\eta$~Cha cluster has an excess similar to a cTTS
(ET~Cha), whereas five other members show only weak IR-excesses.
Interestingly, H$\alpha$ spectroscopic data suggest that among these
six stars, five are accreting, although at a much smaller rate than in
the T~Tauri regime.  The authors suggest that, like in the TW~Hya
Association, we are witnessing a fast transition from a cTTS like disk
to a debris disk.
%For a broader discussion on life-time of dust disks we refer to \cite{Hillenbrand05}.

One of the fundamental motivations to determine disk lifetimes is to
assess the amount of molecular gas available and the time-span
available in which giant planets could be formed. Although a direct
measurement of gas densities and masses is difficult to obtain
\citep{Richter2002}, it can be inferred from accretion signatures.
Using mainly the equivalent width of H$\alpha$ as an accretion
diagnostic, \cite{Ray06} estimate the frequency of accretors in the
$\eta$~Cha cluster, the TW~Hya, the $\beta$~Pic, and the Tuc-Hor
associations.  They find that three out of 11 ($\sim27$\%) late-type
stars in the $\eta$~Cha cluster are accreting, whereas only two of the
32 targets ($\sim6$\%) in the TW~Hya Association show evidence for
accretion.  None of the $\beta$~Pic and the Tuc-Hor Association
members show these signatures.  \cite{Ray06} infer an inner disk
lifetime of about 10~Myr.  Moreover, the notion of long-lived
primordial disks seems weakened since no accreting star was present in
either the $\beta$~Pic Association (12~Myr) or the Tuc-Hor Association
(30~Myr).  These results have been recently corroborated by
\citet[][see also reference therein]{rebull08} who found an
“inside-out” infrared excess reduction with
time, wherein the shorter-wavelength excesses disappear before
longer-wavelength excesses.  Such a decrease is consistent with the
overall decrease of disk frequency with stellar age.  Moreover,
optically thick disks, seen in the younger TW~Hya Association and the
$\eta$~Cha cluster, are entirely absent in the $\beta$~Pic
Association.

Most stars in the nearby young associations studied here show clear
signs that their primordial disks have already evolved into debris
disks.  A debris disk consists of a mixture of smaller and larger
grains, and larger bodies like planetesimals required for the
planetary formation processes.  In these disks, dust grains are
continuously regenerated by collisions and/or evaporation of
planetesimals.  This dust absorbs stellar radiation at visual
wavelengths and re-radiates the energy at infrared to submillimeter
wavelengths.  It is the large emitting surface area of these numerous
grains that makes debris disks around stars observable in the infrared
and submillimeter, while the mass-dominant planetesimals remain
undetected.

%(Figure~\ref{AUMic}).

\begin{figure}[]
\centering
\includegraphics[draft=False,width=0.92\textwidth]{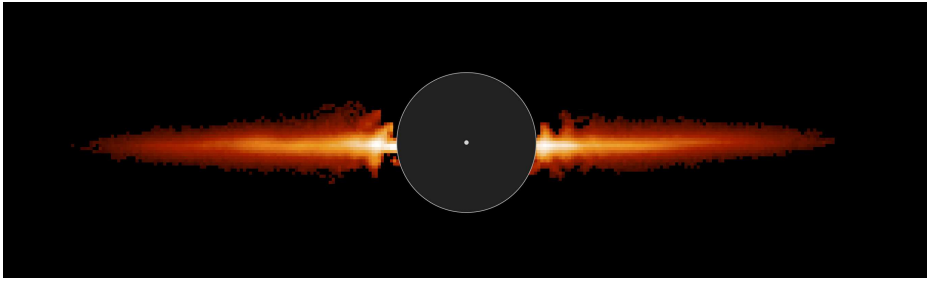}
\caption{The debris disk around AU Mic is seen in this image obtained with the HST. Courtesy STScI.}
\label{aumic}
\end{figure}

Due to the limitation imposed by Earth's atmosphere, space-based
facilities like HST, IRAS, ISO and {\it Spitzer} are primary sources
of information for the study of debris disks (Figure~\ref{aumic}).  A
number of studies have correlated the far infra-red fluxes obtained by
these facilities with catalogs of nearby stars
\citep[e.g.,][]{chen05,moor06,rhee07}.  The main goal is to make a
census of the debris disks around nearby stars of different spectral
types and to understand how these disks evolve in time.  These studies
suffer from the difficulty of determining stellar ages for field
stars.  \cite{moor06} suggest to take advantage of the fact that
several nearby stars are members of young nearby associations to
improve age estimates for the debris disks.  They found that the most
prominent debris disks are found around members of the loose
associations.  Using the age estimates for the associations, they also
found a moderate agreement between the theoretical predictions of the
evolution of the fractional luminosity as function of the age.  In the
same line, \citet{rhee07b} estimate that 13$\%$ of the stars with
spectral type earlier than G0 belonging to one of the nearby young
associations show evidence of possessing a warm ($\ga$~150K) mid-IR
excess related to debris-disks or a mix of debris-disk and
primordial-disk material.

Since the debris disks are generated by collisions of planetesimals,
usually the presence of (undetected) planets is inferred from the  presence of a
debris disk itself.
The existence of planets in the debris disk of Vega-like stars has  been
suggested based on the non-axissymmetric structures observed in many of these disks.
Such  structures are thought to be caused by
the gravitational perturbation of massive bodies (Jupiter and brown  dwarf
masses) orbiting around
the host star \citep[e.g.][and references therein]{mac03,zuck04,oka04}.

So far, attempts to find the perturbing body around stars with debris
disks via adaptive optics systems have failed \citep{mac03,zuck04}.
Concerning the nearby young associations, debris disks have been
detected in a few members of the $\beta$~Pic Association, the TW~Hya
Association and other nearby stars
\citep[][]{zuckerman01a,zuck04,zuckerman04}.  In the TW~Hya
Association, the 10~$\mu$m study carried out by \cite{ray99} showed
that most of the TW~Hya Association stars have little or no disk
emission at 10~$\mu$m.  \cite{low05} using {\em Spitzer} showed that
even at 24~$\mu$m most of the stars show no excess. This implies the
absence of dust warmer than 100~K.  For four other objects (TW~Hya,
TWA~3, TWA~4, and TWA~11A) the excess at 24~$\mu$m is, however, a
factor of $\sim$100 above the photosphere in a clear bimodal
distribution (Figure 1 of Low et al.).  A similar behavior is also
seen at 70~$\mu$m, although less marked than at 24~$\mu$m.
Remarkably, mid-infrared images of the disk around $\beta$~Pic itself
reveal brightness asymmetries that can be interpreted as resulting
from a cataclysmic break-up of planetesimals \citep{telesco05}.

Again, as for the $\eta$~Cha cluster, it seems that disks evolve fast during
this age, and thus a mix of disk characteristics is seen.
The most interesting case is perhaps TW~Hya itself which looks
like a cTTS \citep[e.g.][]{RucinskiKrautter1983, delareza89, gregorio92} but
whose disk shows signs of evolution \citep[]
{krist00,weinberger02,weinberger04,wilner05,low05}.
\cite{wilner05} in particular, based on VLA observations
at $\lambda=3.5$ cm, suggest that the emission observed at this  wavelength can
only be explained if
planetesimals of centimeter size have already been formed in the disk  of TW~Hya.

Direct imaging and spectroscopy of Jupiter mass and brown dwarf  objects also benefit from the young age
and proximity of the associations discussed here since atmospheres of  very low mass objects look
brighter at near-IR wavelength \citep[e.g.][]{burrows97} and  therefore can be imaged by current
adaptive optics systems. For instance, an object of 5~$M_J$ can be  detected around a K0V star at
separations larger than 0.7$\arcsec$ \citep{masciadri05}.
Not surprisingly, \cite{chauvin04,chauvin05a} imaged the first extra-solar planet around a brown-dwarf
member of the TW~Hya Association.
A second case (bearing IAU definition of what is a planet) of a sub-stellar object of 13~$M_J$ was
reported by \cite{chauvin05c} at $\sim260$~AU from AB~Pic, a  member proposed of the  Car Association.

It is interesting to note that \cite{masciadri05}, using adaptive optics
techniques, searched for other giant planets around members of the Tuc-Hor, the $\beta$~Pic and
the TW~Hya associations and young field stars.
They report a null result.
\cite{kasper07} used L-band adaptive optics-assisted imaging to look for
planets around 22 members of Tuc-Hor and $\beta$~Pic associations.
Their observations were sensitive to companions with masses down to 1 to 2~M$_J$
at separations larger than 5 to 30~AU. In spite of the unprecedented sensitivity, no sub-stellar
companions were found.

It is an open question whether the low number of sub-stellar detections beyond
5~AU reported by the direct imaging surveys is real or due to an
overestimated sensitivity related to uncertainties
in the predicted luminosities at early ages \citep{marley07}.
In spite of this, it is expected that the next generation adaptive optics systems
(i.e., the planet finders) will be able to probe fainter objects  closer to the central star
($\Delta M \sim17.5$ at 0.5$\arcsec$). Thus the members of the nearby  young associations described
here are still prime candidates to these future adaptive optics surveys.

Radial velocity surveys usually do not target young stars ($<$10~Myr
or so).  This is due to the fact that accretion, high rotation and
chromospheric activity affect a great deal the quality of the radial
velocity, thus hampering any attempt to reach the few m/s precision
needed to find Jovian planets.  In spite of these difficulties, a few
groups have been trying to conduct radial velocity surveys around
young stars.  \citet{setiawan08} announced the discovery of a massive
(10~$M_J$) hot-Jupiter around TW~Hydrae.  However, the results of
\citet{nuria08b} cast doubts about the existence of TW~Hya's
hot-Jupiter. Both results, in spite of being contradictory, are
exciting and will certainly trigger a number of studies in the optical
and in the near-infrared.  The discovery of a population of
hot-Jupiters already at a few Myr would lead to strong constraints on
their formation and evolutionary time-scales.

\vspace{0.5cm}

{\bf Acknowledgements.}
C. A. O. Torres was an ESO Visiting Scientist and
thanks the CNPq Brazilian Agency for the grant 200256/02.0.
We thank Ramiro de la Reza, Licio da Silva and Nuria  Hu{\'e}lamo
for discussions and suggestions, and Subu Mohanty for useful comments.
We thank the Centre de Donn\'{e}es Astronomiques de Strasbourg (CDS), the
U. S. Naval Observatory and NASA for the use of their electronic facilities,
specially SIMBAD, the Washington Double Star Catalog and  ADS.

%%% THE BIBLIOGRAPHY
%%%
%%% CONSULT SECTION 3 OF "INSTRUCTIONS FOR AUTHORS" FOR HOW TO USE NATBIB.
%%% AUTHORS ARE ENCOURAGED TO USE EITHER THE "THEBIBLIOGRAPY" ENVIRONMENT
%%% BY UNCOMMENTING (DELETING THE "%" SYMBOL) THE COMMANDS BELOW, OR BY
%%% USING THE BIBTEX ENVIRONMENT. TO FIND OUT WHICH IS APPLICABLE TO YOUR
%%% CONTRIBUTION, CONSULT THE VOLUME EDITORS FOR YOUR PROCEEDINGS.
%%%

%\bibliographystyle{natbib}
%\bibliography{mybib}

\end{document}